\documentclass[11pt,a4paper]{emulateapj}
\usepackage{graphicx}
\usepackage{textcomp}
\usepackage{amsmath}
\usepackage{amssymb}
\usepackage{gensymb}
\usepackage{rotating}
\usepackage{epsfig}
\usepackage{amsmath}
\usepackage{longtable}
\usepackage{color}
 \def\apjl{ApJL}

 \def\aap{A\&A}
 
 \def\apjs{ApJS}

 \def\gs{\mathrel{\raise0.35ex\hbox{$\scriptstyle >$}\kern-0.6em\lower0.40ex\hbox{{$\scriptstyle \sim$}}}}
 \def\ls{\mathrel{\raise0.35ex\hbox{$\scriptstyle <$}\kern-0.6em\lower0.40ex\hbox{{$\scriptstyle \sim$}}}}

 \def\Lsol{\mathrel{\rm L_{\odot}}}
 
 \def\Wm2{\,\hbox{W}\,\hbox{m}^{-2}}
 \def\gsim{\mathrel{\raise0.35ex\hbox{$\scriptstyle >$}\kern-0.6em\lower0.40ex\hbox{{$\scriptstyle \sim$}}}}
 \def\lsim{\mathrel{\raise0.35ex\hbox{$\scriptstyle <$}\kern-0.6em\lower0.40ex\hbox{{$\scriptstyle \sim$}}}}
 
 \def\pc{\%}

\lefthead{Simpson et al.}  \righthead{Photometric redshift distribution of ALESS SMGs}

\begin{document}

\title{An ALMA Survey of Submillimeter Galaxies in the Extended Chandra Deep Field South: The Redshift Distribution and Evolution of Submillimeter Galaxies}

\author{
J.\,M.\ Simpson,\altaffilmark{1}
A.\,M.\ Swinbank,\altaffilmark{1}
Ian Smail,\altaffilmark{1}
D.\,M\ Alexander,\altaffilmark{1}
W.\,N.\ Brandt,\altaffilmark{2,3}
F.\ Bertoldi,\altaffilmark{4}
C.\ de Breuck,\altaffilmark{5}
S.\,C.\ Chapman,\altaffilmark{6}
K.\,E.\,K.\ Coppin,\altaffilmark{7}
E.\ da Cunha,\altaffilmark{8}
A.\,L.\,R.\ Danielson,\altaffilmark{1}
H.\ Dannerbauer,\altaffilmark{9}
T.\,R.\ Greve,\altaffilmark{10}
J.\,A.\ Hodge,\altaffilmark{8}
R.\,J.\ Ivison,\altaffilmark{11}
A.\ Karim,\altaffilmark{4}
K.\,K.\ Knudsen,\altaffilmark{12}
B.\,M.\ Poggianti,\altaffilmark{13}
E.\ Schinnerer,\altaffilmark{8}
A.\,P.\ Thomson,\altaffilmark{1}
F.\ Walter,\altaffilmark{8}
J.\,L.\ Wardlow,\altaffilmark{14,15}
A.\ Wei${\ss}$,\altaffilmark{16}
P.\,P.\ van der Werf,\altaffilmark{17}}

\setcounter{footnote}{0}
\altaffiltext{1}{Institute for Computational Cosmology, Department of Physics, Durham University, South Road, Durham DH1 3LE, UK; email: j.m.simpson@dur.ac.uk}
\altaffiltext{2}{Department of Astronomy \& Astrophysics, 525 Davey Lab, The Pennsylvania State University, University Park, PA16802, USA}
\altaffiltext{3}{Institute for Gravitation and the Cosmos, The Pennsylvania State University, University Park, PA 16802, USA}
\altaffiltext{4}{Argelander-Institute for Astronomy, Bonn University, Auf dem H{\"u}gel 71, D-53121 Bonn, Germany}
\altaffiltext{5}{European Southern Observatory, Karl-Schwarzschild Stra{\ss}e, D-85748 Garching bei M{\"u}nchen, Germany}
\altaffiltext{6}{Department of Physics and Atmospheric Science, Dalhousie University, Halifax, NS B3H 3J5 Canada}
\altaffiltext{7}{Centre for Astrophysics Research, Science and Technology Research Institute, University of Hertfordshire, Hatfield AL10 9AB, UK}
\altaffiltext{8}{Max-Planck Institute for Astronomy, K{\"o}nigstuhl 17, D-69117 Heidelberg, Germany}
\altaffiltext{9}{Universit{\"a}t Wien, Institut f{\"u}r Astrophysik, T{\"u}rkenschanzstra{\ss}e 17, 1180 Wien, Austria}
\altaffiltext{10}{Department of Physics and Astronomy, University College London, Gower Street, London WC1E 6BT, UK} 
\altaffiltext{11}{Institute for Astronomy, University of Edinburgh, Blackford Hill, Edinburgh EH9 3HJ}
\altaffiltext{12}{Department of Earth and Space Science, Onsala Space Observatory, Chalmers University of Technology, SE-43992 Onsala, Sweden}
\altaffiltext{13}{INAF-Astronomical Observatory of Padova, I-35122 Padova, Italy}
\altaffiltext{14}{Department of Physics \& Astronomy, University of California, Irvine, CA 92697, USA}
\altaffiltext{15}{Dark Cosmology Centre, Niels Bohr Institute, University of Copenhagen, Denmark}
\altaffiltext{16}{Max-Planck-Institut f{\"u}r Radioastronomie, Auf dem H{\"u}gel 69, D-53121 Bonn, Germany}
\altaffiltext{17}{Leiden Observatory, Leiden University, P.O. Box 9513, 2300 RA Leiden, The Netherlands}
\begin{abstract} 
We present the first photometric redshift distribution for a large sample of 870\,$\mu$m SMGs with robust identifications based on observations with ALMA. In our analysis we consider 96 SMGs in the ECDFS, 77 of which have 4--19 band photometry. We model the SEDs for these 77 SMGs, deriving a median photometric redshift of $z_{phot}\,=\,2.3\,\pm\,0.1$. The remaining 19 SMGs have insufficient photometry to derive photometric redshifts, but a stacking analysis of {\it Herschel} observations confirms they are not spurious. Assuming that these SMGs have an absolute $H$--band magnitude distribution comparable to that of a complete sample of $z$\,$\sim$\,1--2 SMGs, we demonstrate that they lie at slightly higher redshifts, raising the median redshift for SMGs to $z_{phot}\,=\,2.5\,\pm\,0.2$. Critically we show that the proportion of galaxies undergoing an SMG-like phase at $z\,\ge\,3$ is at most $35\,\pm\,5$ per cent of the total population. We derive a median stellar mass of $M_{\star}\,=\,(8\,\pm\,1)\,\times\,10^{10}$\,M$_{\odot}$, although there are systematic uncertainties of up to 5\,$\times$ for individual sources. Assuming that the star formation activity in SMGs has a timescale of $\sim$\,$100$\,Myr we show that their descendants at $z$\,$\sim$\,$0$ would have a space density and $M_{H}$ distribution which are in good agreement with those of local ellipticals. In addition the inferred mass-weighted ages of the local ellipticals broadly agree with the look-back times of the SMG events. Taken together, these results are consistent with a simple model that identifies SMGs as events which form most of the stars seen in the majority of luminous elliptical galaxies at the present day.

\end{abstract}


\section{Introduction}\label{sec:intro}
%
%
\begin{figure*}
 \centerline{ \psfig{figure=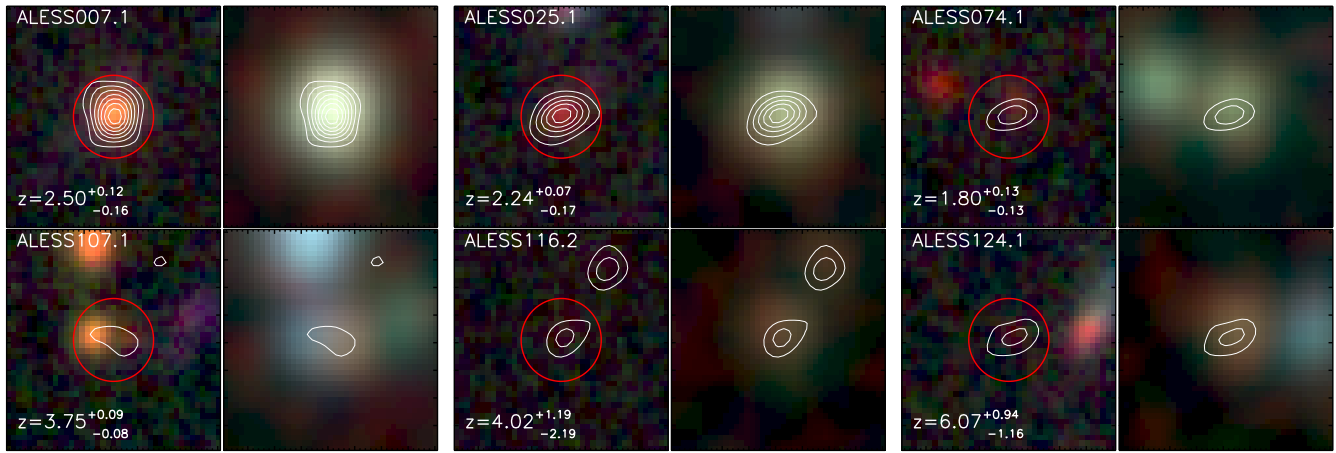,width=0.99\textwidth}}
\caption{ In our analysis we make use of extensive archival imaging of the ECDFS, covering 19 optical through near-infrared wave-bands (see Table~\ref{table:depths}). Here we present $8'' \times 8''$ optical (coadded $B$, $I$ and $K_S$; left) and near-infrared (coadded 3.6, 4.5 and 8.0\,$\mu$m; right) false color images for six example SMGs, from our sample of 96 SMGs, which span the full range of 870\,$\mu$m flux (thumbnails of all sources are shown in Appendix~B). Contours represent ALMA 870\,$\mu$m detections at 3, 5, 7\,....\,$\times \sigma$. The ALMA observations pinpoint the 870\,$\mu$m emission to $<$\,0.3$''$ precision, identifying the optical/near-infrared counterpart without the requirement for indirect statistical associations. Using the precise locations we measure aperture photometry for each SMG (the size of the aperture is indicated with a red circle) and derive photometric redshifts using the SED-fitting code {\sc Hyperz}, see \S\,\ref{sec:photz}.
}
 \label{fig:subimages}
\end{figure*}

In the local Universe $\sim$\,$60$ per cent of the total stellar mass is in early-type and elliptical galaxies~\citep{Bell03}. These galaxies lie on a tight ``red--sequence''~\citep{Sandage78,Bower92,Blanton03}; follow well-defined scaling relations (the fundamental plane); and show correlations between the age, and velocity dispersion ($\sigma$) of their stellar population. Typically, the most massive ellipticals have velocity dispersions of $\sigma$\,$\sim$\,$200$--400\,km\,s$^{-1}$, with estimated luminosity-weighted stellar ages of $\sim 10-13$\,Gyr~\citep{Nelan05}. Recently, near--infrared spectroscopy of quiescent, red, galaxies at $z$\,$\sim$\,1.5--2, the potential progenitors of elliptical galaxies (see\,\citealt{vanDokkum04}), has suggested the stellar populations in these galaxies have a typical age of $\sim$\,$1$--2\,Gyr (e.g.\,\citealt{Whitaker13,Bedregal13}). Taken together, these results suggest that the bulk of the stellar mass in elliptical galaxies formed early in the history of the Universe, at redshifts $z>2$. However it has proved challenging to study the progenitors of these galaxies as the most massive, star forming galaxies at $z >2$ are also the most dust obscured~\citep{dole04,lefloch09}. One route to uncovering these dusty starbursts is to search at submillimeter (submm) wavelengths, where the shape of the spectral energy distribution (SED) of the far-infrared (FIR) dust emission means that cosmological fading is negated by the strongly increasing flux density of the SED. For sources at a fixed luminosity, this ``negative k-correction'' results in an almost constant apparent flux density in the submm over the redshift range $z$\,=\,0.5--7 (see the review by~\citealt{Blain02}).

The earliest surveys aimed at searching for distant submillimeter galaxies (SMGs), particularly with the SCUBA camera on the James Clerk Maxwell Telescope (JCMT), uncovered moderate numbers of submm sources with 850-$\mu$m flux densities of $S_{\nu}$\,=\,5--15\,mJy (e.g.~\citealt{Smail97,Hughes98,Barger98,Eales99,Coppin06}). However, the coarse beam size of single dish submm telescopes ($\sim$\,$15''$ for the JCMT at 850$\mu$m) meant that resolving these submm sources into their constituent SMGs (and so determining their basic properties, such as redshift and luminosity) was impossible without significant assumptions about the properties of their multi-wavelength counterparts.  For example, the correlation between the far-infrared and radio flux density of star-forming galaxies could be employed (e.g.~\citealt{Ivison98,Ivison00}), as deep 1.4\,GHz radio imaging with the Very Large Array (VLA) provides the sub-arcsecond resolution required to accurately locate the counterpart to the submm emission~\citep{Ivison02,Ivison04,Ivison07,Bertoldi07,Biggs11,Lindner11}. However in typical surveys, radio imaging only identifies $\sim$\,50--60 per cent of the SMGs brighter than $S_{\rm 850\mu m} >$\,5\,mJy and furthermore is expected to miss the counterparts of the most distant SMGs due to the disadvantageous radio k-correction. Despite this low identification rate, this technique has facilitated extensive follow-up of the counterparts of SMGs, and spectroscopy has shown that the radio-identified subset of the population have a redshift distribution which peaks at $z$\,$\sim$\,2.3~\citep{Chapman05}.  These observations confirmed that SMGs have luminosities comparable to local ultra luminous infrared galaxies (ULIRGs), but crucially demonstrated that the space density of ULIRGs at $z$\,$\sim$\,2 is $\sim$\,1000$\times$ higher than at $z$\,=\,0. With implied star formation rates of 100--1000\,M$_{\odot}$\,yr$^{-1}$, SMGs brighter than 1\,mJy may contribute up to half of the co-moving star-formation rate density at $z$\,$\sim$\,$2$~\citep{Hughes98,Blain99b,Smail02,Wardlow11,Casey13}.

Extensive multi-wavelength follow-up of the radio-identified subset of the SMG population, particularly with the Plateau de Bure Interferometer, measured the kinematic and structural properties of high-redshift SMGs, suggesting that SMGs have morphologies and gas kinematics consistent with merging systems (e.g.\ \citealt{Tacconi08,Engel10,Swinbank10,susie12,Menendez13}).  Moreover, their large molecular gas reservoirs (which comprise $\sim$\,50 per cent of the dynamical mass in the central few kpc;~\citealt{Greve05,Riechers10,Carilli10,Bothwell13}) and star-formation rates mean they have the potential to form a significant proportion of the stars in a $\sim$\,10$^{11}$\,M$_{\odot}$ galaxy in only 10$^8$\,yr.  Taken with their space densities ($\sim$\,10$^{-5}$\,Mpc$^{-3}$;~\citealt{Chapman05,Wardlow11}), large black hole masses ($\sim$\,10$^{8}$\,M$_{\odot}$; \citealt{alexander05a,alexander08}) and clustering (e.g.~\citealt{Hickox12}) it appears likely that, like local ULIRGs, the luminous starbursts in SMGs are frequently triggered by major mergers of gas-rich galaxies (e.g.~\citealt{Ivison12}). 

Comparison with numerical simulations (e.g.~\citealt{Granato04,Dimatteo05,Hopkins06}) also suggests that the starburst SMG phase will be followed by a dust enshrouded AGN phase, which evolves through an optically bright QSO phase before evolving passively into an elliptical galaxy.  Moreover, assuming the timescales for the AGN and QSO phases are short and that SMGs do not undergo significant gas accretion at much lower redshift, it has been shown via simple dynamical arguments that the SMGs can evolve onto the scaling relations observed for local, early-type galaxies at $z$\,=\,0 (e.g.~\citealt{Nelan05,Swinbank06b}). It has thus been speculated that SMGs are the progenitors of local elliptical galaxies~\citep{Lilly99, Genzel03, Blain04a, Swinbank06b, Tacconi08, Hainline11, Hickox12}.

These 850\,$\mu$m-selected samples remain the best-studied SMGs.  However, by necessity, the samples from which most of the follow-up has so far concentrated have been biased to the radio-identified and UV-bright subset of the population where their counterparts and redshifts could be measured. In 2009 we undertook a 310\,hr survey of the 0.5\,$\times$\,0.5 degree Extended {\it Chandra} Deep Field South (ECDFS) at 870$\mu$m, with the LABOCA camera on APEX.  This ``LESS'' survey~\citep{Weiss09} detected 126 submm sources with 870--\,$\mu$m fluxes $S_{870} > 4.4$\,mJy, but still relied on radio and mid-infrared imaging~\citet{Biggs11} to statistically identify probable counterparts to $\sim$\,$60$ per cent of the sources, with the remaining $\sim$\,$40$ per cent remaining unidentified~\citep{Wardlow11}.

To characterize the {\it whole} population of bright SMGs in an unbiased manner, we have subsequently undertaken an ALMA survey of these 126 LESS submm sources. The ALMA data resolve the submm emission into its constituent SMGs, {\it directly} pin-pointing the source(s) responsible for the submm emission to within $<$\,0.3$''$~\citep{Hodge13}, removing the requirement for statistical radio\,/\,mid-IR associations. Crucially, one of the first results from our survey demonstrated that just $\sim$\,70 per cent of the ``robust'' counterparts from~\citet{Biggs11} were correct and that the radio and 24\,$\mu$m identifications only provide $\sim$\,50 per cent completeness~\citep{Hodge13}, highlighting the potential biases in previous surveys (see also~\citealt{Younger09,Barger12,Smolcic12}). These ALMA identifications allow us for the first time to make basic measurements, such as the redshift distribution, for a complete and unbiased sample of SMGs.

In this paper, we exploit the extensive optical and near--infrared imaging of the ECDFS to derive the photometric redshift distribution, stellar mass distribution, and evolution of the ALMA-LESS (ALESS) SMGs. The paper is structured as follows. In \S\,\ref{sec:observations} we present the multi-wavelength data used in our analysis, followed by a description of our method for measuring aperture photometry for the ALESS SMGs, and sources in the field. In \S\,\ref{sec:photz} we discuss the technique of SED fitting to determine photometric redshifts for the ALESS SMGs. Finally, in \S\,\ref{sec:results} we discuss the derived properties of the ALESS SMGs, such as redshift and stellar mass, and their comparison to similar high--redshift studies, concluding with remarks on their comparison to low redshift populations. We give our conclusions in \S\,\ref{sec:conclusions}. Throughout the paper, we adopt a cosmology with $\Omega_{\Lambda}$\,=\,0.73, $\Omega_{\rm m}$\,=\,0.27, and $H_{\rm 0}$\,=\,71\,km\,s$^{-1}$\,Mpc$^{-1}$ and unless otherwise stated, error estimates are from a bootstrap analysis. All magnitudes quoted in this paper are given in the AB magnitude system.

\section{Observations \& Analysis}
\label{sec:observations}

\subsection{Sample Selection}
\label{sec:sample} In this study we undertake a multi-wavelength analysis of the ALMA detected submm galaxies from the catalog presented by~\citet{Hodge13} [see also~\citealt{Karim13}]. To briefly summarize the observations, we obtained 120s integrations of 122 of the original 126 LESS submm sources, initially identified using the LABOCA camera on the APEX telescope~\citep{Weiss09}. These Cycle 0 observations used the compact configuration, yielding a median synthesized beam of $\sim$\,$1.6'' \times 1.2''$. The observing frequency was matched to the original LESS survey, 344\,GHz (Band 7), and we reach a typical RMS across our velocity-integrated maps of $0.4$\,mJy\,beam$^{-1}$. The observations are a factor of $3\,\times$ deeper than LESS, but crucially the angular resolution is increased from $\sim$\,$19''$ to $\sim$\,$1.5''$. The primary beam of ALMA is $\sim$\,$17''$,  which encompasses the original LESS error circles of $\lsim$\,$5''$. For full details of the data reduction and source extraction we refer the reader to~\citet{Hodge13}. From the observations~\citet{Hodge13} construct a {\sc main} source catalog consisting of all detected SMGs obeying the following criteria; primary-beam-corrected map RMS\,$<0.6$\,mJy\,beam$^{-1}$, S/N\,$>3.5$, beam axial ratio\,$<2.0$ and lying within the ALMA primary beam. The resulting catalog contains 99 SMGs, extracted from 88 ALMA maps, which form the basis of the sample used in this paper. The positional uncertainty on each SMG is $<0.3''$. ~\citet{Karim13} demonstrate that the {\sc main} catalog is expected to contain one spurious source, and to have missed one SMG. We remove three SMGs from our sample which lie on the edge of the ECDFS and so only have photometric coverage in two IRAC bands. Our final sample thus consists of 96 SMGs with precise interferometrically-identified positions. 

A supplementary catalog is also provided comprising sources extracted from outside the ALMA primary beam, or in lower quality maps (i.e.\ primary-beam-corrected map RMS $>0.6$\,mJy\,beam$^{-1}$ or axial ratio\,$>2.0$; see~\citealt{Hodge13}). In contrast to the {\sc main} catalog \citet{Karim13} demonstrate that up to $\sim$\,$30$ per cent of the supplementary sources are likely to be spurious, and as such we do not consider them in the main body of this work. However, we present the photometry of these supplementary sources with detections in more than three wave-bands (14 out of 31 sources) in Appendix~C, along with their photometric redshifts and derived properties.

%
%
\begin{table}
\footnotesize
{\footnotesize
\begin{center}
{\centerline{\sc Table 1: Summary of Photometry}}
\vspace{0.1cm}
\resizebox{\columnwidth}{!}{%
\begin{tabular}{lccc}
\hline 
\noalign{\smallskip}
Filter                          & $\lambda$\,$_{\rm effective}$ & Detection limit    & Reference \\
                                  & ($\mu$m)                        & (3\,$\sigma$; AB mag)   &        \\
\hline \\ [-1.ex]                                                                  
MUSYC WFI $U$                   & 0.35                                &    26.2                          &  Taylor et al.\,(2009) \\
MUSYC WFI $U_{38}$               & 0.37                                &    25.3                          &  Taylor et al.\,(2009) \\
VIMOS $U$                          & 0.38                                &    28.1                          &  Nonino et al.\,(2009) \\
MUSYC WFI $B$                   & 0.46                                &    26.5                          &  Taylor et al.\,(2009) \\
MUSYC WFI $V$                   & 0.54                                &    26.3                          &  Taylor et al.\, (2009) \\
MUSYC WFI $R$                   & 0.66                                &    25.5                          &  Taylor et al.\, (2009) \\
MUSYC WFI $I$                    & 0.87                                &    24.7                          &  Taylor et al.\, (2009) \\
MUSYC Mosaic-II $z$          & 0.91                                 &    24.3                          &  Taylor et al.\, (2009) \\
MUSYC ISPI $J$                   & 1.25                                 &    23.2                          &  Taylor et al.\,(2009) \\
HAWK-I $J$                        & 1.26                                 &    24.6                          & Zibetti et al.\,(in prep) \\
TENIS WIRCam $J$              & 1.26                                 &    24.9                          &  Hsieh et al.\,(2012) \\
MUSYC Sofi $H$                  & 1.66                                 &    23.0                          &  Taylor et al.\,(2009) \\
MUSYC ISPI $K$                   &  2.13                                 &    22.4                          &  Taylor et al.\,(2009) \\
HAWK-I $K_s$                      & 2.15                                 &    24.0                          &  Zibetti et al.\,(in prep)  \\
TENIS WIRCam $K_s$            & 2.15                                 &    24.4                          & Hsieh et al.\,(2012) \\
SIMPLE IRAC 3.6\,$\mu$m          &3.58                       &    24.5                          & Damen et al.\,(2011) \\
SIMPLE IRAC 4.5\,$\mu$m          &4.53                       &    24.1                          & Damen et al.\,(2011) \\
SIMPLE IRAC 5.8\,$\mu$m          & 5.79                      &    22.4                          & Damen et al.\,(2011) \\
SIMPLE IRAC 8.0\,$\mu$m          & 8.05                      &    23.4                          & Damen et al.\,(2011) \\
\hline \hline
\end{tabular}}
\vspace{-0.1cm}
\end{center}
\refstepcounter{table}\label{table:depths}
\begin{flushleft}
 \footnotesize{ }
\end{flushleft}
}

\end{table}

\subsection{Optical \& NIR Imaging}
\label{sec:data} The majority of our optical\,--\,near-infrared data comes from the MUltiwavelength Survey by Yale-Chile (MUSYC;~\citealt{Gawiser06}), which provides $U$--$K$-band imaging~\citep{Taylor09} of the entire $0.5 \times 0.5$ degree ECDFS region  (detection limits are given in Table~\,\ref{table:depths}). We supplement this with $U$--band data from the GOODS/VIMOS imaging survey~\citep{Nonino09}, covering $\sim$\,$0.17$\,deg$^2$ of the ECDFS. Although the additional $U$-band imaging only covers $\sim$\,60 per cent of the ALESS SMGs, it is $\sim$\,$2$ magnitudes deeper than the MUSYC $U$-band imaging, and provides a valuable constraint on SMGs undetected in the shallower imaging.
 
In addition, we include deep near-infrared $J$ and $K_{S}$ imaging from both the ESO--VLT/HAWK--I survey by Zibetti et al.\,(in prep) and the Taiwan ECDFS NIR Survey (TENIS;~\citealt{Hsieh12}), taken using CFHT/WIRCAM. Both surveys are $\sim$\,$1.5$--2.0 magnitudes deeper than the MUSYC $J$ or $K_S$ imaging (Table~\ref{table:depths}). We include all three sets of $J$ and $K_S$ imaging in our analysis, however where multiple observations exist we quote, or plot, a single value in order of the detection limit of the original imaging. 

Finally, we include data taken as part of the {\it Spitzer} IRAC/MUSYC Public Legacy in ECDFS (SIMPLE;~\citealt{Damen11}) survey, which provides imaging at 3.6, 4.5, 5.8 and 8.0\,$\mu$m over the entire field. We note that the 5.8\,$\mu$m imaging is $\sim$\,$2$ magnitudes shallower than the other IRAC imaging. 

To highlight the optical\,--\,near-infrared imaging, in Figure~\ref{fig:subimages} we show $BIK_S$ and $3.6/4.5/8.0$\,$\mu$m false color images for six example ALESS SMGs, spanning the full range of ALMA 870\,$\mu$m flux. Figure~\ref{fig:subimages} demonstrates that the SMGs typically have counterparts in the near-infrared, and where detected appear red in the $BIK_S$ color images. The full sample of 96 sources are shown in the Appendix in Figure~B1.

%
%
\begin{figure*}
\centerline{ 
\psfig{figure= 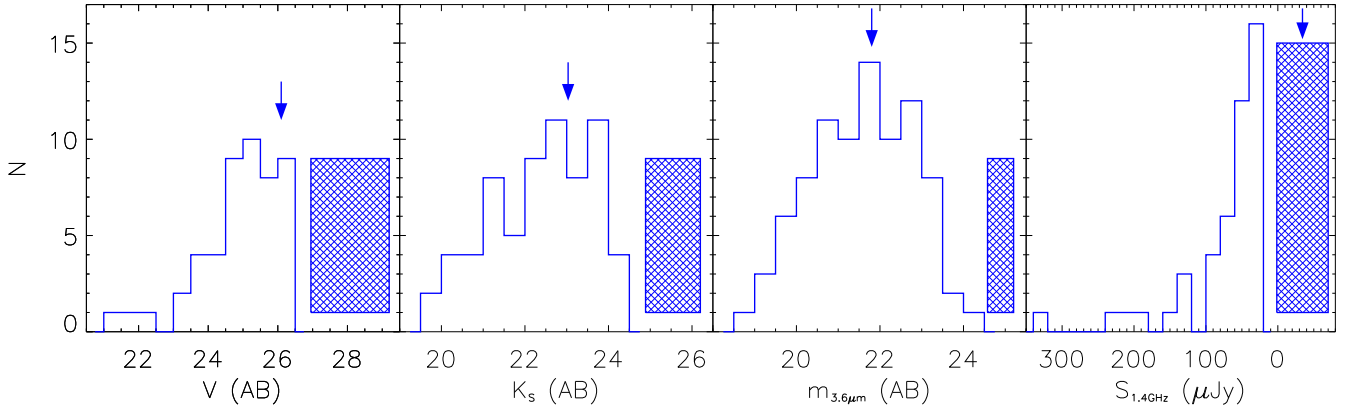,width=0.99\textwidth}
}
\caption{
 The apparent magnitude distributions of ALESS SMGs in the $V$, $K_s$ and IRAC 3.6\,$\mu$m bands, along with the flux density distribution at 1.4\,GHz. On each panel a checked region indicates the undetected sources (see Table 1 for magnitude limits). The median SMG $V$, $K_S$ \& $3.6$\,$\mu$m--band magnitudes, including non-detections, are $V = 26.09\,\pm\,0.19$, $K_{S} = 23.0\,\pm\,0.3$ and $m_{3.6} = 21.80\,\pm\,0.17$ and we mark these on each panel with an arrow. The radio data reaches a depth of 19.5\,$\mu$Jy at its deepest (3\,$\sigma$ detection limit), however only 45 per cent of the ALESS SMGs are detected at this level, and the median 1.4\,GHz flux for SMGs is thus constrained at $\lsim$\,19.5\,$\mu$Jy.
}
 \label{fig:maghist}
\end{figure*}

\subsubsection{Photometry}
\label{subsubsec:alessphot}
To derive photometric redshifts we need to measure seeing-- and aperture--matched multi-band aperture photometry across all 19 filters available (see Table~\ref{table:depths}). First, we align all imaging to the ALMA astrometry. We use {\sc SExtractor}~\citep{Bertin96} to create a source catalog for each image, and match this catalog to the ALESS SMGs. The measured offsets in R.A.\ and Dec.\ are $<0.3''$ in all cases, and correspond to approximately a single pixel shift in the optical imaging, and a sub-pixel shift in the near-infrared imaging. 

After aligning all data to a common astrometric frame, we next seeing match the optical\,--\,near-infrared images. The resolution of the $U$--$K_S$ imaging is $\le\,1.5''$, and we convolve each image to the lowest resolution. We then measure photometry in a $3''$ diameter aperture using the {\sc iraf} package {\sc apphot}. We initially center the aperture at the ALMA identified position, but allow {\sc apphot} to re-center the aperture up to a shift of $<0.5''$ from the original position. To correct for residual resolution differences in the $U$--$K_S$ imaging we aperture correct our measurements to total magnitudes. We create a composite PSF, from 15 unsaturated point sources in each image, and derive the aperture correction as the ratio of the total flux in the composite PSF, to the flux in the original $3''$ diameter aperture. The derived aperture corrections range from $ f_{\rm tot}(\lambda)$\,/\,$f_{\rm ap}({\lambda}) = 1.18$\,--\,$1.27$. We assume sky noise is the dominant source of uncertainty for these faint galaxies, and estimate photometric errors by measuring the uncertainty in the flux in 3$''$ apertures placed randomly on blank patches of sky in each image. 

The resolution of the IRAC imaging is considerably poorer than the $U$--$K_S$ data, $2.2''$ at $8.0\,\mu$m. We therefore match the resolution of all the IRAC imaging to $2.2''$ FWHM, and measure photometry in the same manner as the $U$--$K_S$, using a $3.8''$-diameter aperture. To correct for the resolution difference between the IRAC and $U$--$K_S$ imaging, we again convert the IRAC photometry to total magnitudes. Following the same procedure as above, we measure the aperture correction from a composite PSF of 15 unsaturated point sources in each IRAC image. We measure aperture corrections of $ f_{\rm tot}(\lambda)$\,/\,$f_{\rm ap}({\lambda}) = 1.49$\,--\,$1.89$, in the 3.6--8.0\,$\mu$m wave-bands, which are consistent with those estimated by the SWIRE team~\citep{Surace05}.

In all of the following analysis, we define detections if the flux is 3\,$\sigma$ above the background noise. The median number of filters covering each SMG is 14, and of the 96 SMGs in our sample, 77 are detected in $\ge\,4$ wave-bands. Of the remaining 19 sources, 10 are detected in 2 or 3 wave-bands, and 9 are detected in $\le\,1$ wave-band. We discuss these 19 sources in \S\,\ref{subsubsec:missing}, where we show that a stacking analysis of IRAC $\&$ {\it Herschel} fluxes confirms that they correspond to far-infrared luminous sources, on average. We note that we do not perform any deblending of our photometry, and that we derive redshifts for 12 SMGs which are within $4''$ of a 3.6-\,$\mu$m source of comparable, or greater, flux. In Table~\ref{table:observed} we highlight sources which suffer significant blending, and discuss the effects of blending in \S\,\ref{subsubsec:caveats}.

The photometry for the ALESS SMGs is given in Table~\ref{table:observed}, and in Figure~\ref{fig:maghist} we show the $V$, $K_S$ and 3.6\,$\mu$m magnitude histograms. The ALESS SMGs have median magnitudes of $V = 26.09\,\pm\,0.19$, $K_s = 23.0\,\pm\,0.3$ and $m_{3.6} = 21.80\,\pm\,0.17$ (58, 76\ and 90\,per cent detection rates in each band). We note that at 3.6\,$\mu$m the \citet{Chapman05} sample of radio-detected SMG are a magnitude brighter than the ALESS SMGs ($m_{3.6} = 20.63\,\pm\,0.18$;~\citealt{Hainline09})

\subsection{Herschel/SPIRE}
In this work we make use of observations at 250, 350 and 500\,$\mu$m using the Spectral and Photometric Imaging Receiver (SPIRE;~\citealt{Griffin10}), onboard the {\it Herschel Space Observatory}~\citep{Pilbratt10}. The ECDFS was observed for 32.4 ks at 250, 350 and 500\,$\mu$m in $\sim$\,1.8 ks blocks as part of the Herschel Multi-tiered Extragalactic Survey (HerMES;~\citealt{Oliver12}). These data are described in~\citet{Swinbank13}, the companion paper to this work studying the far-infrared properties of the ALESS SMGs. The final co-added maps reach a 1-$\sigma$ noise level of 1.6, 1.3 and 1.9\,mJy at 250, 350 and 500\,$\mu$m (see also~\citealt{Oliver12}), although source confusion means that the effective depth of these data is shallower than these noise levels imply. Deblended 250, 350 and 500\,$\mu$m fluxes for each ALESS SMG, along with the FIR-properties, are presented in~\citet{Swinbank13}.

\subsection{VLA/1.4\,GHz}
To study the radio properties of the ALESS SMGs, we utilize the VLA 1.4-GHz imaging of the ECDFS. The observations come from~\citet{Miller08} and we use the catalog described in~\citet{Biggs11}.  These data reach an rms of $6.5$\,$\mu$Jy in the central regions, and a median rms of $8.3$\,$\mu$Jy across the entire map. \citet{Biggs11} extract a source catalog, complete to $3\sigma$, and we obtain radio fluxes for the ALESS SMGs by cross-correlating the catalogs with a matching radius of $1''$. In our 1.4\,GHz stacking analysis we use the 1.4\,GHz map from~\citet{Miller13}, a re-reduction of the original data achieving an improved typical map rms of $7.4$\,$\mu$Jy. We note that the $5\,\sigma$ catalog from~\citet{Miller13} does not match any more SMGs than the catalog from ~\citet{Biggs11}, and that the 1.4\,GHz fluxes for individual sources all agree within their 1--$\sigma$ errors.

\section{Photometric redshifts}
\label{sec:photz} The first step in our analysis is to derive photometric redshifts for the ALESS SMGs in our sample, and so determine the first photometric distribution for a large submm-identified population of SMGs. To derive photometric redshifts, we use the SED fitting code {\sc hyperz}~\citep{Bolzonella00}, which computes the $\chi^2$ statistic for a set of model SEDs to the observed photometry. In the case of non-detections we adopt a flux of zero during the SED fitting, but with an uncertainty equal to the 1-$\sigma$ limiting magnitude in that filter. The model SEDs are characterized by a star formation history (SFH), and parametrized by age, reddening and redshift. {\sc hyperz} returns the best-fit parameters for the model SED corresponding to the lowest $\chi^2$. We use the spectral templates of~\citet{Bruzual03}, with solar metalicities, and consider four SFHs; a single burst (B), constant star formation (C) and two exponentially decaying SFHs with timescales of 1\,Gyr (E) and 5\,Gyr (Sb). Redshifts from $z=$\,0--7 are considered and we allow reddening ($A_V$) in the range 0--5, in steps of 0.1, following the~\citet{Calzetti00} dust law. We also include the constraint that the age of the galaxy must be less than the age of Universe. Finally we follow the same prescription as~\citet{Wardlow11} for handling of Lyman--$\alpha$ absorption in {\sc hyperz}; the strength of the intergalactic absorption is increased, but we also allow a wider range of possible optical depths (see \citealt{Wardlow11}).

%
%
\begin{figure*}
\begin{minipage}[t]{\columnwidth}
\centering
\centerline{ \psfig{figure= 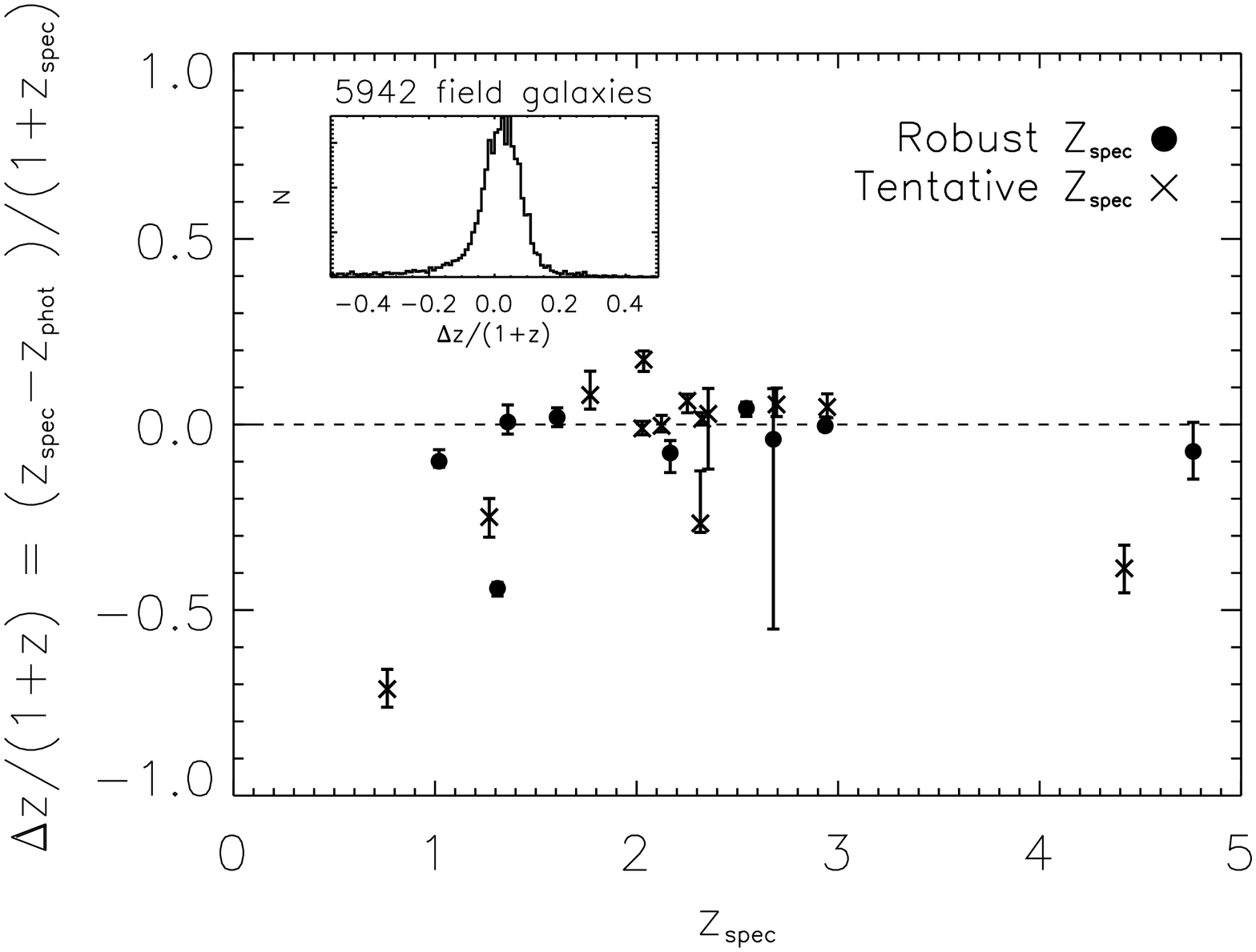,angle=90,width=\columnwidth}}
\caption{ A comparison of the photometric redshifts derived in this work, to spectroscopically confirmed redshifts drawn from the literature and from our redshift follow-up of the original LESS sources (zLESS; Danielson et al.\,in prep). The inset shows the distribution of $\Delta z\,/\,(1+z_{\rm{spec}})$ for a 3.6\,$\mu\rm{m}$ selected training sample with spectroscopic redshifts. For the field sample we find good agreement between the photometric and spectroscopic redshifts, with a median $\Delta z$\,/\,$(1+z_{\rm{spec}})$ of $0.011\,\pm\,0.002$, and a 1-$\sigma$ dispersion of $0.06$. In the main panel we compare the photometric redshifts for 22 ALESS SMGs with confirmed spectroscopic redshifts. We again find good agreement, with a median $\Delta z\,/\,(1+z_{\rm{spec}})$ of $-0.004\,\pm\,0.026$. We identify spectroscopic redshifts as robust where they are calculated from multiple strong emission lines, and tentative where multiple weak lines, or single line IDs are used.  We identify three outliers, at $|\,\Delta z\,/\,(1+z_{\rm{spec}})\,|\,>\,0.3 $. Of the three sources only one, ALESS\,66.1, has a robust spectroscopic redshift, and is an optically bright QSO. The remaining two sources have spectroscopic redshifts drawn from single line identifications.}
 \label{fig:delz}
\end{minipage}
\hfill
\begin{minipage}[t]{\columnwidth}
\centering
\centerline{ \psfig{figure= 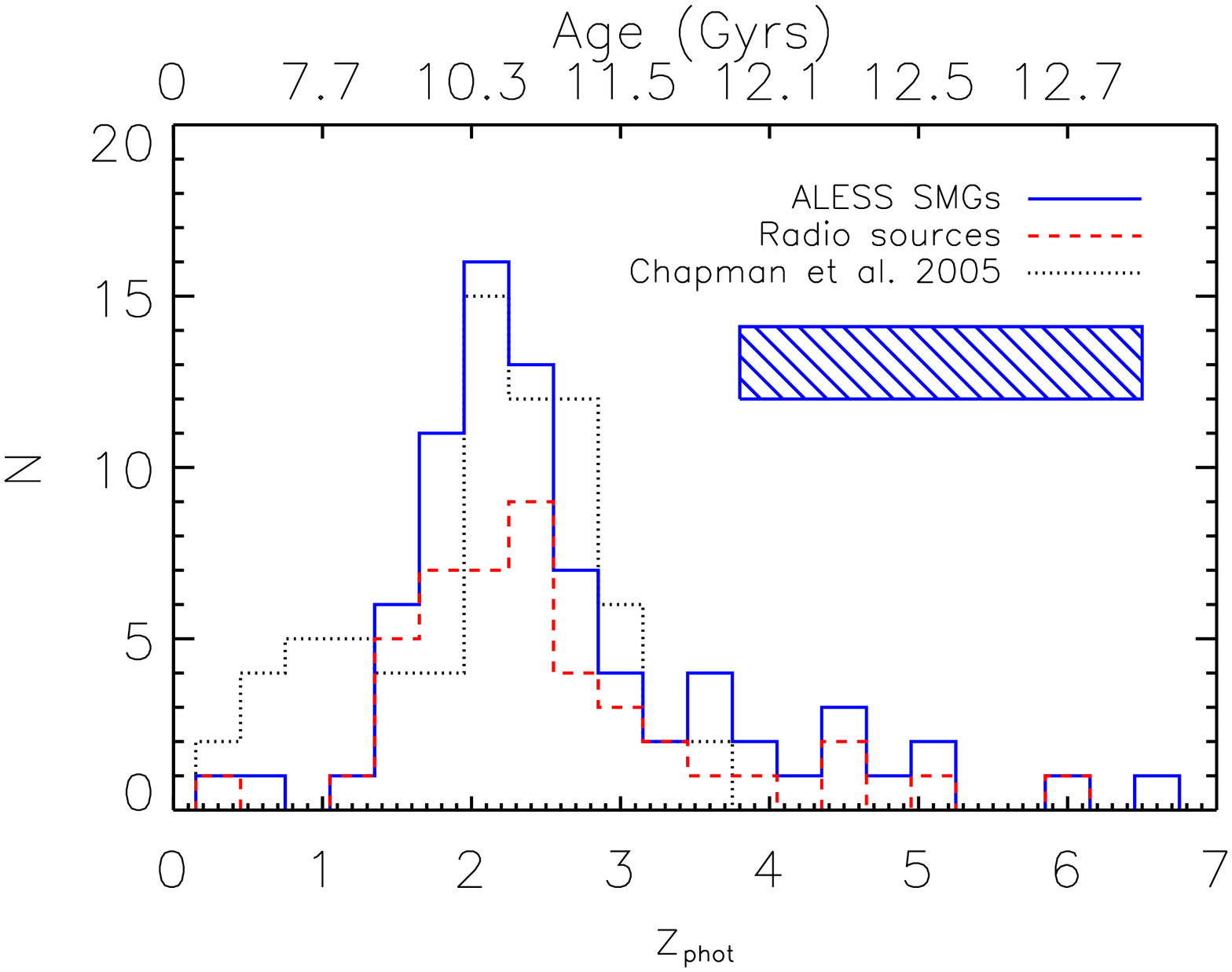,angle=90,width=\columnwidth}}
\caption{ The photometric redshift distribution of ALESS SMGs with individually derived photometric redshifts. For comparison we show the spectroscopic redshift distribution from~\citet{Chapman05}, a radio-selected sample of SMGs. We find the ALESS SMGs lie at a median redshift of $z\,=\,2.3\,\pm\,0.1$, consistent with the result from~\citet{Chapman05}. In contrast to~\citet{Chapman05} we do not find a significant number of SMGs at $z\,\lsim\,1$, and we identify a high redshift tail at $z\,\gsim\,3.5$, not seen in~\citet{Chapman05}. The hatched box represents the area missing from the ALESS histogram due to 19 SMGs with insufficient photometry to derive photometric redshifts. In \S\,\ref{subsubsec:missing} we identify these sources as belonging to the high-redshift tail of the distribution (i.e.\ $z \gsim 3$). Including these 19 SMGs raises the median redshift to $z\,=\,2.5\,\pm\,0.2$, see Figure~\ref{fig:zages}.
}
 \label{fig:zphot}
\end{minipage}
\end{figure*}

%
%
\begin{figure*}
\centerline{ \psfig{figure= 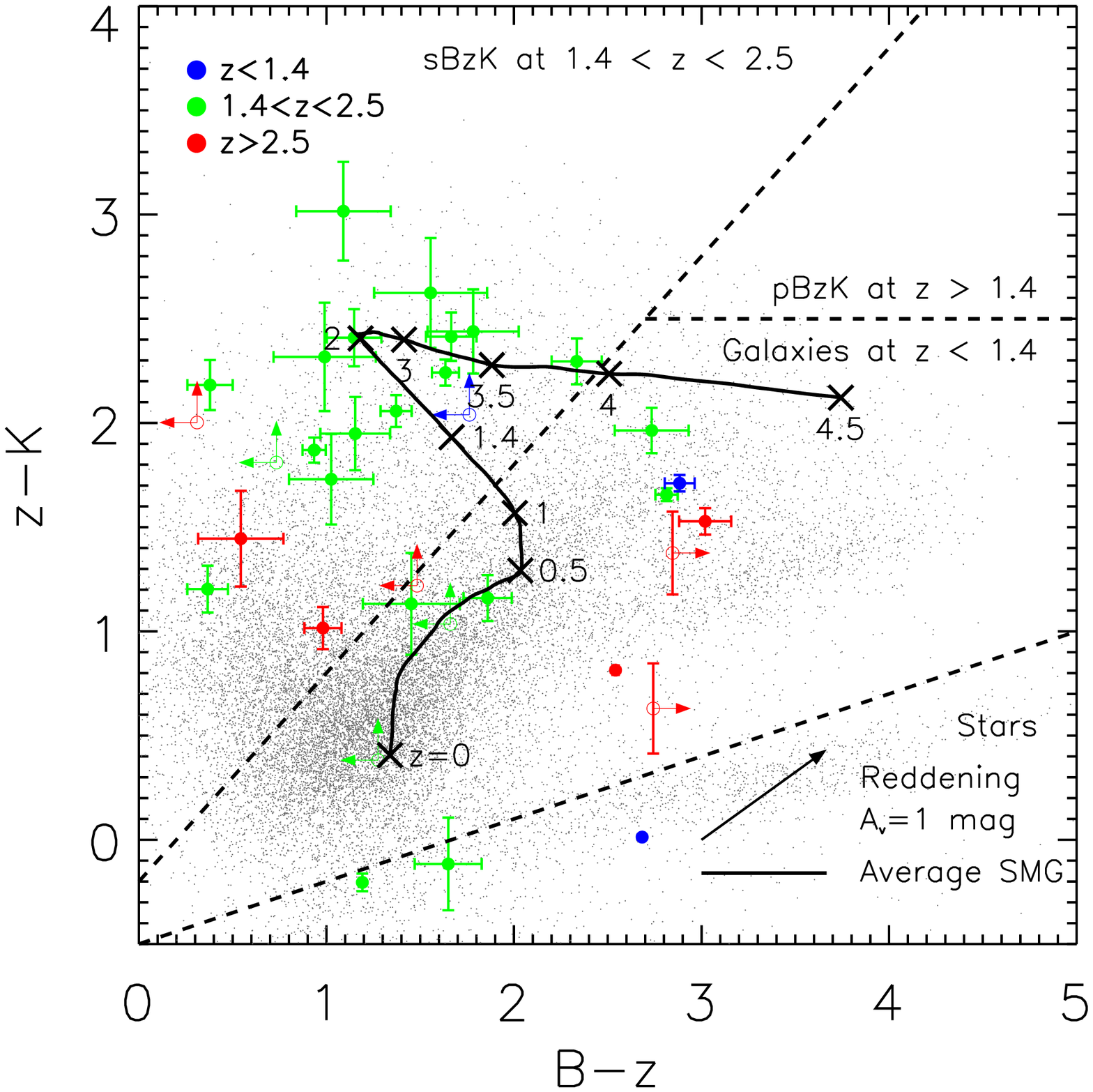,width=\columnwidth} 
\hfill
\psfig{figure= 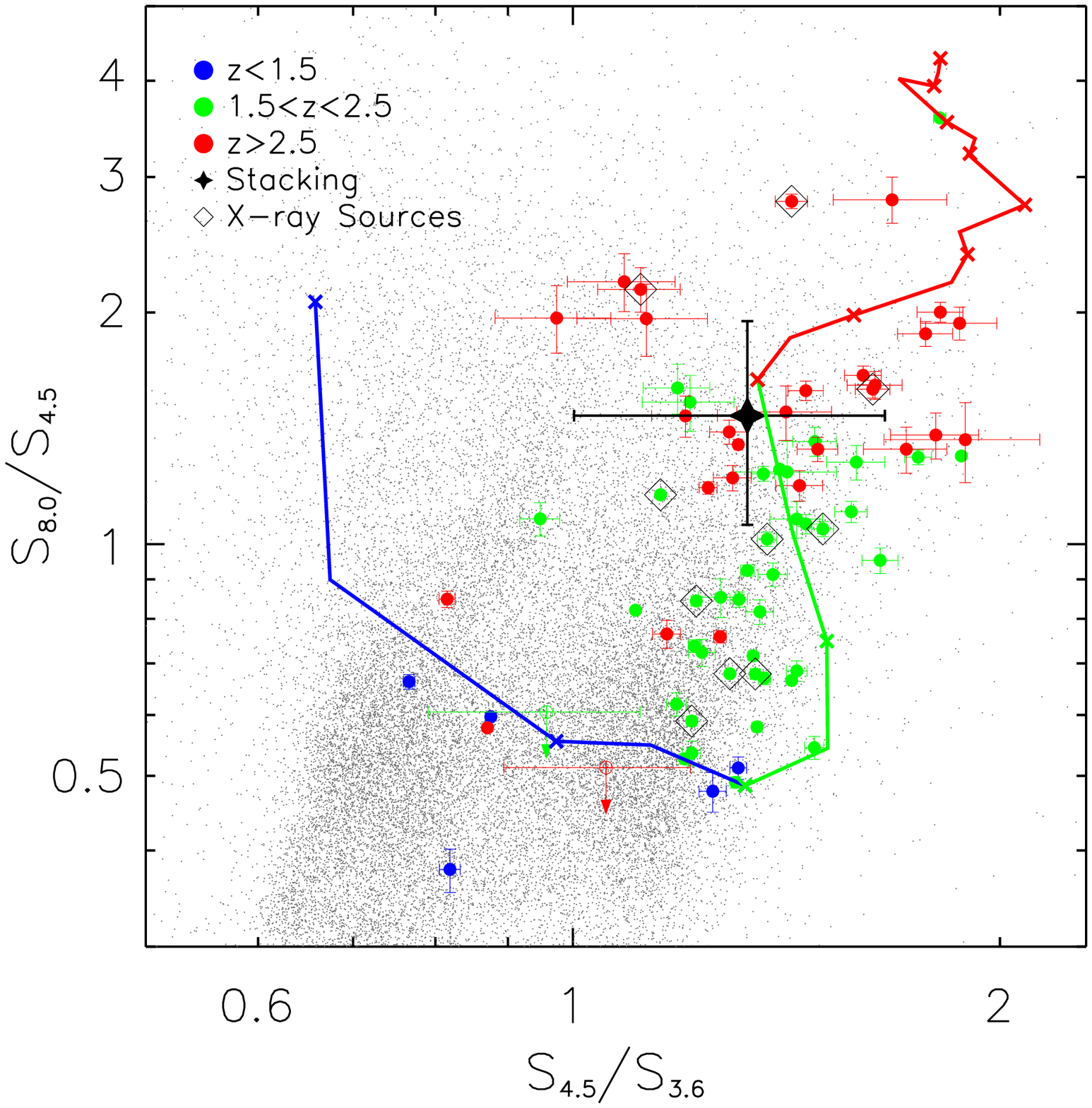,width=\columnwidth}}
\caption{ {\it Left:}  The $(B-z)$\,--\,$(z-K_S)$ colors for the ALESS SMGs, color-coded by photometric redshift. The regions where star-forming and passive BzK galaxies at $z=1.4$--2.5 are expected to lie (sBzK and pBzK respectively) are shown, and we plot the distribution of field galaxies in grey. We find that $\sim$\,$70$ per cent of SMGs have photometric redshifts which are consistent with the predictions from the $BzK$ diagram. The solid line shows the track of the average ALESS SMG, derived by de-redshifting the observed photometry (see Figure~\ref{fig:composite}). From this track we expect the $BzK$ diagnostic to fail for SMGs at $z \gsim 4$. In the lower right we plot the reddening vector for one magnitude of extinction. {\it Right:} The ratio of the IRAC fluxes for the ALESS SMGs, color-coded by redshift. For comparison we plot the track of SMM\,J2135--0102, a $z= 2.3$ SMG~\citep{swinbank10Nature}, color-coded in the same manner as the data points. Similar to the $BzK$ analysis we find our photometric redshifts provide a good match to the expected colors of the ALESS SMGs. We also plot the average IRAC color from a stacking analysis of SMGs detected in only 2 or 3 wave-bands (the error represents the variance in the measured flux). The measured flux from our stacking is clearly noisy, and although we cannot derive photometric redshifts for these SMGs, their colors appear consistent with the bulk of the SMG population at $z$\,$\sim$\,$2.5$. We highlight ALESS SMG detected in X-ray emission~\citep{Wang13}. We find that one X-ray detected SMG, ALESS\,57.1, and six SMGs in the complete sample show evidence of an 8\,$\mu$m excess suggestive of AGN emission. 
}

 \label{fig:bzk}
\end{figure*}

\subsection{Training sample}
Before deriving photometric redshifts for the ALESS SMGs, we first calibrate our photometry to the template SEDs used in the photometric redshift calculation. To do so, we use {\sc SExtractor} to create a 3.6-$\mu$m selected catalog designed to test the reliability of our photometric redshifts against archival spectroscopic surveys. The spectroscopic sample is collated from a wide range of sources (\citealt{Cristiani00, Croom01b, Cimatti02, Teplitz03, Bunker03, LeFevre04, Zheng04, Szokoly04, Strolger04, Stanway04, vanderwel05, Mignoli05, Daddi05, Doherty05, Ravikumar07, Vanzella08, Kriek08, Popesso09, Treister09, Balestra10, Silverman10, Casey11, Cooper12, Bonzini12, Swinbank12}; Koposov et al.\ in prep; Danielson et al.\ in prep), yielding 5942 spectroscopic redshifts with a median $z_{\rm spec} = 0.67$, an interquartile range of 0.45--0.85 and 1077 galaxies at $z>1$. We measure photometry for these sources in the same manner as the ALESS SMGs (see \S\,\ref{subsubsec:alessphot}). For reference, the spectroscopic sample has 10\,--\,90 percentile magnitude ranges of $V = 21.6$\,--\,24.4, and $m_{3.6} = 19.3$\,--\,22.8.

To test for small discrepancies between the observed photometry and the template SEDs we run {\sc hyperz} on our training set of 5942 galaxies with spectroscopic redshifts, fixing the redshift to the spectroscopic value. We then measure the offset between the observed photometry and that predicted from the best-fit model SED. We apply the measured offset to the observed photometry and then repeat the procedure for three iterations. After the final iteration we derive, and apply, significant offsets in the MUSYC $U$ ($-$0.16), $U_{38}$ ($-$0.12), MUSYC $J$ ($-$0.10), $H$ ($-$0.14), HAWK $K_S$ ($-$0.10), TENIS $K$ (0.10), 5.8\,$\mu$m (0.19) and 8.0\,$\mu$m (0.40) photometry. Offsets in the remaining bands are $<\,0.06$, and the typical uncertainty is $\pm 0.02$. The largest offset is an excess in the IRAC 8.0\,$\mu$m, which may be due to a hot dust component in the SEDs which is not included in the {\sc hyperz} templates. We test whether the 8.0\,$\mu$m data drives systematic offsets at other wavelengths by omitting the IRAC 5.8 and 8.0\,$\mu$m data and repeating the procedure, but find the magnitude offsets are consistent with those determined when these wave-bands are included. 

To determine the accuracy of our photometric redshifts we initially compare the results for the 5942 galaxies in the ECDFS with spectroscopic redshifts. We calculate $\Delta z = z_{\rm spec} - z_{\rm phot}$ for each galaxy and plot the histogram of $\Delta z$\,/\,$1+z_{\rm spec}$ in Figure~\ref{fig:delz}. We find excellent agreement between the photometric and spectroscopic redshifts, measuring a median $\Delta \,z\,$\,/\,$( 1+z_{\rm spec} ) = 0.011\,\pm \,0.002 $, with a $1\,\sigma$ dispersion of 0.057 and a Normalized Median Absolute Deviation (NMAD) of $\sigma _{\rm NMAD} = 1.48\times \rm{median} ( | \Delta z - \rm{median} ( \Delta z ) | / 1+z_{\rm spec}  ) = 0.073$\,\footnote{We also derived photometric redshifts for our training sample using the SED fitting code {\sc EAZY}~\citep{Bramer08}. We find the photometric redshifts derived by {\sc EAZY} are comparable with those from {\sc HYPERZ}, with a median $\Delta z$\,/\,$1+z_{\rm spec} = 0.020\,\pm \,0.006$; consistent with~\citet{Dahlen13} who find comparable performance between photometric redshift estimation codes.}. 

Previous studies indicate that the majority of the ALESS SMGs lie at $z>1.0$ (see \citealt{Wardlow11}), and so we also investigate the accuracy of our photometric redshifts limiting just to this redshift range. For the $z>1.0$ sources in the training sample the median $\Delta \,z$\,/\,$ ( 1+z_{\rm spec} )$ is $0.033\,\pm\,0.005$, marginally higher than for the training sample as a whole. We define catastrophic failures as sources with $\Delta \,z$\,/\,$( 1+z_{\rm spec} ) > 0.3 $, and we find the failure rate for the 1077 sources at $z>1.0$ is 4 per cent. Importantly the $z>1.0$ training sample has a median 3.6\,$\mu$m magnitude of $m_{3.6} = 21.2 \pm 0.1$, which is similar to the median  3.6\,$\mu$m magnitude of the ALESS sample, $m_{3.6} = 21.8 \pm 0.2$.

Although {\sc Hyperz} returns a best-fit model and $1$\,$\sigma$ error, for our sample of field galaxies we determine that the {\sc Hyperz} ``99 per cent'' confidence intervals provide the best estimate of the redshift error; yielding $\sim$\,$68\pc$ agreement between the photometric and spectroscopic redshifts at $1\sigma$ and so we adopt these as our 1--$\sigma$ error estimates (see also~\citealt{Luo10,Wardlow11}).

%
%
\begin{figure*}

\begin{minipage}[t]{\columnwidth}
\centering
\centerline{ \psfig{figure= 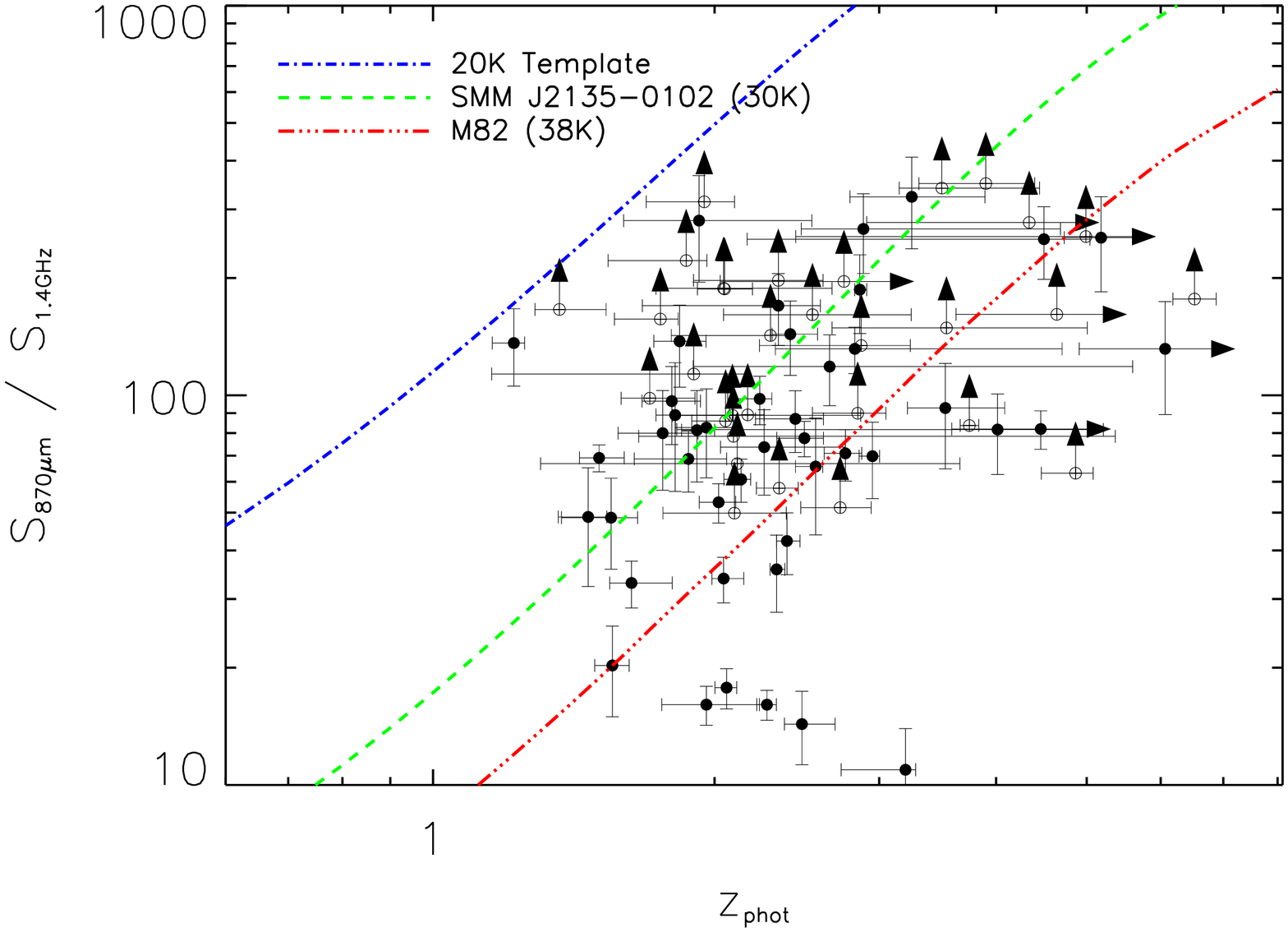,angle=90.,width=\columnwidth}}
\caption{ The variation in $S_{870\mu \rm{m}}$\,/\,$S_{1.4 \rm{GHz}}$ as a function of redshift for the 77 SMGs with photometric redshifts. We overlay tracks for the local star-forming galaxy M\,82 ($T_{\rm{d}} = 38$\,K), SMM\,J2135--0102 a $z$\,$\sim$\,$2.3$ SMG ($T_{\rm{d}}$\,$ \sim$\,$30$\,K) and a cool dust template ($T_{\rm{d}} = 20$\,K;~\citealt{Chary01}). The tracks for M\,82 and SMM\,J2135--0102 pass through the bulk of the population, however we find a large dispersion in S$_{870\mu \rm{m}}$\,/\,S$_{1.4 \rm{GHz}}$; around 1.5 dex at a fixed redshift. For the 32 SMGs which are not detected at 1.4\,GHz we adopt a 3-$\sigma$ upper limit (open symbols with arrows), corresponding to 3\,$\times$ the VLA map rms at the SMG position. The 32 ALESS SMGs which are not detected in available radio data have a range of photometric redshifts from $z_{phot} > 1$. We note that the $S_{870\mu \rm{m}}$\,/\,$S_{1.4 \rm{GHz}}$ flux ratios of these undetected, low redshift ($z \lsim 2.5$), SMGs can be adequately reproduced using a ``cool'' dust template (T$_{\rm{d}} =20$--30\,K) consistent with previous studies~\citep{Magnelli12,weiss13}.
}
 \label{fig:firradio}
\end{minipage}
\hfill
\begin{minipage}[t]{\columnwidth}
\centering
\centerline{ \psfig{figure= 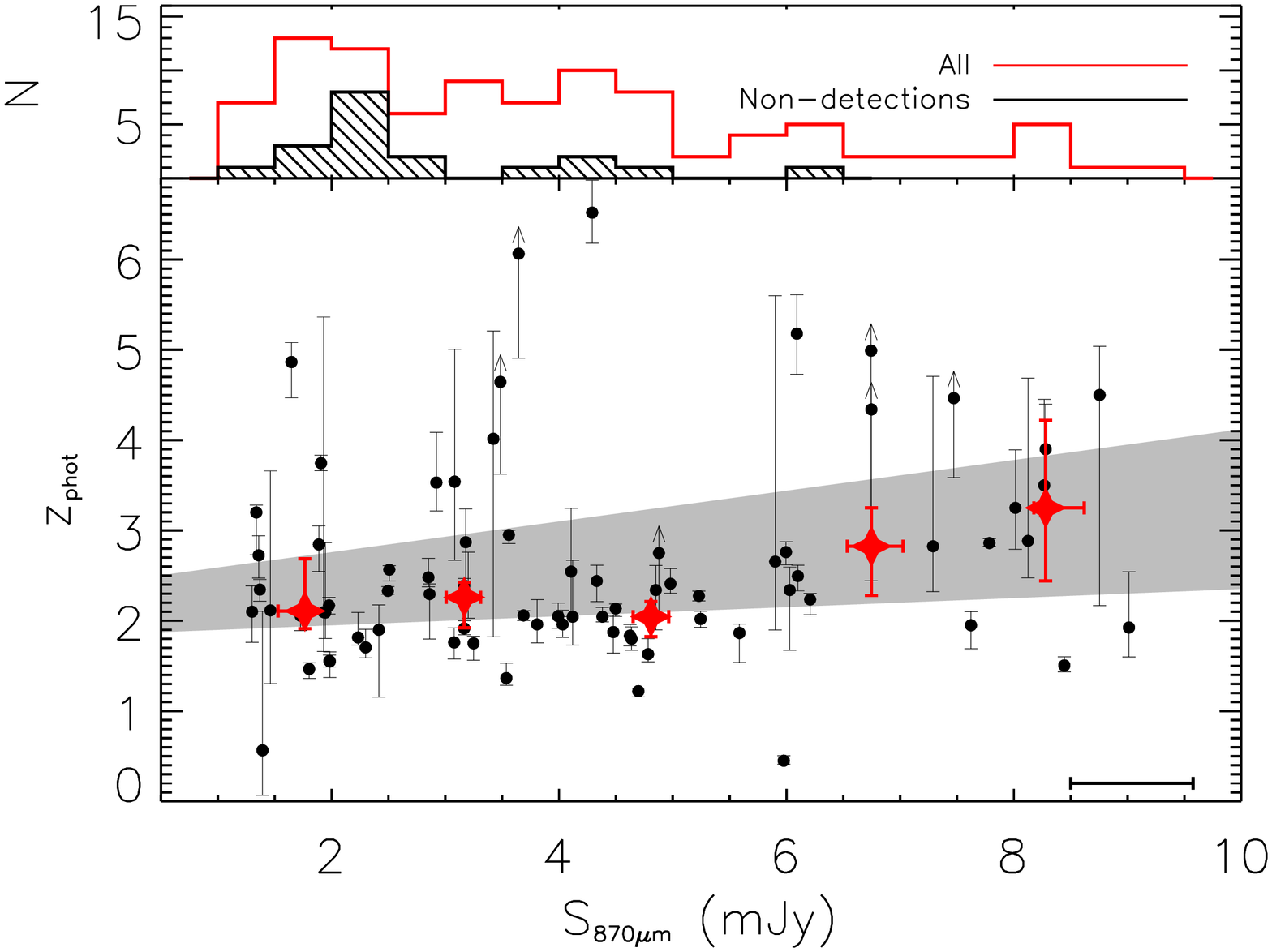,angle=90.,width=\columnwidth} }
\caption{ The photometric redshifts of the ALESS SMGs versus their 870\,$\mu$m flux density. We also split the data into 2-mJy bins and plot the median $S_{870\mu \rm{m}}$ and redshift for each bin, with 1--$\sigma$ error bars. We test for a trend of higher $S_{870\mu \rm{m}}$ sources lying at higher redshift, but a linear fit to the data shows the deviation from a constant with redshift is not significant, at $<\,1.5\,\sigma$. A grey region shows the linear fit and associated 1--$\sigma$ uncertainty. In the upper panel we highlight the 870-$\mu$m flux distribution of those SMGs for which we cannot derive photometric redshifts. As we show in \S\,\ref{subsubsec:missing} these SMGs are likely to lie at $z \gsim 3$, and the weak trend we see with redshift is therefore likely driven by incompleteness in our results: including these photometrically-faint SMGs would further weaken any trend. The error bar on the median flux density is shown in the lower right. 
}
 \label{fig:870z}
\end{minipage}

\end{figure*}

\subsection{ALESS Photometric Redshifts}
\label{subsec:aless_photz}
Before deriving the redshift distribution for all ALESS SMGs, we next make use of the existing spectroscopy of ALESS sources to test the reliability of our photometric redshifts for SMGs. Combining our results with a small number from the literature we have spectroscopic redshifts for 22 ALESS SMGs (\citealt{Zheng04,Kriek08,Coppin09,Silverman10,Casey11,Bonzini12,Swinbank12}; Danielson et al.\ in prep). We run {\sc hyperz} on these SMGs to derive their photometric redshifts, and in Figure~\ref{fig:delz} we compare the spectroscopic results to our photometric redshifts (Figure~\ref{fig:delz}) and find a median $\Delta\,z$\,/\,$ ( 1+z_{\rm spec} ) =-0.004\,\pm\,0.026$, and $\sigma _{\rm NMAD} = 0.099$. The spectroscopically confirmed ALESS SMGs have a median redshift $z_{\rm spec} = 2.2\pm 0.2$ and a median $3.6$\,$\mu$m magnitude of $m_{3.6} = 20.5 \pm 0.5$. Together with the results for the 5924 galaxies in the spectroscopic training sample we can therefore be confident that the photometric redshifts we derive provide a reliable estimate of the SMG population.

\subsubsection{Reliability of SMG redshifts}
\label{subsubsec:caveats}
Running {\sc Hyperz} on the photometry catalog of 77 ALESS SMGs, with detections in $>3$ wave-bands, we derive a median photometric redshift of $z_{\rm phot} = 2.3 \pm 0.1 $, with a tail to $z$\,$\sim$\,$6$ (Figure~\ref{fig:zphot}) and a 1--$\sigma$ spread of $z_{phot} = 1.8$--3.5. In Table~\ref{table:derived} we provide the redshifts for individual sources. We note that we will return to discuss the 19 SMGs detected in fewer than four wave-bands in \S\,\ref{subsubsec:missing}. We caution that five SMGs (ALESS\,5.1, 6.1, 57.1, 66.1 and 75.1) have best-fit solutions with anomalously high values of $\chi^2_{\rm{red}}$ ($\,>\,10$). We inspect the photometry for each of these and find ALESS\,57.1, 66.1 and 75.1 have an 8.0\,$\mu$m excess consistent with obscured AGN activity (ALESS\,66.1 is an optically identified QSO, ALESS\,57.1 is an X-ray detected SMG and ALESS\,75.1 has excess radio emission consistent with AGN activity, \citealt{Wang13}). As we do not include AGN templates in our model SEDs it is unsurprising that we find poor agreement for these sources. For the remaining two sources, the photometry of ALESS\,5.1 is dominated by a large nearby galaxy, while ALESS\,6.1 is a potential lensed source; the 870\,$\mu$m emission is offset by $\sim$\,$1.5''$ from a bright optical source at $z_{\rm phot}$\,$\sim$\,$0.4$. We therefore advise that the photometric redshifts for ALESS\,5.1 and 6.1 are treated with caution and we highlight these SMGs in Table~\ref{table:derived}.

For six ALESS SMGs we derive photometric redshifts from detections in only four wave-bands, our enforced minimum (although we note the SED fit is constrained by sensitive upper limits in the remaining wave-bands). To test if this introduces a bias in our following analysis we take the photometry for 37 SMGs in our sample detected in $>8$ wave-bands and make each source intrinsically fainter until only four of the photometry points remain above our detection limits. We then repeat the SED fitting procedure on these ``faded'' SMGs. We find a median offset in $( z_{4} - z_{\rm All} )$\,/\,$ ( 1+z_{\rm All} ) = -0.098\, \pm \,0.050 $, and agreement at 3$\,\sigma$ for all sources. Crucially, whilst we find increased scatter between the original and faded photometric redshifts, and larger associated uncertainties, we do not find any bias towards higher, or lower, redshifts\,\footnote{We also test the likely effect of emission lines on the SED fitting using a young/blue template, with emission lines of similar equivalent width to SMGs~\citep{Swinbank04}, provided with the {\sc eazy} SED fitting code ~\citep{Bramer08}. We run {\sc hyperz} on the ALESS SMGs, using both the emission line template, and the same template with all emission lines removed. The resulting photometric redshifts are in agreement to within $\Delta\,z$\,/\,$ ( 1+z ) = 0.000 \, \pm \, 0.001 $. We observe a small increase in scatter at $z$\,$\sim$\,$2.5$, which we attribute to H$\alpha$ falling in/out of the $K_s$--band. The effect is small and over the redshift range $z=2.2$--2.8 and we measure $\Delta\,z$\,/\,$ ( 1+z ) = 0.009 \, \pm \, 0.009 $. Due to the modest magnitude of the effect of H$\alpha$ on the photometric redshifts we do not make any attempt to correct for it in our SED fitting.}.

Five SMGs in our sample are covered by IRAC imaging alone. To test the reliability of redshifts for ALESS SMGs derived from such photometry, we take the same sub-sample of 37 SMGs, remove {\it all} other photometric data, including upper limits, and repeat the SED fitting. We find a median offset in $(z_{\rm IRAC} - z_{\rm All})$ \,/\,$ ( 1+z_{\rm All} ) =0.015\, \pm \,0.031 $, and agreement at 3$\,\sigma$ for 36/37 SMGs. If we restrict our comparison to the ALESS SMGs with spectroscopic redshifts then we find $(z_{\rm IRAC} - z_{\rm spec})$ \,/\,$ ( 1+z_{\rm spec} ) = -0.09\, \pm \,0.13$, with a median error on each photometric redshift of $\sigma_{\rm z} = 0.6$. We note that if we only use 3 photometric data points in the SED fitting then the photometric redshifts are unconstrained, with a median 1--$\sigma$ error of $\sigma_{\rm z} = 2.0$. We therefore can be confident in the reliability of photometric redshifts derived from detections in four photometric bands, and adopt this limit throughout our analysis.

Finally we investigate the effect of source blending on our results. We re-measure aperture photometry for all of the ALESS SMGs, in the same manner described in \S\,\ref{subsubsec:alessphot}, but with a 2$''$ diameter aperture across all wavelengths. A smaller aperture means the effects of blending are reduced, especially in the IRAC data. We repeat the SED fitting procedure described in \S\,\ref{sec:photz}, to derive photometric redshifts from the new, small aperture, photometry. Considering all ALESS SMGs we find good agreement between the photometric redshifts, with $( z_{\rm Original} - z_{\rm Small Aperture} )$\,/\,$ ( 1+z_{\rm All} ) = -0.012\, \pm \,0.009$. Of the 12 SMGs we flag as blended with a nearby bright IRAC source (see Table~\ref{table:observed}), nine have a photometric redshift derived from photometry measured in a smaller aperture which is consistent with the original redshift to within 1--$\sigma$. Of the remaining three SMGs, ALESS\,5.1 has been discussed already as a possible lens system, and two other sources ALESS\,75.4 $\&$ 83.4 are not detected in $>3$ wave-bands in the smaller apertures. We highlight these SMGs in Table~\ref{table:derived} and note that their redshifts should be treated with caution. In Figure~\ref{fig:mh_z} we highlight these three SMGs, along with ALESS\,6.1, another potential lens system, as having suspicious photometry. We conclude that blending of sources does not have a significant effect on the bulk of the redshifts we derive.

%
%
\begin{figure*}
\centerline{ \psfig{figure= 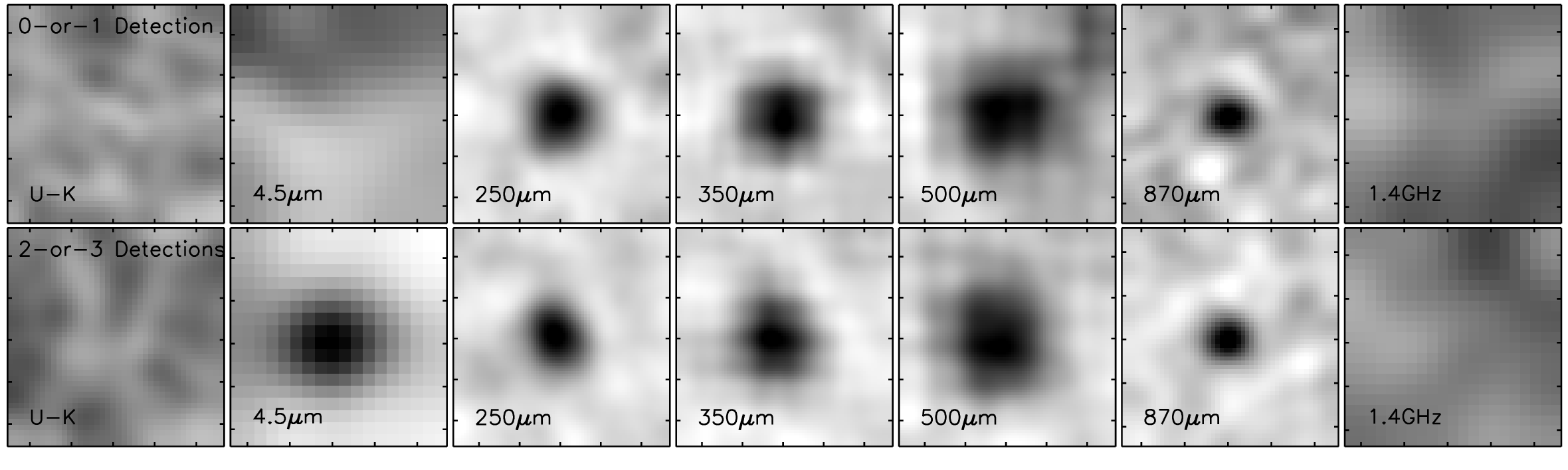,angle=90,width=2\columnwidth}}
\centerline{ \psfig{figure= 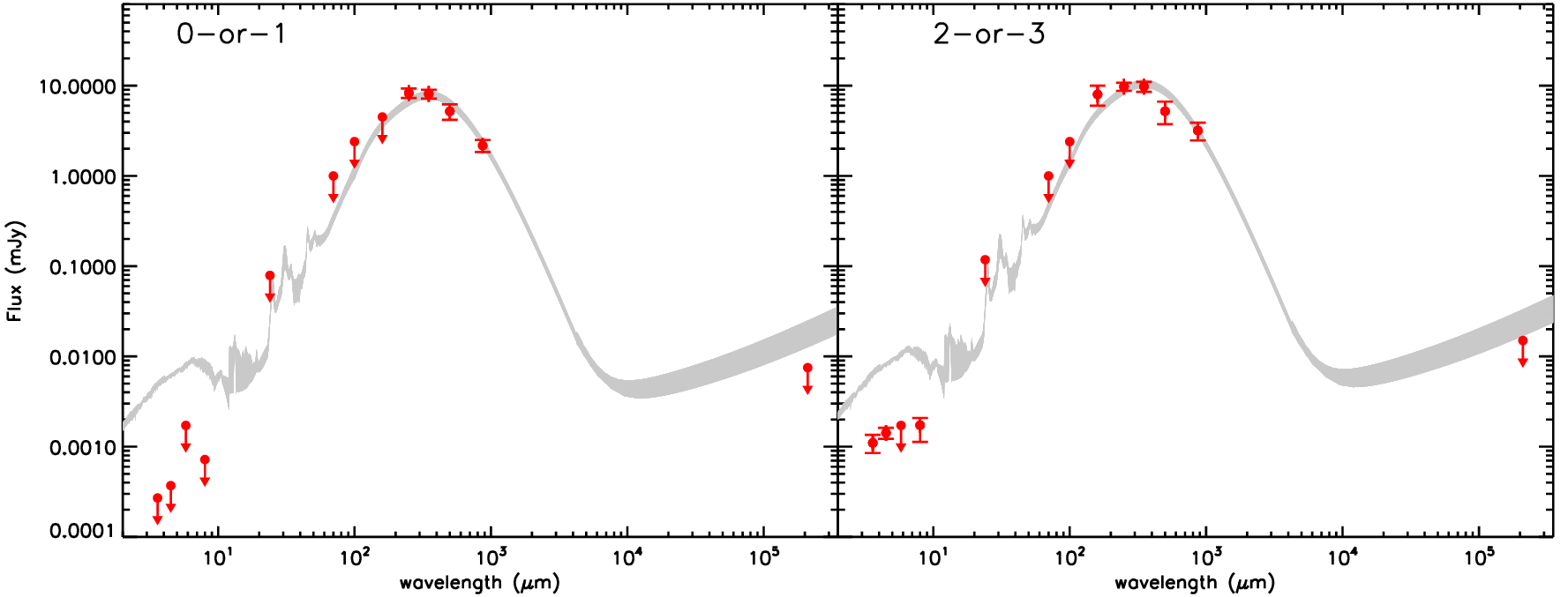,width=2\columnwidth}}
\vspace{0.2cm}
\caption{ 19 SMGs from our sample are detected in fewer than four optical\,--\,near-infrared wave-bands (see Table~\ref{table:observed}). We sub-divide these 19 sources into those detected in either 0-or-1 (9) or 2-or-3 (10) wave-bands and stack the available imaging at the ALMA position to test if they are real or spurious. In particular we stack the {\it  Herschel} 250, 350, and 500\,$\mu$m imaging, and the 1.4\,GHz VLA data. The {\it Herschel} 250, 350, and 500\,$\mu$m greyscale images are 80$'' \times 80''$, and the other greyscale images are 10$'' \times 10''$. Both subsets are detected in stacks of all three {\it Herschel} bands, confirming, on average, that they do represent far-infrared sources. We do not find a detection for either subset in a stack of the $U$\,--\,$K_{s}$ imaging, or the 1.4\,GHz data. The subset of SMGs with 2-or-3 detections does yield a detection in the IRAC 3.6, 4.5 and 8.0\,$\mu$m bands, however this is unsurprising as their 2-or-3 photometric detections are usually in the IRAC bands. In the lower panel we show the far-infrared properties of these 19 SMGs. We also show the results of stacking the MIPS 24\,$\mu$m and PACS 70, 100 and 160\,$\mu$m imaging, for details of this data see~\citet{Swinbank13}. The stacked FIR emission for both subsets appears to peak between 250--350\,$\mu$m, suggesting that these SMGs do not lie, on average, at very high redshifts ($z \gsim 5$). On each panel the grey line represents the average ALESS SMG SED, plotted at a redshift of $z= 3.6$ and 3.2, for the  0-or-1 and 2-or-3 subsets respectively (see also \citealt{Swinbank13}). We note that the line is not a fit to the data points, but is simply scaled to match the peak of the far-infrared SED. We highlight that the shape of the SED appears well matched to the stacked FIR emission, but that the NIR properties of these SMGs are an order of magnitude fainter than the average SED. Similarly, at the nominal redshift plotted the composite SED over-predicts the radio emission from both subsets of SMGs.
}
 \label{fig:proof}
\end{figure*}

\subsubsection{Redshift indicators}
\label{subsubsec:trends}
A number of color--color diagnostics have been suggested to identify star-forming galaxies. We consider three of these as simple tests of the reliability of our photometric redshifts. The first we consider is the $BzK$ diagram, which has been proposed as a tool to separate star-forming and passive galaxies at $z$\,$\sim$\,$1.4$--2.5, by means of identifying the Balmer/4000\AA\ break. In Figure~\ref{fig:bzk} we show the $BzK$ diagram for the ALESS SMGs with suitable photometric detections. We find that within the photometric errors 65 per cent are correctly identified as star-forming at $z > 1.4 $ and 25 per cent are incorrectly classed as lying at $z<1.4$. One ALESS SMG is correctly classified as a galaxy at $z < 1.4$, and no ALESS SMGs are classed as passive galaxies at $z > 1.4$. Three ALESS SMGs have $BzK$ colors consistent with stars, one of which, ALESS$\,66.1$, is an optically identified QSO. We caution that half of the ALESS SMGs incorrectly classified as galaxies at $z<1.4$ have photometric redshifts greater than the upper range of the $BzK$ diagnostic, i.e.\ $z > 2.5$. We plot the SED for the composite ALESS SMG in Figure~\ref{fig:bzk}, which shows that we expect the BzK diagram to classify SMGs at redshifts from $z$\,$\sim$\,1--4, as star-forming BzKs at $z$\,$\sim$\,$1.4$--2.5, and SMGs at redshifts greater than $z>4$ as galaxies at $z<1.4$.

We find that the ALESS SMGs display a clear trend with redshift in $S_{8.0}$\,/\,$S_{4.5}$ versus $S_{4.5}$\,/\,$S_{3.6}$ color (Figure~\ref{fig:bzk}), with sources at high redshift tending to have higher ratios of $S_{8.0}$\,/\,$S_{4.5}$. As a further test of our photometric redshifts we overlay the predicted colors of SMM\,J2135--0102 (a well-studied SMG at $z = 2.3$;~\citealt{swinbank10Nature}) as a function of redshift on Figure~\ref{fig:bzk}. We find that the derived photometric redshifts for the ALESS SMGs are in good agreement with the predictions from this SED track. 

Ten ALESS SMGs are detected in data taken with the {\it Chandra} X-ray Observatory (see ~\citealt{Wang13}). This X-ray emission is often indicative of an AGN component in the host galaxy, which can affect the SED shape. As such, we now investigate whether the X-ray detected SMGs~\citep{Wang13} are distinguishable from the parent sample of SMGs in terms of their IRAC fluxes. We identify one X-ray detected SMG, ALESS\,57.1, which has a high $S_{8.0}$\,/\,$S_{4.5}$ ratio, relative to $S_{4.5}$\,/\,$S_{3.6}$, suggestive of a power-law AGN component in the SED. A further inspection of the SED fits in Appendix~A shows that only two SMGs display a clear enhancement in IRAC flux (ALESS\,57.1 \& 75.1), which is often attributed to AGN heated dust emission\,\footnote{The low rate of near--infrared excess in the ALESS SEDs is in stark contrast to the SEDs seen in previous SMG samples, where a large fraction show restframe near--infrared excesses whose amplitude appears to correlate with AGN luminosity~\citep{Hainline11}. This may reflect differences in the sample selection between the predominantly radio-pre-selected, spectroscopically confirmed, SMGs in~\citet{Hainline11} and the purely submm-flux-limited sample analyzed here.}. The remaining X-ray sources appear well-matched to the complete SMG sample. We perform a two-sided Kolmogorov-Smirnov (KS) test between the X-ray detected SMGs, and the parent sample, in terms of both $S_{8.0}$\,/\,$S_{4.5}$ and $S_{4.5}$\,/\,$S_{3.6}$. The KS test returns a probability of 85 per cent that the samples are drawn from the same parent distribution, in terms of both $S_{8.0}$\,/\,$S_{4.5}$ and $S_{4.5}$\,/\,$S_{3.6}$. This suggests that in terms of IRAC color the X-ray detected SMGs do not represent a distinct subset of SMGs.

Finally we consider the link between 870\,$\mu$m and 1.4\,GHz emission, which has been used to identify the optical\,--\,near-infrared counterpart to submm emission.  We first stress that we see an order of magnitude of scatter in $S_{870\mu \rm m}$\,/\,$S_{1.4{\rm GHz}}$ at a fixed redshift (Figure~\ref{fig:firradio}). We now compare the ALESS SMGs to three template SEDs with varying characteristic dust temperatures. We use templates for two well-studied dusty galaxies, SMM\,J2135--0102\footnote{The best-fit far-infrared SED to the photometry of SMM\,J2135--0102 is a two component dust model at 30\,K and 60\,K. The dust masses of each component are $M_{\rm d}^{{\rm warm}} = 10^6$\,M$_\odot$ and $M_{\rm d}^{{\rm cold}} = 4 \times 10^8$\,M$_\odot$~\citep{Ivison10eyelash}} ($\sim$\,30\,K) and M\,82 (38\,K). In addition we also use a 20\,K template drawn from the~\citet{Chary01} template SED library. These templates span typical dust temperatures for SMGs~\citep{Magnelli12,weiss13}, and we find they are sufficient to describe the majority of ALESS SMGs. Previous studies have suggested redshift solutions below $z \lsim 2.5$ are incorrect for radio-non-detected SMGs~\citep{Smolcic12}. We find templates with a characteristic temperature of 20--30\,K are a plausible explanation for similar ALESS SMGs, and we therefore do not discard redshift solutions at $z \lsim 2.5$ (see also~\citealt{Swinbank13}).

%
\begin{figure}
\hspace{-0.75cm}
\centerline{ \psfig{figure=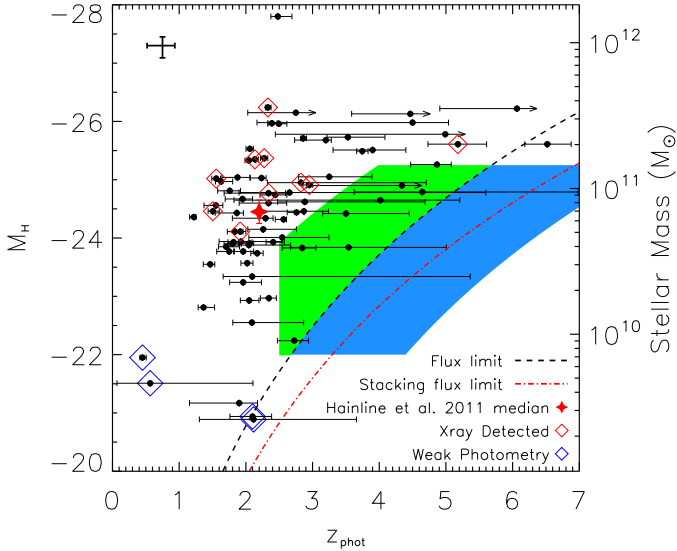,width=0.9\columnwidth}}
\caption{ The absolute $H$-band magnitude distribution, derived from SED fits to the observed photometry. The median M$_{H}$ of the ALESS SMGs is $-24.56\,\pm\,0.15$, which agrees with that derived for the SMG sample presented by~\citet{Hainline11}, M$_{H} =-24.45\,\pm\,0.20$. The dashed line illustrates the flux limited nature of our survey, and the red line the limit in our IRAC stacking (flux limit taken from the IRAC 4.5\,$\mu$m limiting magnitude). By requiring that the $M_{H}$ distribution is not bimodal, and using the IRAC selection limits we can estimate the redshift distribution for those ALESS SMGs without sufficient photometry to derive photometric redshifts. Shaded regions represent the area populated by ALESS sources detected in 0-or-1 (blue) or 2-or-3 (green) wave-bands. For each source we determine a mass-to-light ratio, $M$\,/\,$L_{H}$, from the SFH returned in the SED fitting, and estimate a median stellar mass for the complete sample of 96 SMGs of $M_{\star} = (8 \pm 1) \times 10^{10}$\,M$_{\odot}$, for a Salpeter IMF. We caution that due to the unknown SFHs, the stellar masses of the ALESS SMGs are very poorly constrained (see \S\,\ref{subsec:masses}). Sources which are possible gravitational lenses, or with questionable photometry, are highlighted with blue diamonds. 
}
 \label{fig:mh_z}
\end{figure}

%
%
%

%
\begin{figure*}
\centerline{\psfig{figure=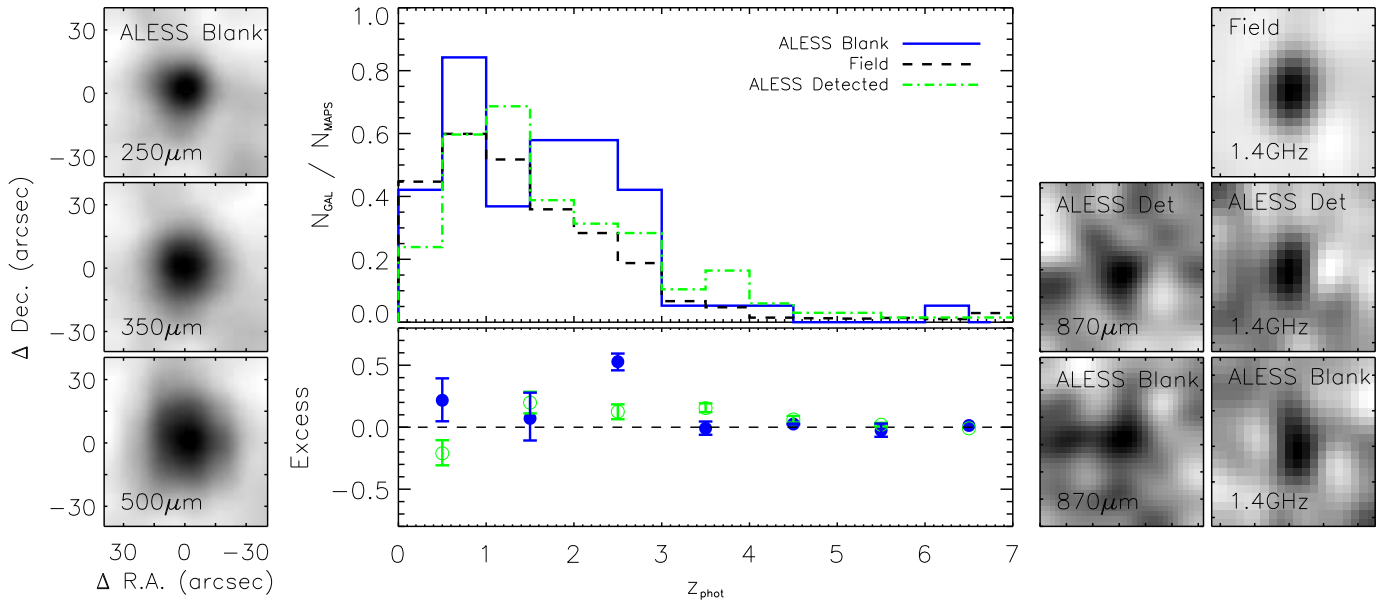,angle=90.,height=0.9\columnwidth}}

\vspace{0.2cm}
\caption{ {\it Left:} 19 of the 86 ALMA maps of LESS submm sources considered in this work do not contain an SMG brighter than our 870-$\mu$m detection threshold of $\sim$\,$1.4$\,mJy, within the ALMA primary beam. We show the results of stacking the {\it Herschel}\,/\,SPIRE 250, 350 and 500\,$\mu$m maps at the position of these 19 LABOCA detections in the left hand column. We detect emission at $>8$\,$\sigma$ at all three SPIRE wavelengths, confirming that at least on average these sources are real. {\it Middle:} We compare the redshift distribution of IRAC sources in ALMA maps without SMGs to the field, and to those in areas covered by ALMA maps containing detected SMGs. The field sample is drawn from random apertures with the same size as the ALMA primary beam, and any ALESS SMGs are removed from all samples. In the maps with detections we find that the redshift distribution is consistent with the field, but in the blank ALMA maps we find an excess of sources at $z$\,$\sim$\,$2.5$. This suggests that the LESS SMGs have fragmented into multiple components below our detection threshold, and that they have a redshift distribution consistent with the $S_{870\mu \rm m}$--brighter ALESS sample. {\it Right:} Stacked maps of  3.6-$\mu$m selected galaxies with photometric redshifts between $z=1$--3. We stack the 870\,$\mu$m emission for sources covered by our ALMA observations, and the 1.4\,GHz emission for all galaxies. We again split the sample into subsets based on whether they are in an ALMA map or not, and furthermore into ALMA blank maps and ALMA maps containing SMGs. We obtain a $\sim$\,$4$\,$\sigma$ detection of the IRAC samples in both 870\,$\mu$m stacks. The ALMA maps lacking SMGs have a primary--beam corrected flux of $S_{870} = 0.36 \pm 0.09$\,mJy, and a number density of sources $\sim$\,$2\times$ higher than the field over the range $z=1$--3. We detect all subsets at 1.4\,GHz, and we find the IRAC sources in the ALMA maps are $\sim$\,$2.5 \times$ brighter at 1.4\,GHz than those in the general field population at 2.8\,$\sigma$. This tentative result suggests that IRAC-selected sources are typically brighter at $870$\,$\mu$m when in the vicinity of a submm source.
}
 \label{fig:stacking_blank}
\end{figure*}

\subsubsection{Undetected or Faint Counterparts}
\label{subsubsec:missing}

For the 77 ALESS SMGs which have counterparts in at least four optical or near-infrared bands, we are able to estimate reliable photometric redshifts. However, this leaves 19 ALESS SMGs ($\sim$\,20 per cent of the sample) which do not have sufficient detections to derive a photometric redshift.  To test whether these sources are spurious or simply fainter than the rest of the population, we divide the SMGs into subsets compromising 0-or-1 and 2-or-3 detections in both the optical ($U$\,--\,$K_S$) and IRAC wave-bands, and stack their emission in these wave-bands using a clipped mean algorithm. Fig.~\ref{fig:proof} shows that only the 2-or-3 wave-band subset yields a stacked detection in the IRAC wave-bands at the 7$\sigma$ level, whereas the optical stacks of both subsets, and the IRAC stack of the 0-or-1 subset, all yield non-detections at the $< 3 \sigma$ level.

Next, we stack the emission from these SMGs in the far-infrared {\it Herschel}\,/\,SPIRE maps at 250, 350 and 500\,$\mu$m and show these in Figure~\ref{fig:proof}. The SMGs are clearly detected at $> 4\sigma$ in all SPIRE bands in both the 0-or-1- and 2-or-3 subsets, with 250, 350 and 500\,$\mu$m flux densities between 4--16\,mJy\,\footnote{We use the deblended SPIRE maps described in~\citet{Swinbank13} but to account for the clustering, we use a deblended map where the ALESS SMGs are not included in the a-priori catalog.}. We note that four of the SMGs are detected individually at 250\,$\mu$m, two of which are detected at 350 and 500\,$\mu$m. The SEDs for these stacks peak between 250 and 350\,$\mu$m for both the 0-or-1 and 2-or-3 subsets, as shown in Figure~\ref{fig:proof}.  In this figure we also overlay the composite ALESS SMG SED, see \S\,\ref{subsubsec:sfh}, redshifted to $z=3.2$ and $z=3.6$ for the 0-or-1 and 2-or-3 subsets respectively, to match the peak of the far-infrared SED. The redshifted template appears to roughly reproduce the far-infrared properties of these SMGs, although we note that their near-infrared properties are approximately an order of magnitude fainter than the composite ALESS SMG SED. We caution that variation in the dust temperature or redshift of the SMGs in the 0-or-1 and 2-or-3 subsets would smear the peak wavelength of the stacked far-infrared SED. Thus the far-infrared SED of the 0-or-1 and 2-or-3 subsets peaking at longer wavelengths is only tentative evidence that they lie at higher redshift. A full discussion of the far-infrared properties of these SMGs is presented in~\citet{Swinbank13}.

In Figure~\ref{fig:mh_z} we plot the $H$-band absolute magnitude ($M_H$) versus redshift for the 77 ALESS SMGs where we have derived a photometric redshift.  We also highlight the survey selection limits, which show that between $z=0$ and $z$\,$\sim$\,$2.5$ the near-infrared survey limits should be complete at magnitudes brighter than $M_{H}=-22$ (equivalent to a stellar mass of M$_{\star}$\,$\sim$\,$10^{10}$\,M$_{\odot}$ for a light-to-mass ratio of L$_{H}$\,/\,M$_{\star}$\,$\sim$\,$3.8$; \citealt{Hainline11}).  However, above $z$\,$\sim$\,$2.5$, the optical\,--\,near-infrared survey limits mean that only the brightest SMGs are detected, despite the 870\,$\mu$m selection ensuring we have an unbiased sample of SMGs from $0<z \lsim 6$. We make the assumption that the absolute $H$-band magnitude distribution of the ALESS SMGs is complete at $z<2.5$ and that incompleteness in the distribution at $z>2.5$ is due to our near-infrared selection limits, i.e. that the 19 SMGs detected in $<4$ wave-bands lie at $z>2.5$ and that the absolute $H$-band magnitude distribution is not bimodal. Our assumption is in agreement with Fig.~\ref{fig:proof}, which shows the stacked far-infrared SED of these SMGs peaks at longer wavelengths than the average ALESS SMGs, and indeed one of the SMGs detected in $<4$ wave-bands, ALESS\,65.1, has been spectroscopically confirmed to be at $z=4.4$~\citep{Swinbank12}. We caution that an alternative explanation is that the SMGs detected 0-or-1 and 2-or-3 wave-bands are either significantly more dust obscured ($A_V>4$) or lower stellar mass (M$_{\star}$\,$<$\,$10^{10}$\,M$_{\odot}$) than the optical--near-infrared detected ALESS SMGs, but note that this would mean both properties have a bimodal distribution.

To estimate the likely redshift distribution of the 19 ALESS SMGs which are detected in $<4$-bands, we first assume that the survey is complete in $M_{H}$ at $z< 2.5$ (Figure~\ref{fig:mh_z}).  To determine incompleteness in the magnitude distribution at $z >2.5$, we construct $\sim$\,$1000$ realizations of the ALESS SMG $H$-band absolute magnitude distribution over the range $z=0$--2.5 and compare this to the $H$-band absolute magnitude at $z>2.5$. We then assign values of $M_{H}$ to the ten ALESS SMGs detected in 2-or-3 wave-bands to minimize the incompleteness in the $H$-band absolute magnitude above $z > 2.5$. These ten sources have a stacked flux close to our photometric selection limit, and hence are assigned redshifts based on the selection limit at the corresponding value of their $M_{H}$.  We repeat this procedure for the remaining nine ALESS SMGs detected in the 0-or-1 wave-bands.  Since these SMGs are not detected in our optical or near-infrared stacking, we assume these SMGs must lie below (or close to) the detection limit in our stacked IRAC maps.  We caution that, in both cases, this may underestimate the redshift of these SMGs, although since both subsets peak at $\sim$\,$350\mu$m in the {\it Herschel} stacks in Figure~\ref{fig:proof} it appears on average they do not lie at very high redshifts ($z \gsim 5$).  Using this approach, the median redshift of the ALESS SMGs is $z$\,$\sim$\,$3.5$ and $z$\,$\sim$\,$4.5 $ for sources detected in 2-or-3 and 0-or-1 wave-bands respectively, similar to the redshifts derived from the composite SMG SED (Figure~\ref{fig:proof}). Including these redshifts in our redshift distribution, the median photometric redshift for our complete sample of 96 ALESS SMGs is then $z_{phot} = 2.5 \pm 0.2$ (Figure~\ref{fig:zages}). The distribution has a tail to high redshift, and $35 \pm 5$ per cent of the ALESS SMGs lie at $z_{phot} > 3$.

\subsubsection{ALMA Blank Maps}
\label{subsubsec:blankmaps}
We have now discussed the redshift distribution of {\it all} SMGs in the ALESS {\sc main} catalog. Before continuing it is important to consider the ALMA maps in which we do not detect any SMGs. In total we obtained high quality ALMA observations of 88 LABOCA submm sources. Of these 88 ALMA observations, 19 are blank maps and do not contain an SMG above a $S/N > 3.5$ within the primary beam~\citep{Hodge13}. The ALMA blank maps are predominantly faint LABOCA detections, and comprise 14 out of 24 LABOCA detections with $S_{870} < 5.5$\,mJy. To verify the reliability of the original LABOCA detections we stack the FIR-emission from all 19 sources in the far-infrared {\it Herschel}\,/\,SPIRE maps at 250, 350 and 500\,$\mu$m. We detect emission at $>8$\,$\sigma$ in our stacks of all three SPIRE wave-bands, and show the images of each stack in Figure~\ref{fig:stacking_blank}. Furthermore, we split the sample at a detection significance of $4.2\,\sigma$ in the original LABOCA map, yielding subsets containing 10 and 9 sources respectively. We again stack the FIR-emission for both subsets and detect emission at $>6$\,$\sigma$ at 250, 350 and 500\,$\mu$m in both subsets, again confirming that on average both subsets contain real sources. The results of our stacking analysis are consistent with \citet{Weiss09}, who state that only $\sim$\,$3$ of the 88 LESS sub-mm sources are expected to be false detections\,\footnote{\citet{Weiss09} predict that the complete LESS sample of 126 sub-mm sources contains five false detections. In addition \citet{Weiss09} consider the effects of map noise on measured source fluxes, which boosts otherwise faint sources above the nominal flux limit of their catalog. However, they do not account for source clustering in their analysis, which may result in a higher flux boost.}.

Our ALMA observations have demonstrated that single-dish-detected submm sources often fragment into multiple SMGs in interferometric observations (\citealt{Karim13,Hodge13}; see also ~\citealt{Barger12}). We now test if it is possible that the ALMA blank maps similarly contain multiple SMGs, each below the 870-$\mu$m flux limit of the ALESS survey, but which together appear as a single, blended, source in the LABOCA observations. First, we use the photometric redshifts derived for the 3.6\,$\mu$m training set to search for an excess of 3.6\,$\mu$m sources in the ALMA blank maps, when compared to the field (see Figure~\ref{fig:stacking_blank}). We construct the redshift distribution for the field by placing 1000 random apertures of equal size to the ALMA primary beam across the ECDFS. We compare the redshift distribution in these random fields to that of sources in the ALMA blank maps, and identify an excess of $0.61 \pm 0.07$ sources per ALMA blank map across the redshift range $z=2$--3. There is also a small excess of $0.15 \pm 0.06$ of 3.6\,$\mu$m selected sources in the ALMA maps containing an SMG, over the same redshift range $z=2$--3, compared to the field, where we have removed the ALESS SMG counterparts from the comparison. The existence of a small excess suggests that the ALMA blank maps contain multiple faint SMGs, and crucially that they have a redshift distribution broadly similar to the ALESS SMGs.

We assess the 870\,$\mu$m flux contribution of these IRAC sources by stacking the primary beam corrected ALMA maps at the position of the 3.6\,$\mu$m sources, again removing all ALESS SMGs in the {\sc main} catalog from the sample. In Figure~\ref{fig:stacking_blank} we show 870\,$\mu$m stacks for 3.6\,$\mu$m sources, over the redshift range $z=1$--3, in both ALMA blank maps (``Blank''), and ALMA maps containing at least one SMG (``Detected''). We choose to stack over the redshift range $z=1$--3 as it covers the observed excess in IRAC sources, and the expected range of redshifts for the bulk of the SMGs. Both ``Blank'' and ``Detected'' subsets are detected at a significance of $\sim$\,$4\sigma$ in the 870\,$\mu$m stacks, and the ALMA blank maps have an average primary beam corrected peak flux of $S_{870} = 0.36 \pm 0.09$\,mJy (the sources in maps with detected SMGs yield $S_{870} = 0.29 \pm 0.08$\,mJy). Using the number density of the 3.6\,$\mu$m sources we calculate that these contribute a total 870\,$\mu$m flux per ALMA blank map of $S_{870} = 0.76 \pm 0.19$\,mJy.

However, we also need to confirm that these 3.6\,$\mu$m sources in the ALMA maps are brighter in the submm than the IRAC population outside the ALMA fields. This is difficult as we only have ALMA coverage of the LABOCA source positions, but we can take advantage of the radio coverage of the whole field to use this as a proxy to estimate the relative brightness of these two samples. We therefore stack the 1.4\,GHz VLA map at the positions of the 3.6\,$\mu$m sources, at  $z=1$--3, in the ALMA blank maps and the surrounding field. We note that we do not account for resolved radio emission in our stacking, but at $z=2$ the resolution of the 1.4\,GHz map is $\sim$\,$25\times14$\,kpc and we do not expect SMGs to be significantly resolved on these scales~\citep{Biggs08}. We measure $\sim$\,$2.5 \times$ higher radio fluxes, at a significance of 2.8\,$\sigma$, for 3.6\,$\mu$m sources at $z=1$--3 in the ALMA ``Blank'' and ``Detected'' maps compared to the field. We note that this analysis is limited by the small number of 3.6\,$\mu$m sources considered in the ALMA maps and the depth of the radio map combined with the expected faint 1.4\,GHz flux distribution of SMGs (Figure~\ref{fig:maghist}). If we instead only consider the ALMA ``Blank'' maps we measure $\sim$\,$3.5 \times$ higher radio fluxes, at a decreased significance of 2.0\,$\sigma$, compared to the field. The significance of these results means they only provide tentative evidence that the ALMA maps contain 3.6\,$\mu$m sources, at  $z=1$--3, which are typically brighter in the submm, compared to the field.

The flux limit of the original LESS survey was 4.4\,mJy\,beam$^{-1}$, and hence our results are insufficient to fully explain the ALMA blank maps. There are two important caveats with this result. The first is that we expect at least three of the LABOCA sources to be spurious detections, which will downweight our stacking results to lower values of $S_{870}$. The second is that although we selected sources at 3.6\,$\mu$m, the requirement for a photometric redshift means each source must be detected in $>4$ wave-bands. If we have the same proportion of sources detected in $<4$ wave-bands as the MAIN sample, i.e.\ 20 per cent, this would explain a further fraction of the missing flux. Although we cannot explain all of the missing flux, these results do indicate that the ALMA blank maps contain multiple faint SMGs, below the detection limit of the ALESS survey. Crucially we find that there is an excess of sources at $z$\,$\sim$\,$2.5$ in these maps, which suggests the redshift distribution of faint SMGs appears to match the ALMA-detected SMGs.

\subsection{Constraints on SFH}
\label{subsubsec:sfh}
The primary use of {\sc hyperz} is to derive photometric redshifts, however in the SED fitting procedure  {\sc hyperz} also determines the best fit star formation history (SFH) for each source. We now investigate the reliability of the returned SFH parameters. We find that 52 (68 per cent), 15 (19 per cent), 6 (8 per cent) and 4 (5 per cent) of the ALESS SMGs have SFHs corresponding to the burst, 1\,Gyr, 5\,Gyr and constant templates respectively. While this appears to indicate a strong preference for the instantaneous burst SFH, we test for degeneracy in our results by re-running {\sc hyperz} allowing just the constant or just the burst SFHs. The SED fits, for the two SFHs, are indistinguishable, with a median $\Delta \chi^2_{\rm red}$ between the constant and burst SFH of $0.34^{+0.16}_{-0.09}$. The SED fits return a median age of $35\pm15$\,Myr and $1.0 \pm 0.4$\,Gyr for the burst and constant SFHs respectively. In \S\,\ref{subsec:masses} we discuss the uncertainties introduced into stellar mass estimates for SMGs from these unconstrained SFHs.

\begin{figure}
\centerline{ \psfig{figure= 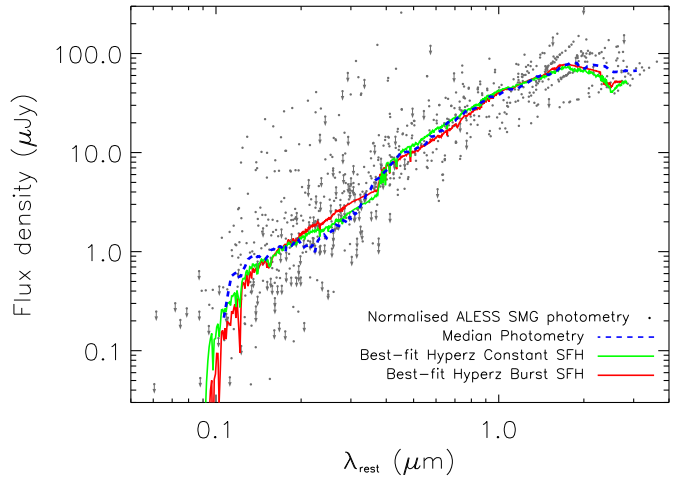,angle=90,width=\columnwidth}}
\caption{ The photometry for the ALESS SMGs, de-redshifted and normalized by their median $H$-band absolute magnitude. We show the running median (dashed line), which represents the SED of an average SMG, and exhibits a steep red spectrum indicative of strong reddening but with a hint of a break at 0.4\,$\mu$m due to Balmer or 4000\,\AA\ break. We indicate non-detections with arrows, and set the flux for these values to zero when calculating the running median. To test whether we can distinguish between different SFHs for the ALESS SMGs, we measure the median photometry through each filter in Table~\ref{table:depths}. We perform SED fitting on the average photometry using {\sc hyperz}, but allow only the two extremes of SFH, a constant SFH and an instantaneous burst. The best-fit corresponds to a constant SFH with an age of 2.3\,Gyr and $A_{\rm{v}} = 1.5$ (best-fit Burst; 30\,Myr and $A_{\rm{v}} = 1.8$), however we find the two SFHs are indistinguishable with $\Delta \chi^{2} = 0.2$. We use the Constant SFH SED to create an average optical\,--\,far-infrared SMG template, which is presented in~\citet{Swinbank13}. 
}
 \label{fig:composite}
\end{figure}

To investigate whether we can extract any further information about the SFHs from the ALESS photometry we construct the SED for the ``average'' ALESS SMG. In Figure~\ref{fig:composite}, we present the de-redshifted photometry for the ALESS SMGs normalized by restframe $H$--band luminosity. The composite SED shows a steep red spectrum consistent with strong dust reddening, as expected for SMGs. However, there may also be a hint of a break at $\sim$\,$4000$\,\AA. If this feature is indeed real it is most likely from a Balmer break, which would suggest the presence of stars with ages $\ge 10^{8}$\,yr.  We derive photometry for the average ALESS SMG by convolving the running median with the photometric filters used in this work. As we observe a hint of a Balmer break in the SED of the typical SMG, which could help differentiate between the SFHs, we also include an extra filter close to the break to provide a stronger test of the similarity of the models to the average photometry in this area (for this we use a $Y$-band filter shifted in wavelength to lie directly between the $z$ and $J$ filters). We then fit the average photometry redshifted to $z = 2.5$ using {\sc hyperz}, and compare to both the constant and instantaneous burst SFHs.

The constant SFH provides the best-fit to the median SED ($\chi_{r} = 1.0$), however we cannot reliably distinguish the models, which have $\Delta \chi_{\rm r} = 0.2$. The constant SFH has a burst age of 2.3\,Gyr, $A_{V} = 1.5$ and corresponding L$_{H}$\,/\,M$_{\star}$ of $\sim$\,$3$ while the instantaneous burst has an age of 30\,Myr, $A_{V} = 1.8$ and corresponding L$_{H}$\,/\,M$_{\star}$ of $\sim$\,$15$. The derived L$_{H}$\,/\,M$_{\star}$ and ages are very different and we conclude that even for limited selection of SFHs we consider for the ALESS SMGs it is not possible to distinguish between each SFH in a statistically robust manner. This is in agreement with previous work which demonstrates the difficulty in constraining the individual SFHs of high-redshift SMGs with SED fitting~\citep{Hainline11, Michalowski12}.

Although we find it is not possible to distinguish between the SFHs of the ALESS SMGs, the reddening correction returned by {\sc hyperz} appears consistent. Considering all SFHs we find a median reddening correction of $A_V^{\rm all} = 1.7 \pm 0.1$, and $A_V^{\rm const} = 2.0 \pm 0.1$ for the constant SFH alone. The reddening correction is an average correction across the entire galaxy, however, the dust in SMGs is likely to be clumpy~\citep{swinbank10Nature,Danielson11,Hodge12,Menendez13}, and as such it is likely to be considerably higher in the star-forming regions. To confirm this we derive SFRs from the dust-corrected restframe UV emission, at 1500\,\AA, of each ALESS SMG, following~\citet{Kennicutt98}, and compare these values to the far-infrared SFRs derived by~\citet{Swinbank13}. To bring the UV-derived SFR into agreement with SFR$_{\rm FIR}$, requires a median reddening correction of $A_{V} =  2.4 \pm 0.1$, or an additional $\sim$\,$0.7$\,mag to the $A_V$ derived from SED fitting, indicating that star formation in the SMGs is occurring in highly obscured regions. We note that the UV-derived SFR indicator is only likely to be reliable for a constant SFR SFH at ages of $>100$\,Myr, and as such it is likely that our UV-derived SFR is overestimated, and that the reddening correction is higher than $A_{V}$\,$\sim$\,$2.4$.

%
%

%
%
\begin{figure*}
\centerline{ 
\psfig{figure= 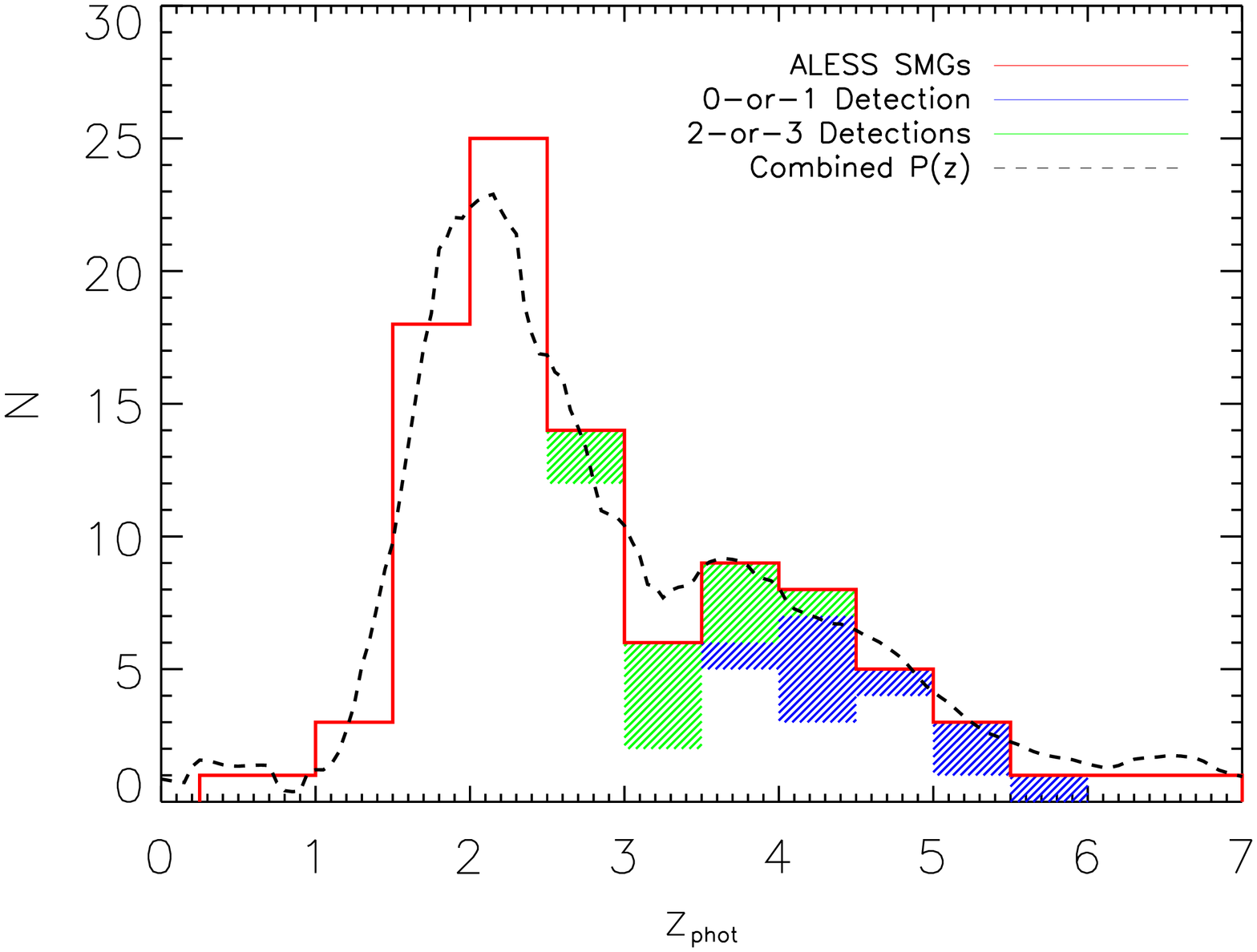,angle=90.,width=\columnwidth}
\hfill
\psfig{figure= 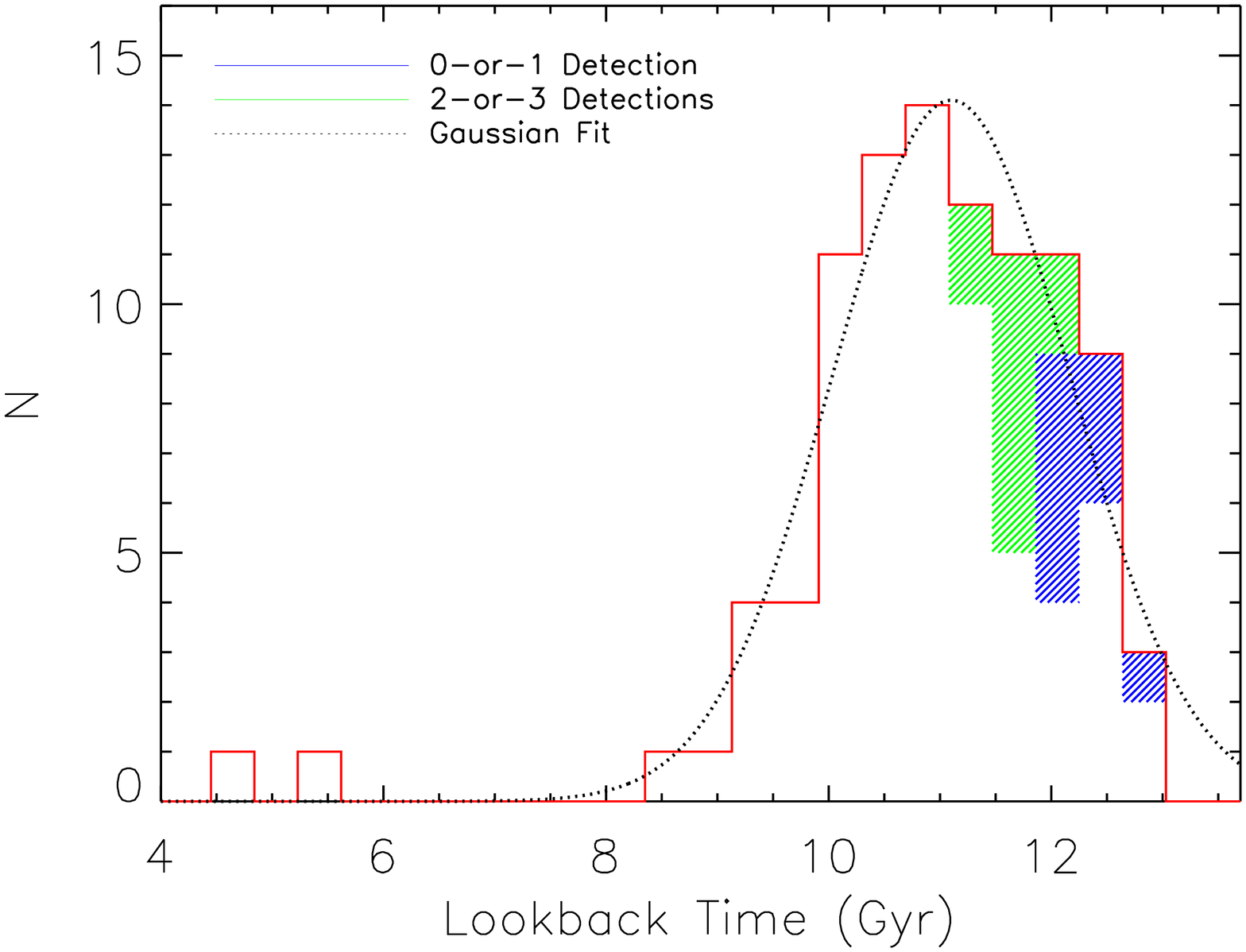,angle=90.,width=\columnwidth}}
\centerline{\psfig{figure= 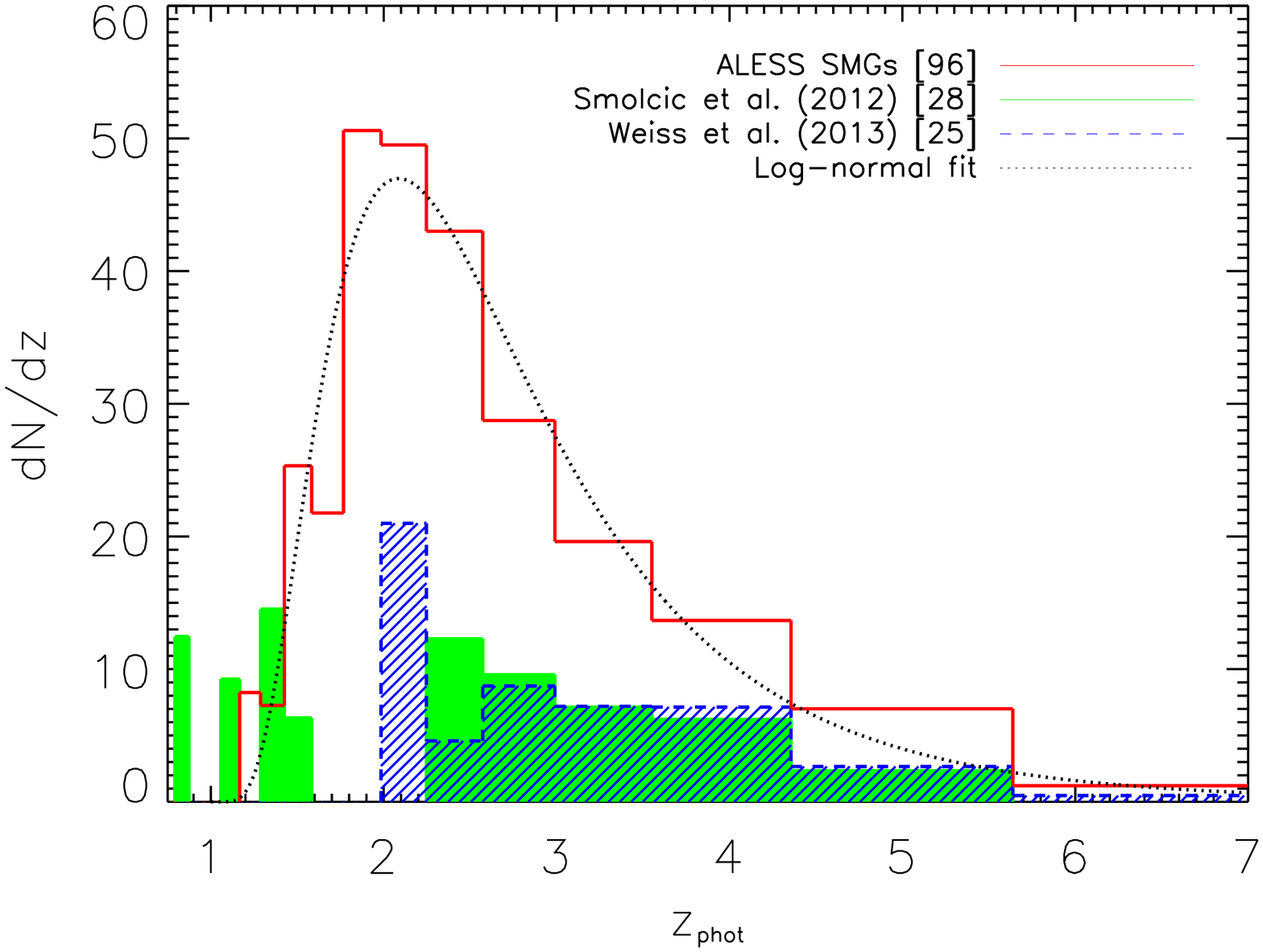,angle=90.,width=\columnwidth} }
\caption{ {\it Top Left:} The complete redshift distribution of the ALESS SMGs. We assign redshifts to SMGs detected in 0-or-1 and 2-or-3 wave-bands by completing the $H$--band absolute magnitude distribution at $z > 2.5$, as described in \S\,\ref{subsubsec:missing}. We combine the probability distribution for the photometric redshift of each SMG, and overlay this as a dashed line. The combined probability distribution is in close agreement with the shape of the redshift distribution, indicating that the distribution is not sensitive to the uncertainties on individual photometric redshifts, or secondary redshift solutions. {\it Top Right:} The complete distribution of ALESS SMGs as a function of time. We find the distribution is well-described by a Gaussian centered at $11.10 \pm 0.05$\,Gyr (equivalent to $z = 2.6 \pm 0.1$), with a width of $1.07 \pm 0.05$\,Gyr.  {\it Bottom:} The redshift distribution of the ALESS SMGs, binned uniformly in time, and normalized by the width of each bin. We find the redshift distribution is well-represented by a log-normal distribution (see Eqn\,2) with $\mu = 1.53 \pm 0.02$ and $\sigma_{\rm z} = 0.59 \pm 0.01$. For comparison we show the redshift distribution from ~\citet{Smolcic12}, an interferometric study of 28 millimeter-selected SMGs, containing spectroscopic and photometric redshifts. We also show the spectroscopic redshift distribution from a similar interferometric study of 25 millimeter-selected {\it lensed} SMGs from Wei$\ss$ et al.\,(2013; hatched), choosing the robust or best-guess redshifts from their analysis. We note that we have included the lensing probability as function of redshift, given in Wei$\ss$ et al.\,(2013), in the distribution. The SMG samples presented here have selection functions that are difficult to quantify (especially the lensed sample of Wei$\ss$ et al.\,(2013)), and hence do not have a well defined survey area. As such, we present the redshift distributions in terms of raw number counts but provide the number of sources in each sample in the legend in the top right. In contrast to these previous studies, the redshift distribution of the ALESS SMGs does not show evidence of a flat distribution between $z$\,$\sim$\,$2$--6, and displays a clear peak in the distribution at $z = 2$.
}
\label{fig:zages}
\end{figure*}

\section{Discussion}
\label{sec:results}
\subsection{Redshift Distribution}
\label{subsec:zdist}

The complete redshift distribution of the 96 ALESS SMGs in our sample has a median redshift of $z_{phot} = 2.5 \pm 0.2$ and a tail to high redshift, with $35 \pm 5$ per cent of sources lying at $z > 3$. As an initial comparison to the ALESS SMGs we use the spectroscopically confirmed, radio-identified, SMG sample from Chapman et al.\ (2005; C05). The C05 sample has a median redshift of $z = 2.2 \pm 0.1$, in agreement with our results, however there are notable discrepancies between the samples. The C05 sources are radio-selected, which due to the positive K-correction is likely to bias their results to lower redshifts. Indeed the highest redshift SMG in the C05 sample is $z = 3.6$, and the distribution does not show such a pronounced tail to high redshifts as we observe in the ALESS SMG distribution. We also note differences between the samples at $z < 1.5$. The C05 sample contains a significant number of SMGs at these redshifts (25 per cent), but only five ALESS SMG, or 6 per cent, lie at $z < 1.5$. A two-sided Kolmogorov-Smirnov (KS) test between the ALESS SMGs and C05 indicates that there is a 13 per cent probability that the samples are drawn from the same parent distribution. A fairer comparison is to only consider the ALESS SMGs with radio fluxes $S_{1.4} > 40$\,$\mu$Jy, roughly the selection limit of the C05 sample. Here, the median redshift of $S_{1.4} > 40$\,$\mu$Jy ALESS SMGs is $z_{phot} = 2.3 \pm 0.1$, and the above analysis remains unchanged. We note the median redshift of the radio-detected ALESS SMGs is $ z_{phot} = 2.3 \pm 0.1$, and is lower than the radio-non-detections which have a median redshift of $z_{phot} = 3.0 \pm 0.3$.

We caution that the ALESS sample is selected from the original LABOCA survey, which had a detection threshold of 4.4\,mJy. Our ALMA observations reach a typical depth of 1.4\,mJy (3.5\,$\sigma$) and so we have SMGs in our sample below the original LABOCA limit. These SMGs are biased in their selection, and are only in our SMG sample due to their on-sky clustering with other SMGs. It is difficult to quantify the effect of these SMGs on our redshift distribution, but we note that we do not see any significant trend between redshift and 870\,$\mu$m flux density (see Figure~\ref{fig:870z}). If we split the ALESS SMGs into sub-samples based on the LABOCA detection limit, we find the median redshift for SMGs above 4.4\,mJy is $z_{phot} = 2.5 \pm 0.2$, and $z_{phot} = 2.6 \pm 0.3$ below 4.4\,mJy. However, we should also consider the ALMA maps where the original LABOCA source has not fragmented into multiple components. The median redshift of these 45 ``isolated'' SMGs is $z_{phot} = 2.3 \pm 0.2$, consistent with the complete sample of 96 SMGs.

A number of SMGs in our sample have secondary redshift solutions (which correspond to secondary minima in $\chi^2$, e.g.\ Figure\,A1) or have large uncertainties in their photometric redshifts. To investigate whether these could significantly affect the shape of the redshift distribution we calculate the redshift probability distribution for each SMG,
and normalize the integral of the distribution. For the SMGs detected in $<3$ wave-bands we assign a uniform probability distribution between the detection limits described in \S\,\ref{subsubsec:missing}. We combine the redshift probability distributions for each SMG and show the combined redshift distribution in Figure~\ref{fig:zages}. We find that the redshift distribution derived from the combined probability distributions is in excellent agreement with the ``best--fit'' redshift distribution, indicating that while secondary minima and large redshift uncertainties are important for individual sources, they do not significantly affect the shape of the redshift distribution.

In Figure~\ref{fig:zages} we show the redshift distribution of the ALESS SMGs as a function of look-back time. The distribution is well-described by a Gaussian ($\chi^{2}_{r} = 0.99$) of the form 
\begin{equation}
  N(T) = A\,e[\,- ( T-T{_0} )^2 \,/\, 2\sigma^2_{\rm T}\,],
\end{equation} 

where $A = 14.10 \pm 0.55$, $T_{0} = 11.10 \pm 0.05$ and $\sigma_{\rm T} = 1.07 \pm 0.05$, (of course, this function extends beyond the Hubble time and hence must be truncated at 13.7\,Gyr). We note that the high redshift tail to the distribution is a less pronounced feature when the distribution is parametrized, linearly, by age\footnote{We note that the ALESS SMG redshift distribution is well described by a  log-normal distribution of the form: 
\begin{equation} 
 \frac{dN}{dz} = \frac{B}{ (z-1) \sigma_{\rm z}} e^{ - [   ( {\rm ln}(z-1) - \mu)^2  / 2 \sigma_{\rm z}^2    ]}
\end{equation}
where $B = 89.2 \pm 1.7$, $\mu = 1.53 \pm 0.02 $ and $\sigma_{\rm z} = 0.59 \pm 0.01$ (see also~\citealt{Yun12})}.

One of the main results from our ALESS survey is that the ``robust'' Radio/1.4\,Ghz and MIPS/24$\,\mu$m identifications for the multiwavelength counterpart to the original LABOCA detection were only 80 per cent correct, and 45 per cent complete (\citealt{Hodge12}). As such, we do not compare our results to redshift distributions derived from single-dish sub-mm/mm surveys (i.e.~\citealt{Aretxaga07,Chapin09,Wardlow11,Yun12,Casey13}). Instead, we restrict the comparison to recent millimeter interferometric observations of other, albeit small, samples of SMGs. First we compare to the 28 SMGs from \citet{Smolcic12}, which can be split into two distinct subsets: 1) 17 1.1\,mm--selected sources, with follow-up observations at 890\,$\mu$m with the Submm Array (SMA) and 2) 16 870\,$\mu$m--selected sources, with follow-up observations at 1.3\,mm with the Plateau de Bure Interferometer (PdBI). Five sources are duplicated in both samples. 

The 1.1\,mm selected sample from \citet{Smolcic12} has a median redshift of $z = 2.8 \pm 0.4$, which is comprised of a mixture of seven spectroscopic redshifts, seven photometric redshifts and three redshifts derived from the mm\,--\,radio relation. (see Figure~\ref{fig:firradio}). Due to the shape of the FIR SED we might expect samples selected at longer wavelength to lie at higher redshift, and indeed we observe this for the ALESS SMGs when the sample is split into detections which peak at 250, 350 and 500\,$\mu$m~\citep{Swinbank13}. As such it is unsurprising that the $1.1$\,mm selected sample from~\citet{Smolcic12} has a marginally higher median redshift, although we note that within the errors it is in agreement with the median of the ALESS SMGs. The second sample consists of 870\,$\mu$m selected galaxies, with interferometric observations at 1.3\,mm. The initial selection criteria at 870\,$\mu$m means the sample is a closer match to the ALESS SMGs (although they must still be brighter than $\sim$\,$1.5$\,mJy at 1.3\,mm) and indeed the median redshift is $z = 2.6 \pm 0.6$ (five spectroscopic/eight photometric/three mm--radio redshifts), in good agreement with the results presented here.

Overall the combined mm and submm samples from \citet{Smolcic12} contains 28 SMGs with a median redshift of $z = 2.6 \pm 0.4$, in agreement with the ALESS SMGs. We note that redshifts for five SMGs from the~\citet{Smolcic12} sample, are derived from the mm\,--\,radio relation and are claimed to lie at $z > 2.6$. As we have noted, this relation displays an order of magnitude scatter at a fixed redshift (Figure~\ref{fig:firradio}), however these sources are not detected in the photometry employed by~\citet{Smolcic12} and hence are indeed likely to lie at high redshifts. We note two interesting features of the~\citet{Smolcic12} redshift distribution: firstly there is a deficit of SMGs at $z$\,$\sim$\,$2$ which lies close to the peak of the ALESS SMG redshift distribution (see Figure~\ref{fig:zages}). Secondly, a further possible discrepancy between the samples is the shape of the distribution from $z = 2.5$--4.5, where the ALESS SMG redshift distribution declines whereas the~\citet{Smolcic12} distribution remains relatively flat. However, given the limited number of sources in the comparison we caution against strong conclusions.

We can also compare to another ALMA sample.~\citet{weiss13} recently used ALMA to search for molecular emission lines from a sample of 28 strongly lensed SMGs (see also~\citealt{Vieira13}), selected from observations at 1.4\,mm with the South Pole Telescope (SPT). Given the large beam size ($\sim$\,$1'$) of the SPT the sources were also required to be detected at $870$\,$\mu$m with LABOCA. \citet{weiss13} obtain secure redshifts for 20 SMGs in their sample and provide tentative redshifts, derived from single line identification, for five sources (three sources are not detected in emission). Considering the lower estimates for the tentative redshifts the sample has a median redshift of $z = 3.4 \pm 0.5$, and for the upper limits on the tentative redshifts $z = 3.8 \pm 0.4$, with the true median likely lying between the two values. 

The median redshift for the SPT sources is higher than that of the ALESS SMGs, although the two are formally in agreement at a $\sim$\,$2$--$\sigma$ confidence level. However, the most noticeable discrepancy between the samples lies in the shape of the distributions.  Firstly, there are no robust spectroscopic redshift SPT sources at $ z < 2 $, whereas $\sim$\,$25$ per cent of the ALESS SMGs lie at $z_{phot} < 2$, of which 7 are spectroscopically confirmed to lie at $ z < 2 $ (see Figure~\ref{fig:delz}; Danielson et al.\ in prep). Secondly, the ALESS photometric redshift distribution has a tail to high redshift ($z_{phot}$\,$\sim$\,$6$), however the distribution declines steadily between $z_{phot} = 2$--6. In contrast the SPT distribution is relatively flat between $z = 2$--6. As stated by~\citet{weiss13}, their bright 1.4mm flux selection criteria, $S{_{1.4 {\rm mm}}} > 20$\,mJy ensures they only select lensed sources. This potentially introduces two biases into the redshift distribution: 1) The lensing probability is a function of redshift; for example, from $z = 1.5$ (where there are no SPT sources) to $z = 6$ the probability of strong gravitational lensing ($\mu$\,$\sim$\,$10$) increases from $P(z)=0.6\times 10^{-4}$ to $3 \times 10^{-4}$ (i.e.\ a factor of 5 increase; see Figure~6 from~\citealt{weiss13}). In Figure~\ref{fig:zages} we show the redshift distribution for the SPT sample, corrected by the lensing probability function given in Wei$\ss$ et al.\,(2013; see also~\citealt{Hezaveh12}). We note this has a significant effect on the shape of the distribution, bringing it into closer agreement with the ALESS sample at $z>2.0$. The weighted median of the corrected Wei$\ss$ et al.\,(2013) sample is $z$\,$\sim$\,$3.1 \pm 0.3$, which is in agreement with the median redshift of the $z>2.0$ ALESS SMGs of $z_{phot} = 3.0 \pm 0.4$.  2) Evolution in the source size with redshift will affect the lensing magnification, as increasingly compact sources are more highly amplified (\citealt{Hezaveh12}, but see discussion in~\citealt{weiss13})

Given the different selection wavelengths between the ALESS SMGs and the SPT sample, and the potentially uncertain effects of lensing, we caution against drawing far-reaching conclusions between these two redshift distributions.  To resolve any tension between these two samples, we require spectroscopic redshifts for an unlensed sample of SMGs, selected at both 870\,$\mu$m and 1.4\,mm, however given the optical properties of these sources this will only be feasible using a blind redshift search of molecular emission lines, similar to that employed by~\citet{weiss13}.

\subsection{Pairs \& Multi-component SMGs}
\label{subsec:pairs}
At least 35 per cent of the LESS submm sources fragment into multiple SMGs. Using our photometric redshifts we can now test whether these multiple SMGs are physically associated or simply due to projection effects. In total we derive photometric redshifts for 18 SMG pairs, of which the photometric redshifts of all but one agree at a 3--$\sigma$ confidence level. However, as the median combined uncertainty on the photometric redshift of each pair is $\sigma_z = 0.3$, this simply highlights these large uncertainties. Furthermore this uncertainty on each pair is similar to the width of the redshift distribution of the whole population and so we expect SMGs to appear as pairs, irrespective of whether they are associated.  

A more sensitive method to test for small scale clustering of SMGs is to investigate if there is a significant excess of ALESS SMGs, at similar redshifts, and in the same ALMA map, compared to pairs of SMGs drawn from different ALMA maps. To test for any excess we initially create random pairs of SMGs, drawn from different ALMA maps, and measure $\Delta z = z^{phot}_1 - z^{phot}_2$. We then compare this to the distribution of $\Delta z$ we measure between SMGs in the same ALMA map. To take into account the errors on each photometric redshift we Monte Carlo the redshift for each SMG within the associated error bar, and repeat the entire procedure 1000 times. We identify a tentative excess of $2.8 \pm 1.5$ pairs, from the sample of 18, at $0 < \Delta z < 0.5 $ in the ALMA maps containing multiple sources. However this is not a significant result\,\footnote{We note that including sources from the Supplementary ALESS catalog in this analysis does not increase the significance of the result.}. 

The strongest candidates for an associated pair of SMGs are ALESS\,55.1 and 55.5. These SMGs are separated by $\sim$\,$2''$, have merged 870\,$\mu$m emission and straddle a single optical\,--\,near-infrared counterpart; the photometric redshift of this source indicates it is not a lensing system. We note that the photometry for these SMGs is drawn from the same optical\,--\,near-infrared source. A further two LABOCA sources, LESS\,67 and LESS\,116, fragment into multiple ALESS SMGs with similarly small on-sky separations ($< 3''$) however we cannot verify if they are physically associated. 

%
%
\begin{figure*}
\centerline{
\psfig{figure= 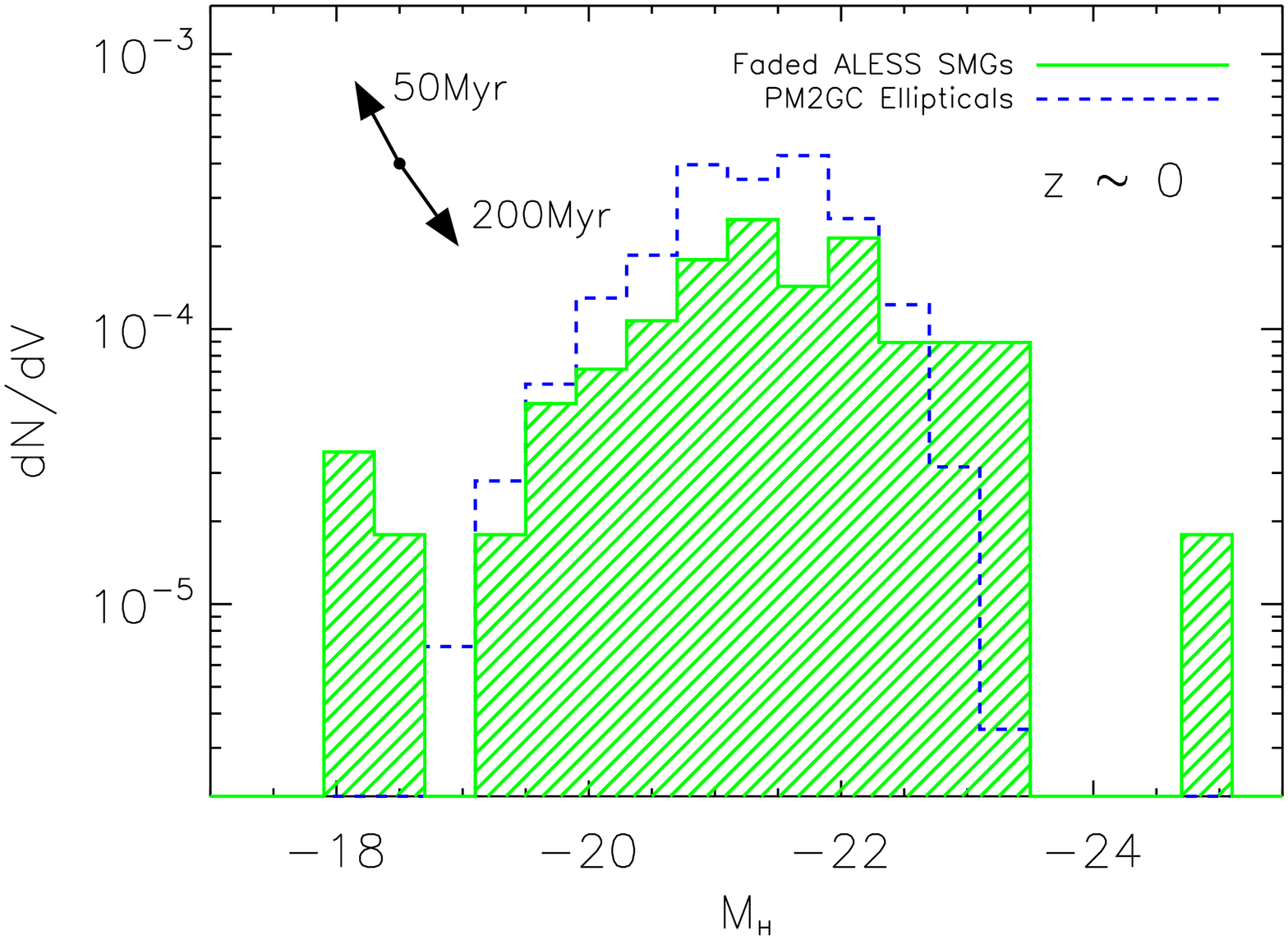,angle=90.,width=\columnwidth}
\hfill
\psfig{figure= 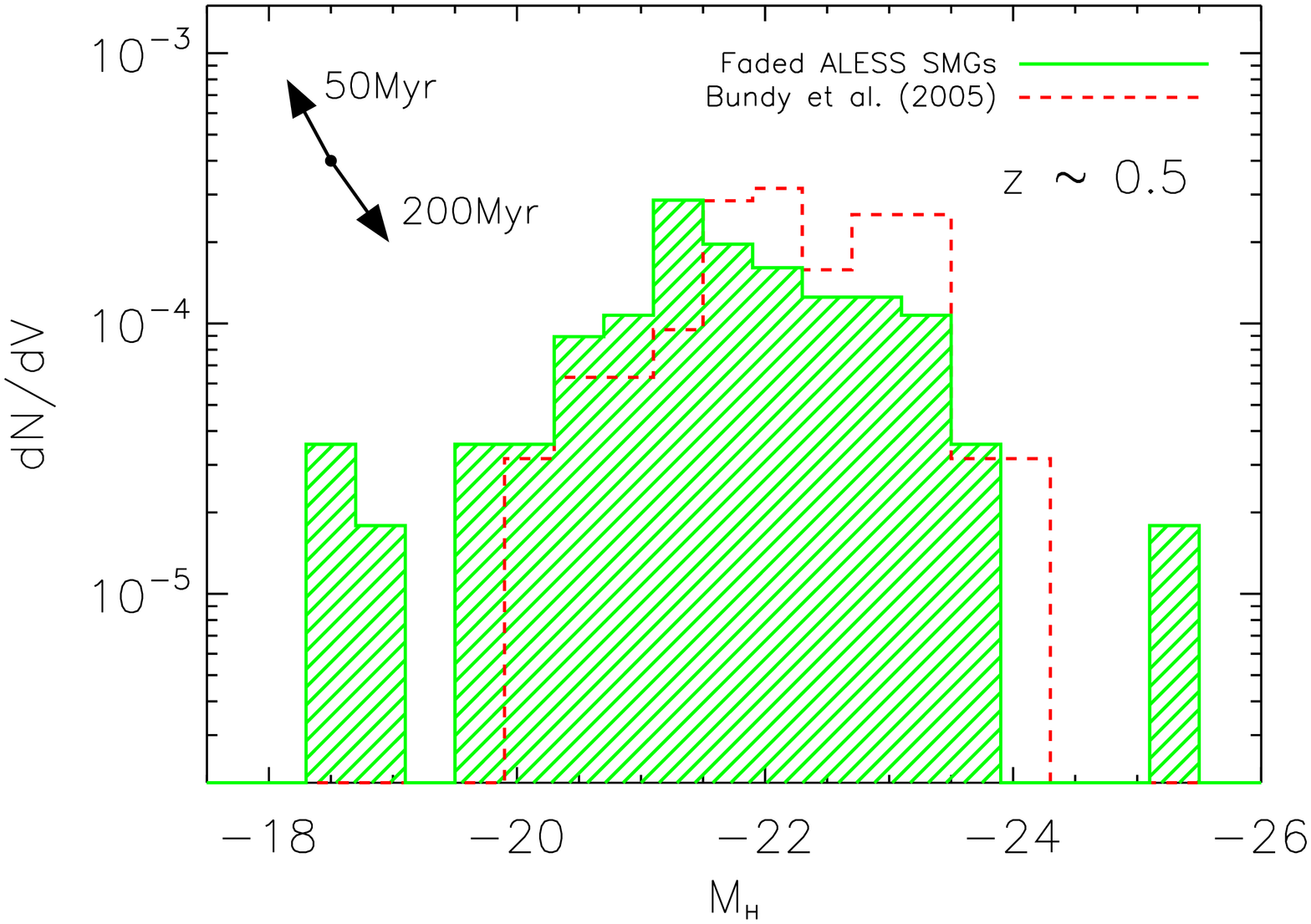,angle=90.,width=\columnwidth}}
\centerline{\psfig{figure= 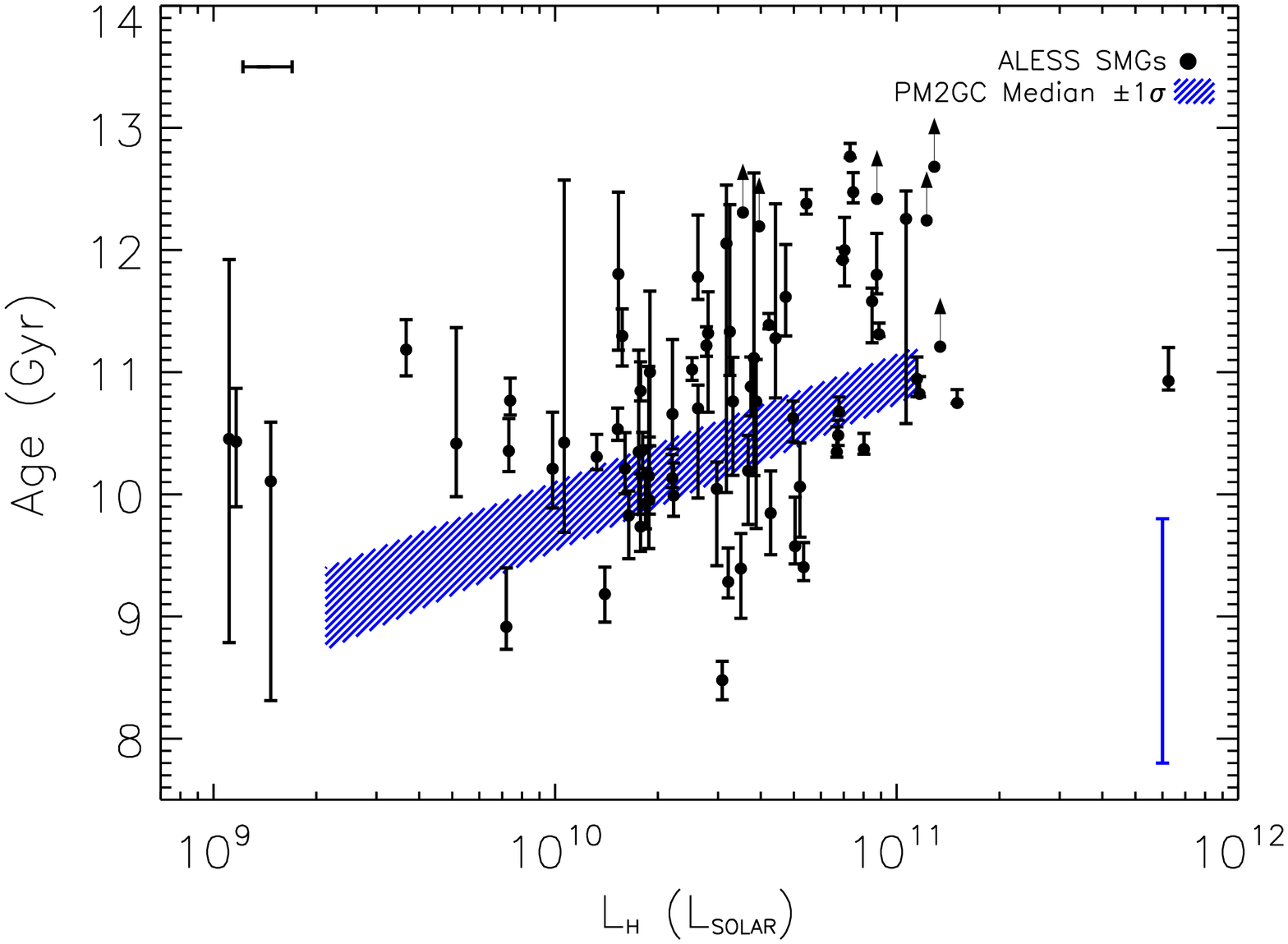,angle=90,width=\columnwidth}}
\caption{ {\it Top Left:} The absolute $H$--band magnitude distribution of the ALESS SMGs, faded to the present day assuming a 100\,Myr burst duration SFH. We adopt the SMG number counts from~\citet{Karim13} to extrapolate the ALESS SMG sample to 1\,mJy, and duty cycle correct the volume density. In the upper left, vectors indicate the effect of adopting either a 50\,Myr or 200\,Myr burst. In comparison we show the absolute $H$--band magnitude distribution of a morphologically classified, volume-limited sample, of elliptical galaxies over the redshift range 0.03--0.1 (PM2GC;~\citealt{Calvi11,Calvi13}). We conclude good agreement in both typical luminosity range, and space density, of faded SMGs to local ellipticals. {\it Top Right:} We again show the ALESS SMGs, faded to $z = 0.5$. We compare this to an absolute $H$--band magnitude distribution for a morphologically classified sample of elliptical galaxies at $z$\,$\sim$\,$0.5$~\citep{bundy05}. We find the number density of the SMGs and intermediate redshift ellipticals are in agreement, however the ALESS SMGs are on average $\sim$\,$0.5$\,magnitudes fainter. {\it Bottom:} A comparison of the mass-weighted ages of the PM2GC sample of elliptical galaxies to the current look-back age of the ALESS SMGs. The error bar in the bottom right of the figure shows the typical MW-age error at these old ages, as derived from the spectrophotometric modeling~\citep{Poggianti13}. Despite the large systematic uncertainty, the PM2GC ellipticals have mass-weighted ages which are in broad agreement with the ALESS SMGs. This is consistent with a simple evolutionary model where SMGs are the progenitors of local elliptical galaxies. The typical error in $L_{H}$ for the ALESS SMGs is shown in the upper left of the figure.
}

 \label{fig:mhfading}
\end{figure*}

\subsection{Stellar Masses}
\label{subsec:masses}
We estimate stellar masses for the ALESS SMGs from their absolute $H$--band magnitudes, which we note are calculated from the best-fit SED and take into account the effects of the k-correction. We select this wave-band as a compromise between limiting the effects of dust extinction (the correction decreases with increasing wavelength), and the potential contribution of thermally pulsating asymptotic giant branch (TP--AGB) stars (which increases at shorter wavelengths [\citealt{Henriques11}]).

The median absolute $H$--band magnitude for the 77 ALESS SMGs detected in $>4$ wave-bands is $-24.56 \pm 0.15$. As discussed in \S\,\ref{sec:photz}, by assuming the ALESS SMGs detected in $<4$ wave-bands are missed due to our photometric selection limits, we can complete the $M_{\rm H}$ distribution by enforcing the condition that the distribution is not bimodal.  Using the complete $M_{\rm H}$ distribution we measure a median absolute $H$--band magnitude for the ALESS SMGs of $-24.33 \pm 0.15$. We note that this is corrected for a median reddening of $A_V = 1.7 \pm 0.1$ (\S~\ref{subsubsec:sfh}). The median value of $M_{\rm H}$ for the ALESS SMGs is in agreement with previous work by \citet{Hainline11}, who measure a median $M_{\rm H} = -24.45\,\pm\,0.20$ for the stellar emission from a sample of 65 spectroscopically confirmed, radio-identified, SMGs from the~\citet{Chapman05} sample.

To convert these absolute $H$--band magnitudes to stellar masses we must next adopt a mass-to-light ratio. As we discussed in \S\,\ref{sec:photz} the SFHs for the ALESS SMGs are highly degenerate, and it is not possible to accurately distinguish between the model SFHs. To determine a mass-to-light ratio we therefore consider the range spanned by the best-fit Burst and Constant SFHs (the two extremes of SFH we consider). We use the~\citet{Bruzual03} Simple Stellar Populations (SSPs) to construct an evolved spectrum from the best-fit constant and burst SFHs for each SMG, and measure the absolute $H$--band magnitude\,\footnote{We enforce the condition that the age of the star-formation event is $20$\,Myr\,$\le$\,$t_{\rm age}$\,$\le$\,$1$\,Gyr.}. We then define the stellar mass as the total mass in stars and stellar remnants, using {\sc Starburst99} to determine the mass lost due to winds and supernovae~\citep{Leitherer99,Vazquez05, Leitherer10}.

The median mass-to-light ratio for the ALESS SMGs is $M$\,/\,$L_{H} = 0.08 \pm 0.02$ for the burst SFH, and $M$\,/\,$L_{H} = 0.25 \pm 0.05$ for the constant SFH, however we caution the mass-to-light ratios between the burst and the constant SFH solutions for individual SMGs vary by $> 3 \times$ for $\sim$\,$40$\, per cent  of the sample. Nevertheless, we apply the best fit mass-to-light ratios for each of the 77 ALESS SMGs detected in $> 3$ wave-bands to their dust-corrected absolute $H$--band magnitudes, and determine median stellar masses of $M_{\star} = (7.4 \pm 1.0) \times 10^{10}$\,M$_{\odot}$ for the Burst SFH, $M_{\star} = (9.2 \pm 0.8) \times 10^{10}$\,M$_{\odot}$ for the Constant SFH, and $M_{\star} = (8.9 \pm 1.4) \times 10^{10}$\,M$_{\odot}$ if we take the average of the mass estimates for each SMG. For the 19 SMGs detected in $ < 4$ wave-bands we do not have sufficient information on the SFH to determine a mass-to-light ratio. If we adopt the median mass-to light ratio for the detected SMGs, the stellar mass of the non-detected SMGs is $M_{\star} = (2.9 \pm 0.4) \times 10^{10}$\,M$_{\odot}$ for the Burst SFH, $M_{\star} = (8.5 \pm 1.3) \times 10^{10}$\,M$_{\odot}$ for the Constant SFH, and $M_{\star} = (5.7 \pm  0.8) \times 10^{10}$\,M$_{\odot}$ for the average of the mass estimates. Combining the samples we derive a median stellar mass for the 96 ALESS SMGs of $M_{\star} = (8 \pm  1) \times 10^{10}$\,M$_{\odot}$, when taking the mass as the average of the the Burst and Constant values. We note that all the stellar masses quoted here are for a Salpeter Initial Mass Function (IMF), and the median mass-to-light ratio is $M$\,/\,$L_{H} = 0.15 \pm 0.01$\,\footnote{The mass-to-light ratio for the instantaneous burst SFH is sensitive to changes on the order $\sim$\,$10$\,Myr, and over the range 10\,--\,40\,Myr varies from $M$\,/\,$L_{H}$\,$\sim$\,$0.02$--$0.1$. However, when considering the range 10\,--\,40\,Myr the median stellar mass remains stable at $M_{\star}$\,$\sim$\,$( 8 \pm 1 ) \times 10^{10}$\,M$_{\odot}$.} (the average mass-to-light ratio between a 100\,Myr Burst and Constant SFH is $M$\,/\,$L_{H} = 0.14$).
 
The median stellar mass for the ALESS SMGs is lower than that found for the C05 sample of SMGs by Hainline et al. (2011; $M_{\star} = 1.6 \pm 0.3 \times 10^{11}$\,M$_{\odot}$) [see also Michalowski et al. 2010; $M_{\star} = 3.5 \times 10^{11}$\,M$_{\odot}$]. Given the uncertainty surrounding stellar mass estimates it is more informative to compare the absolute $H$--band magnitudes of the ALESS SMGs to the C05 sample. As stated earlier these are in agreement, and so any difference in the median stellar mass is due to differences in the mass-to-light ratios adopted. We note that the C05 sample of SMGs have significant contamination in $M_H$ due to AGN activity~\citep{Hainline11}, which we do not see for the ALESS SMGs. However, the median $M_H$ and masses for the C05 SMGs quoted here are corrected for that AGN contamination.

Although we highlight that the stellar masses for the ALESS SMGs are highly uncertain, we can crudely test their accuracy by comparing them to the dynamical masses, and CO--derived gas masses, of similar SMGs. \citet{Bothwell13} recently obtained observations of $^{12}$CO emission from 32 SMGs, drawn from the C05 sample. These SMGs have typical single-dish derived 870\,$\mu$m fluxes of 4--20\,mJy, and were found to have a median gas mass of $M_{gas} = (3.5 \pm 1.1) \times 10^{10}$\,M$_{\odot}$. We note that \citet{Swinbank13} used the dust masses of the ALESS SMGs to derive a median gas mass of $M_{gas} = 4.2 \pm 0.4 \times 10^{8}$\,M$_{\odot}$, comparable to the result from \citet{Bothwell13}. Combining the median gas mass from \citet{Bothwell13} with the median stellar mass of the ALESS SMGs, and assuming a dark matter contribution of $\sim$\,$25$ per cent, suggests that SMGs have typical dynamical masses of $\sim$1--$2 \times 10^{11}$\,$M_{\odot}$. Crucially our estimate of the dynamical mass is consistent with spectroscopic studies of resolved H$\alpha$ or $^{12}$CO emission lines, which demonstrate that SMGs typically have dynamical masses of 1--$2 \times 10^{11}$\,$M_{\odot}$ (\citealt{Swinbank04,susie12,Bothwell13}), inside a 5\,kpc radius.

\subsection{Evolution of SMGs: z=0}
\label{subsec:pathways}
We now investigate the possible properties of the descendants of the ALESS SMGs at the present day by modeling how much their $H$--band luminosity will fade between their observed redshift and the present redshift. First we must make assumptions about the future evolution of the ALESS SMGs, the most crucial of which is the duration of the SMG phase. As stated in \S\,\ref{subsec:masses}, based on existing CO studies of SMGs the ALESS SMGs are likely to have a median gas mass of $M_{gas}$\,$\sim$\,$(4 \pm 1) \times 10^{10}$\,M$_{\odot}$, and from~\citet{Swinbank13} they have a median SFR of $840 \pm 120$\,M$_{\odot}$yr$^{-1}$ for a Salpeter IMF. If the SFR remains constant, and all the gas is converted into stars, this suggests that the SMG phase has a maximum duration on the order of 100\,Myr (see also~\citealt{Swinbank06b,Hainline11,Hickox12}).

To measure the change in $H$--band luminosity of the ALESS SMGs we use the~\citet{Bruzual03} SSPs to model the SED evolution. On average we are seeing each SMG midway through its burst and so, for a SMG duration of 100\,Myr, we calculate the fading in $L_{H}$ between 50\,Myr into the burst, and the required age at the present day. We note that this assumes that the contribution to the fading from a pre-burst stellar population is negligible and that each SMG undergoes only a single burst.

The ALESS SMGs represent a complete survey over 0.25\,degree$^{2}$ and so we can also calculate their co-moving space density. We first extrapolate the ALESS sample to $S_{870} \ge 1$\,mJy, using the ALESS SMG number counts from~\citet{Karim13}, noting that we again make the assumption there is no dependence of $M_{H}$ on $S_{870}$. We also apply a factor of two correction to the number counts to account for the under-density of SMGs in the ECDFS (see~\citealt{Weiss09}). As the SMG phase has a finite duration we duty-cycle correct the number density following:

\begin{equation}
 \phi_{\rm D} = \rho_{{\rm SMG}} ( t_{\rm obs} / t_{\rm burst}  )
\end{equation}

where $\phi_{\rm D}$ is the comoving space density of SMG descendants, $\rho_{\rm SMG}$ is the observed space density of ALESS SMG, $t_{\rm obs}$ is the duration of the epoch that we observe the SMGs over and $t_{\rm burst}$ is the duration of the SMG phase. We estimate $t_{\rm obs}$ from the 10--90th percentiles of the redshift distribution, $1.6 < z < 4.5$, and as stated earlier we assume the SMG phase has a duration of 100\,Myr. Taking these corrections into account we estimate the volume density of the descendants of $S_{870} \ge 1$\,mJy SMG is $\sim$\,$(1.4 \pm 0.4) \times 10^{-3}$\,Mpc$^{-3}$. 

It has been suggested that SMGs may be the progenitors of local elliptical galaxies (e.g.\ \citealt{Lilly99, Genzel03, Blain04a, Swinbank06b, Tacconi08, Swinbank10}). We now test the relation of the descendants of the ALESS SMGs to local ellipticals using a morphologically classified sample of ellipticals galaxies, taken from the Padova Millennium Galaxy and Group Catalog (PM2GC; ~\citealt{Calvi11,Calvi13}). This catalog represents a volume-limited survey ($z$\,=\,$0.03$--0.1) over 38\,degrees$^2$, with morphologies determined by an automatic tool that mimics a visual classification (\citealt{calvi12}, see also~\citealt{Fasano12}). These galaxies were observed in the $Y-, H-$ and $K$--bands by the UKIDSS Large Area Survey~\citep{Lawrence07} and we derive absolute $H$--band magnitudes from the recent data release~\citep{Lawrence12}. This comparison sample of local elliptical galaxies, has a median redshift of $z = 0.08$, median absolute $H$--band magnitude of $M_{H} = -21.1 \pm 0.1$ and a space density of $(2.0 \pm 0.1) \times 10^{-3}$\,Mpc$^{-3}$. 

Using the observed redshift of the SMGs, and our adopted SFH, we individually fade each ALESS SMG to $z=0.08$, and estimate a median faded absolute $H$--band magnitude of $M_{H} = -21.2 \pm 0.2$. We show the ``faded'' distribution in Figure~\ref{fig:mhfading}, where we see very good agreement with absolute $H$--band magnitude distribution of the PM2GC ellipticals. As stated earlier, we estimate the space density of the descendants of ALESS SMGs is $(1.4 \pm 0.4) \times 10^{-3}$\,Mpc$^{-3}$, similar to the PM2GC ellipticals, $(2.0 \pm 0.1) \times 10^{-3}$\,Mpc$^{-3}$. We note that both the fading correction in $M_{H}$, and the number density, of the SMGs are dependent on the duration of the SMG phase. If we instead adopt a burst of 50 or 200\,Myr duration then the median absolute $H$--band magnitude is $M_{H} = -20.9 \pm 0.2$ or $M_{H} = -21.7 \pm 0.2$, and the number density is $(3 \pm 1)$ or $(0.7 \pm 0.2)$ $\times 10^{-3}$\,Mpc$^{-3}$, respectively, and these changes are shown by vectors in Figure~\ref{fig:mhfading}

We note that if the burst duration is indeed $\gsim$\,$200$\,Myr then $>$\,$10$ per cent of the SMG descendants would have a $H$--band absolute magnitude brighter than the brightest elliptical in the PM2GC sample. This excess of bright galaxies assumes SMGs undergo no future interactions, i.e.\ minor mergers, or subsequent star formation, which would only act to increase the total absolute $H$--band magnitude, and so make the discrepancy larger. We suggest that this makes burst durations of $>$\,$200$\,Myr unlikely. For our estimated burst duration of 100\,Myr the space density of SMGs is lower than local ellipticals, indicating the SMG phase could be $<$\,$100$\,Myr. A burst duration shorter than 100\,Myr would make the descendants of the ALESS SMGs fainter than the $z = 0$ elliptical sample, but have a larger space density. If we consider a burst duration of 50\,Myr then a dry-merger fraction of two-thirds would bring the number density into agreement with the PM2GC ellipticals, and the resulting median $M_{H}$ of the SMGs to $M_{H} = 21.4 \pm 0.1$. 

We now consider two further tests of this evolutionary model. First, we consider the mass-weighted stellar ages of the PM2GC ellipticals, calculated with a spectrophotometric model that finds the combination of SSP synthetic spectra that best-fits the observed spectroscopic and photometric features of each galaxy~\citep{Poggianti13}. As can be seen in Figure~\ref{fig:mhfading} these appear broadly consistent with the ages of the ALESS SMGs. The PM2GC ellipticals have a mass-weighted stellar age of $\sim$\,$10$\,Gyr ($z$\,$\sim$\,$2$) but with a systematic uncertainty of $\pm 2$\,Gyr~\citep{Poggianti13}, compared to the median age of the ALESS SMGs of $11.1 \pm 0.1$\,Gyr. This might indicate that the bulk of stellar mass in the ellipticals formed later than the current redshift of the ALESS SMGs, however given the systematic uncertainties and the difficulty in age-dating very old stellar populations we find the similarity in the ages striking. 

As a second comparison, we consider the the mass of the dark matter halos that the PM2GC ellipticals reside in. We use the halo mass catalogs from~\citet{Yang05}, and find that these ellipticals have a typical halo mass of $0.5 \times 10^{13}$\,M$_{\odot}$ with a 1--$\sigma$ range of $0.1$--$8 \times 10^{13}$\,M$_{\odot}$. This is consistent with the typical halo masses of SMGs descendants ($3 \times 10^{13}$\,M$_{\odot}$ with a 1--$\sigma$ range of $0.9$--$7 \times 10^{13}$\,M$_{\odot}$;~\citealt{Hickox12}), although there is clearly a large amount of scatter.

We conclude that by assuming a simple scenario where an SMG undergoes a star formation event with a duration of 100\,Myr, at a constant SFR, and then evolves passively, we determine that the median absolute $H$--band magnitude and number density of the ALESS SMGs are in good agreement with those of $z$\,$\sim$\,$0$ ellipticals. We also find that the shape of the absolute $H$--band distribution (Figure~\ref{fig:mhfading}), the mass-weighted stellar ages and the halo masses of local ellipticals are in good agreement with those predicted for the descendants of SMGs, suggesting that within this simple model SMGs are sufficient to explain the formation of most local elliptical galaxies brighter than $M_{H}$\,$\sim$\,$-18.5$.

\subsection{Evolution of SMGs: Intermediate redshift tests}
We now consider whether intermediate redshift populations agree with the toy model proposed here. We use a catalog from~\citet{bundy05}, which provides morphological classifications for $z < 22.5$\,mag galaxies in the GOODS--South field.  These galaxies are covered by the photometry employed in \S\,\ref{sec:observations} and so we can derive a photometric redshift and measure an absolute $H$--band magnitude for each source from SED fitting in the same manner as the ALESS SMGs. We select galaxies from the catalog that are visually classified as ellipticals, and with a photometric redshift $0.4<z_{phot}<0.6$ (the mid-point in time of evolution from an SMG phase, to the local Universe). The final sample has a median absolute $H$--band magnitude of $M_{H} = -22.0 \pm 0.2$ and a number density of $(1.6 \pm 0.2) \times 10^{-3}$\,Mpc$^{-3}$. The number density of the ALESS SMGs, $(1.4 \pm 0.4) \times 10^{-3}$\,Mpc$^{-3}$, is in agreement with the intermediate population of elliptical galaxies, however fading the absolute $H$--band magnitudes of the ALESS SMGs to $z$\,$\sim$\,$0.5$, we find that they are marginally fainter at $M_{H} = -21.5 \pm 0.2$ (see Figure~\ref{fig:mhfading}). Given the difficulty in morphological classification at $z$\,$\sim$\,$0.5$, and the small number of sources in each sample, we caution against drawing strong conclusions from this result. However it does indicate the ALESS SMGs are broadly consistent with $z$\,$\sim$\,$0.5$ elliptical galaxies.

Recently, near-infrared spectroscopy with the Wide--Field Camera 3 (WFC3) on the {\it Hubble Space Telescope (HST)} has been used to estimate the age of the stellar populations in quiescent spheroidal galaxies at $1.5<z<2$, the likely progenitors of local ellipticals. In particular, \citet{Whitaker13} recently used this technique to estimate stellar ages for a sample of 171 quiescent galaxies at $1.4<z<2.2$, with stellar masses $M_{\star} \gsim 5 \times 10^{10}$\,M$_{\odot}$ (similar to the ALESS SMGs). \citet{Whitaker13} divide their quiescent sample into blue (34) and red (137) subsets, based on $U, V$ and $J$--band colors, and fit absorption line models to the stacked spectra of each subset, deriving median stellar ages of $0.9^{+0.2}_{-0.1}$ and $1.6^{+0.5}_{-0.4}$\,Gyr for the blue and red subsets respectively. We use a weighted average of the number of galaxies in each subset, to determine the complete sample has a median stellar age of $\sim$\,$1.4$\,Gyr, at a median redshift of $z \sim$\,$1.7$ (see Figure 4;~\citealt{Whitaker13}). Combining the median stellar age and redshift of these post--starburst galaxies suggests that they formed at $z$\,$\sim$\,$2.6$ (see also~\citealt{Bedregal13}) This is consistent with the toy model proposed earlier where the ALESS SMGs, with a median redshift of $z_{phot} = 2.5 \pm 0.2$, form the majority of their stellar mass in a single burst at $z$\,$\sim$\,$2.5$, and do not undergo subsequent periods of significant star formation.

 \subsection{Evolution of SMGs: Black Hole Masses}
Finally, given the apparent link between SMGs and elliptical galaxies at $z$\,$\sim$\,$0$ we now investigate whether their black hole masses are consistent with our toy evolutionary model. To estimate black hole masses for the ALESS SMGs we use their X-ray properties, presented in~\citet{Wang13}. However, only ten ALESS SMGs are detected at X-ray energies and so we must first consider whether these SMGs are representative of the entire sample.

In our approach we will assume the black hole masses of the X-ray detected SMGs are related to the total stellar mass of the galaxy; however in our initial analysis we will use the $H$--band luminosity as a proxy for stellar mass. To account for the positive K--correction in the X-ray band we only consider SMGs in the redshift range $z=0$--2.5, which includes seven X-ray detected SMGs. These SMGs have a median $H$--band luminosity of $L_{H} = (1.1 \pm 0.3) \times 10^{12}$\,$\Lsol$, which is brighter than for the population as a whole [$L_{H} = (0.5 \pm 0.1) \times 10^{12}$\,$\Lsol$; Figure~\ref{fig:mh_z}]; a two-sided KS test returns a 3 per cent probability they are drawn from the same parent distribution. If we instead split the ALESS SMG sample at the median $H$--band luminosity, $L_{H} = 0.5 \times 10^{12}$\,$\Lsol$, then the X-ray non-detected SMGs at $L_{H} > 0.5 \times 10^{12}$\,$\Lsol$ have a median $H$--band luminosity of $L_{H} = (1.0 \pm 0.2) \times 10^{12}$\,$\Lsol$ and a two-sided KS test returns a probability of 85 per cent that the X-ray detected and non-detected SMGs with $L_{H} > 0.5 \times 10^{12}$\,$\Lsol$ are drawn from the same parent distribution (see also~\citealt{Wang13}).

A possible explanation for the higher $H$--band luminosities of the X-ray detected SMGs is that they suffer contamination in $L_{H}$ from an AGN power-law component (see~\citealt{Hainline11}). However, to bring the $H$--band luminosity of the X-ray detected SMGs into agreement with the complete sample requires a power-law fraction of $\sim$\,$50$ per cent which is not seen in the SEDs of the ALESS SMGs (see Appendix~A).

We now investigate whether these X-ray detected SMGs are preferentially detected in X-ray emission due to higher star formation rates. Using the far-infrared derived SFRs for the ALESS SMGs from~\citet{Swinbank13}, we find the X-ray detected SMGs have a median star formation rate of ${\rm SFR_{\rm FIR}} = 570 \pm 140$\,M$_{\odot}$yr$^{-1}$, which is consistent with the X-ray non-detected, $L_{H} > 0.5 \times 10^{12}$\,$\Lsol$, SMGs which have a median ${\rm SFR_{\rm FIR}}$ of $590 \pm 130$\,M$_{\odot}$yr$^{-1}$. We note that the median ${\rm SFR_{\rm FIR}}$ of the $L_{H} < 0.5 \times 10^{12}$\,$\Lsol$ SMGs is  $220 \pm 40$\,M$_{\odot}$yr$^{-1}$, a factor of $\sim$\,$2.5 \times$ lower. This suggests that the potential X-ray emission from star formation is similar for the X-ray detected and non-detected SMGs with $L_{H} > 0.5 \times 10^{12}$\,$\Lsol$, and is consistent with the results from~\citet{Wang13}, who argue for dominant AGN contributions to the X-ray emission of the X-ray detected SMGs.  

These results suggest that the X-ray detected SMGs have $H$--band luminosities and SFRs comparable to the $L_{H} > 0.5 \times 10^{12}$\,$\Lsol$ SMGs. As we demonstrated in \S\,~\ref{subsubsec:trends} the X-ray detected SMGs are also indistinguishable from the X-ray non-detected SMGs in terms of IRAC flux ratios, and so combining all these results we conclude that other than in X-ray emission, the X-ray detected SMGs are not distinguishable from average $L_{H} > 0.5 \times 10^{12}$\,$\Lsol$ SMGs. Instead we propose that SMGs with $L_{H} \lsim 0.5 \times 10^{12}$\,$\Lsol$ are not detected at X-ray energies due to the selection limits on the X-ray data. As the median $H$--band luminosity of the ALESS SMGs with $L_{H} < 0.5 \times 10^{12}$\,$\Lsol$ is a factor of $3.5 \times$ lower than the brighter half of the sample then, under the simple assumption that $L_X$ and $L_H$ represent $M_{\rm BH}$ and $M_{\star}$, this suggests that X-ray data a factor of $\sim$\,$3.5 \times$ deeper is required to detect these SMGs down to the same Eddington ratio as the X-ray detected SMGs.

Although the X-ray detected SMGs only appear representative of the ALESS SMGs above $L_{H} > 0.5 \times 10^{12}$\,$\Lsol$, we can still estimate their black hole masses, and hence their relation to the black hole masses of local ellipticals. The ten ALESS SMGs detected at X-ray energies have a median absorption corrected X-ray luminosity of ${\rm log} L_{0.5-8\,{\rm keV, corr}} = 43.3 \pm 0.4$\,erg\,s$^{-1}$. We initially convert the luminosity from 0.5--8\,keV to 2--10\,keV, dividing through by a conversion factor of 1.21, and estimate a bolometric luminosity following $L_{bol} = 35 \times L_{2-10}$~\citep{alexander08}. We adopt an Eddington ratio of $\eta = 0.2$, which was calculated for a small number of SMGs with direct black hole mass measurements~\citep{alexander08}, and from this estimate we calculate a median black hole mass of $M_{\rm BH}$\,$\sim$\,$(3 ^{+3}_{-1}) \times 10^{7}$\,M$_{\odot}$. We note that our uncertainty does not include any error in the Eddington ratio or the bolometric luminosity conversion, and assumes that the average Eddington ratio for these SMGs is consistent with the SMGs in~\citet{alexander08}.

Using the black hole masses and stellar masses of the X-ray detected ALESS SMGs, we can estimate the growth required to match the black holes in local ellipticals. The X-ray detected SMGs have $M_{\rm BH}$\,/\,$M_{\star} = 0.2^{+0.2}_{-0.1} \times 10^{-3} $, and require $\sim$\,$9 \times$ growth of their black hole masses, at fixed stellar mass, to match the local relation ($M_{\rm BH}$\,/\,$M_{\star} = 1.7 \pm 0.4 \times 10^{-3} $ at these masses;~\citealt{Haring04}). If we assume the SMG lifetime is $\sim$\,100\,Myr, and that we are seeing each SMG on average half-way through the burst, then following Eqn\,10 from~\citet{alexander12}, and assuming $\eta = 0.2$, the SMBH will have grown by $\sim$\,$20$ per cent at the end of the SMG phase. However, a further factor of $\sim$\,$7 \times$ growth is still required to match the local M$_{\star}$--M$_{\rm BH}$ relation. It has been speculated that this growth may come in the form of a QSO phase (e.g.\ \citealt{Sanders88,Coppin08b, Simpson12}), during which the SMBH would grow at approximately the Eddington limit. If we assume all of the remaining SMBH growth occurs in a QSO phase, then the duration of this phase must be $\sim$\,$100$\,Myr.

A lifetime of 100\,Myr for a QSO phase is high but not unreasonable~\citep{martini01}, however it is highly unlikely that the accretion is Eddington limited for the entirety of this period~\citep{Mclure04,Kelly10}. If the accretion is not Eddington limited during this phase, then this analysis suggests that we have either under-estimated the black hole masses or over-estimated the stellar masses in these SMGs. Indeed, when estimating black holes masses for the X-ray detected SMGs we have made assumptions on the current Eddington ratio, $\eta$, and the conversion from X-ray luminosity to a bolometric luminosity, both of which have significant uncertainties. The black hole masses we estimate are inversely proportional to the initial Eddington ratio, and it is feasible that this is lower than the value we adopted ($\eta = 0.2$;~\citealt{alexander08}). Due to the uncertainties surrounding our estimate of the median black hole mass, we simply conclude that the SMBHs in these SMGs are likely to require an extended period of black hole growth after the SMG phase, and that this is most likely to occur during a high accretion rate QSO-phase. 

\section{Summary \& Conclusions}
\label{sec:conclusions}
In this paper we have presented a multi-wavelength study of the first large sample of $870$\,$\mu$m-selected SMGs with unambiguous identifications based on ALMA interferometry. Crucially, these precise identifications, at the same wavelength as used for the original single-dish survey, mean that our analysis is free from the mis-identification and incompleteness associated with the use of radio--mid-infrared proxies to locate the counterparts of single-dish identified sub-mm sources (see discussion in~\citealt{Hodge13}). The main conclusions from our work are:

\begin{itemize}

\item We measure aperture photometry in 19 wave-bands for 96 ALESS SMGs~\citep{Hodge13}. From this initial sample 77 SMGs are detected in $\ge 4$ wave-bands and have sufficient photometry to derive photometric redshifts from SED fitting. These 77 SMGs have a median redshift of $z_{phot} = 2.3 \pm 0.2$, with  a 1--$\sigma$ spread of $z_{phot} = 1.8$--3.5.

\item Nineteen SMGs in our sample have insufficient photometry to derive photometric redshifts. We initially divide these sources into subsets detected in 0-or-1 and 2-or-3 wave-bands and test whether they are real or spurious by stacking their emission in other wave-bands. Detections at 250, 350 and 500\,$\mu$m for both subsets confirm that these are typically far-infrared bright sources on average.

\item We use the distribution of absolute $H$--band magnitudes at $z < 2.5$ to measure the incompleteness in the distribution at $z > 2.5$, and use this to estimate redshifts for the 19 ALESS SMGs detected in $<4$ wave-bands. We estimate the median redshifts for the SMGs detected in 2-or-3 and 0-or-1 wave-bands as $ z$\,$\sim$\,$3.5$ and $z$\,$\sim$\,$4.5$ respectively.

\item The redshift distribution for the complete sample of 96 ALESS SMGs has a median redshift of $z_{phot} = 2.5 \pm 0.2$ with $35 \pm 5$ per cent of SMGs lying at $z > 3$. In terms of their look-back age, the distribution is well-fit by a Gaussian distribution centered at $11.1 \pm 0.1$\,Gyr, and with a 1--$\sigma$ width of $1.1 \pm 0.1$\,Gyr. 

\item We compare the redshift distribution of the ALESS SMGs to recent interferometric observations of smaller samples of millimeter-selected SMGs from~\citet{Smolcic12} and~\citet{weiss13}. The median redshift of the SMGs presented in~\citet{Smolcic12} is $z = 2.6 \pm 0.4$, in agreement with the ALESS SMGs, however the median redshift millimeter-selected, lensed, SMGs in~\citet{weiss13} is considerably higher, $ z$\,$\sim$\,$3.6 \pm 0.4$, although correcting for lensing effects reduces this to $z$\,$\sim$\,$3.1 \pm 0.3$. Due to differences in the selection wavelength and the difficulties in accurately constraining the lensing selection function we caution against strong conclusions drawn from the discrepancy in the median redshifts of the ALESS SMGs and the SMGs presented in~\citet{weiss13}. We note that in contrast to both of these previous studies, which have suggested the redshift distribution of SMGs is flat above $z$\,$\sim$\,$3$, the ALESS redshift distribution peaks at $z$\,$\sim$\,$2.5$, and declines steadily towards high redshift ($z > 4$).

\item Nineteen of the 86 ALMA maps employed in our analysis do not contain an SMG with an 870--$\mu$m flux density $>1.4$\,mJy.~\citet{Karim13} and \citet{Hodge13} propose this is due to the original LABOCA submm source fragmenting into a number of faint SMGs, and so we measure the number density and redshift distribution of 3.6\,$\mu$m sources in these ``blank'' maps and compare it to the redshift distribution in the field. We identify an excess of sources at $z$\,$\sim$\,$1$--3 indicating that, on average, these ``blank'' maps contain multiple SMGs below our detection threshold. We stack the ALMA 870\,$\mu$m maps at the position of these 3.6\,$\mu$m sources and measure an average flux of $0.36 \pm 0.09$\,mJy per source, and a total contribution per map of $0.76 \pm 0.19$\,mJy. Although this is not sufficient to explain the difference in total flux measured by LABOCA and ALMA in these regions, it suggests that there are faint SMGs in these ``blank'' maps, which lie at a similar redshift to the $S_{870}$ brighter ALESS SMGs. 

\item Using the complete $M_{H}$ distribution for our sample we derive a median stellar mass of $(8 \pm 1) \times 10^{10}$\,M$_{\odot}$ for the ALESS SMGs, but caution that due to the unconstrained SFHs, and hence mass-to-light ratios, this is subject to significant systematic uncertainties (at least $\sim$\,5\,$\times$). 

\item We investigate the possible properties of the descendants of the ALESS SMGs at the present day. Fading the $H$--band luminosities of the ALESS SMGs, assuming a SFH consisting of a 100\,Myr burst with constant SFR, we show the present day descendants will have an absolute $H$--band magnitude of $M_{H}$\,$\sim$\,$-21.2 \pm 0.2$, and a space density of $(1.4 \pm 0.4) \times 10^{-3}$\,Mpc$^{-3}$. These properties are in good agreement with those of local elliptical galaxies, derived from the volume-limited PM2GC survey, which have a median absolute $H$--band magnitude of $M_{H} = -21.1 \pm 0.1$ and a space density of $(2.0 \pm 0.1) \times 10^{-3}$\,Mpc$^{-3}$. We show that the mass-weighted stellar ages of the PM2GC ellipticals ($\sim$\,$10$\,Gyr, but with a systematic uncertainty of $\pm 2$\,Gyr) are in broad agreement with the lookback times to the ALESS SMGs, providing support for our simple evolutionary model. 

\item We test our simple evolutionary model against intermediate- and high-redshift populations of quiescent spheroidal galaxies. We find that a morphologically classified sample of ellipticals at $z$\,$\sim$\,$0.5 $ shows broad agreement in shape, and number density to the ``faded'' ALESS SMGs at this epoch, but are typically $\sim$\,$0.5 \pm 0.3$ magnitudes fainter. In addition, recent near-infrared spectroscopy of quiescent, red, spheroids at $z = 1.5$--2, the likely descendants of SMGs, indicates their stellar populations formed at $z$\,$\sim$\,$2.6$, consistent with the median redshift of the ALESS SMGs, $z_{phot} = 2.5 \pm 0.2$.

\item Finally, we show that the X-ray detected SMGs~\citep{Wang13} are indistinguishable in terms of $H$--band luminosities, SFRs and IRAC flux ratios to the ALESS SMGs brighter than $L_{H} > 0.5 \times 10^{12}$\,$\Lsol$. We use the X-ray properties of these SMGs to estimate a median black hole mass of $M_{\rm BH}$\,$\sim$\,$(3 ^{+3}_{-1}) \times 10^{7}$\,M$_{\odot}$, which combined with stellar mass estimates, indicates the black holes in these SMGs are required to grow by approximately an order of magnitude to match the local black hole--spheroid mass relation.

\end{itemize}

We have presented the redshift distribution of a large sample of $870$\,$\mu$m-selected SMGs with precise interferometrically-determined positions from ALMA. Crucially, the redshift distribution of the ALESS SMGs declines steadily from $z$\,$\sim$\,$2.5$ and does not plateau at high redshift ($z>3$), as has been suggested by smaller samples of millimeter-selected SMGs. Furthermore, we present a simple evolutionary scenario where SMGs undergo a period of intense star-formation, before passively evolving into local elliptical galaxies. We compare the number density and ``faded'' $H$--band luminosity of the ALESS SMGs to local and intermediate-redshift samples of elliptical galaxies, and find that both are in agreement with our simple evolutionary model. Our toy model is consistent with an evolutionary scenario where most of the stars in the majority of local luminous elliptical galaxies are formed through a single starburst SMG event at high redshift.

\section*{Acknowledgments}

JMS and ALRD acknowledge the support of STFC studentships (ST/J501013/1 and ST/F007299/1, respectively). AMS acknowledges financial support from an STFC Advanced Fellowship (ST/H005234/1). IRS acknowledges support from a Leverhulme Senior Fellowship, ERC Advanced Investigator program DUSTYGAL 321334, RS/Wolfson Merit Award and STFC (ST/I001573/1). FB and AK acknowledge support by the Collaborative Research Council 956, sub-project A1, funded by the Deutsche Forschungsgemeinschaft (DFG). RJI acknowledges support from ERC via the Advanced Grant, COSMICISM. JWL acknowledges support for program number HST-GO-12866.13-A provided by NASA through a grant from the Space Telescope Science Institute, which is operated by the Association of Universities for Research in Astronomy, Incorporated, under NASA contract NAS5-26555 and the Dark Cosmology Centre which is funded by the Danish National Research Foundation. WNB acknowledges support from a Chandra X-ray Center grant AR3-14015X and Space Telescope Science Institute grant HST-GO-12866.01-A.

The ALMA data used in this paper were obtained under program ADS/JAO.ALMA$\#$2011.0.00294.S. ALMA is a partnership of ESO (representing its member states), NSF (USA) and NINS (Japan), together with NRC (Canada) and NSC and ASIAA (Taiwan), in cooperation with the Republic of Chile. The Joint ALMA Observatory is operated by ESO, AUI/NRAO, and NAOJ. This publication also makes use of data acquired with the APEX under program IDs 078.F-9028(A), 079.F-9500(A), 080.A-3023(A), and 081.F-9500(A). APEX is a collaboration between the Max-Planck-Institut f{\"u}r Radioastronomie, the European Southern Observatory, and the Onsala Space Observatory.
In this paper we make use of spectroscopic data obtained as part of the ESO/VLT large program zLESS:183.A-0666, and the VLT/XSHOOTER proposal 090.A.0927

This research has made use of data from HerMES project (http://hermes.sussex.ac.uk/;~\citealt{Oliver12}). HerMES is a {\it Herschel} Key Programme utilizing Guaranteed Time from the SPIRE instrument team, ESAC scientists and a mission scientist.

All data used in this analysis can be obtained from the ALMA, ESO, {\it Spitzer} and {\it Herschel} archives.

\bibliographystyle{mn2e} 
\bibliography{ref.bib}

\clearpage

  %
  %
  \begin{sidewaystable}
  \vspace{-10cm}
  \centerline{\sc Table 2: Photometry}
  \resizebox{\linewidth}{!}{%
  \tiny
  \begin{tabular}{lccccccccccccccccc}
  \hline
  \noalign{\smallskip}
  ID & MUSYC $U$ & MUSYC $U_{38}$ & VIMOS $U$ & $B$ & $V$ & $R$ & $I$ & $z$ & $J^{b}$ & $H$ & $K^{b}$ & 3.6\,$\mu$m & 4.5\,$\mu$m & 5.8\,$\mu$m & 8.0\,$\mu$m \\  [0.5ex] 
  \hline 
  ALESS 1.1$^{a}$ &  $>$26.18 & $>$25.29 & $>$28.14 & $>$26.53 & $>$26.32 & $>$25.53 & $>$24.68 & $>$24.29 & $>$24.88 & $>$23.06 & 23.68\,$\pm$\,0.19$^{*}$ & 22.81\,$\pm$\,0.06 & 22.72\,$\pm$\,0.07 & 22.28\,$\pm$\,0.19 & 21.87\,$\pm$\,0.06 \\
  ALESS 1.2 &  $>$26.18 & $>$25.29 & $>$28.14 & $>$26.53 & $>$26.32 & $>$25.53 & $>$24.68 & $>$24.29 & $>$24.88 & $>$23.06 & 24.27\,$\pm$\,0.25 & 22.81\,$\pm$\,0.06 & 22.84\,$\pm$\,0.08 & 21.99\,$\pm$\,0.15 & 22.11\,$\pm$\,0.07 \\
  ALESS 1.3$^{a}$ &  $>$26.18 & $>$25.29 & 27.52\,$\pm$\,0.16 & $>$26.53 & $>$26.32 & $>$25.53 & $>$24.68 & $>$24.29 & $>$24.88 & $>$23.06 & 23.98\,$\pm$\,0.20 & 23.05\,$\pm$\,0.07 & 22.92\,$\pm$\,0.09 & $>$22.44 & 22.19\,$\pm$\,0.08 \\
  ALESS 2.1$^{a}$ &  $>$26.18 & $>$25.29 & 28.05\,$\pm$\,0.25 & $>$26.53 & 26.30\,$\pm$\,0.26 & $>$25.53 & $>$24.68 & $>$24.29 & 24.45\,$\pm$\,0.18 & ... & 23.03\,$\pm$\,0.09 & 21.92\,$\pm$\,0.03 & 21.66\,$\pm$\,0.03 & 21.35\,$\pm$\,0.08 & 21.83\,$\pm$\,0.06 \\
  ALESS 2.2 &  $>$26.18 & $>$25.29 & $>$28.14 & $>$26.53 & $>$26.32 & $>$25.53 & $>$24.68 & $>$24.29 & $>$24.88 & ... & $>$24.35 & 23.32\,$\pm$\,0.09 & 23.00\,$\pm$\,0.10 & $>$22.44 & 22.12\,$\pm$\,0.07 \\
  ALESS 3.1 &  $>$26.18 & $>$25.29 & ... & $>$26.53 & $>$26.32 & $>$25.53 & $>$24.68 & $>$24.29 & $>$24.56$^{*}$ & $>$23.06 & 23.24\,$\pm$\,0.10 & 22.34\,$\pm$\,0.04 & 21.71\,$\pm$\,0.03 & 21.33\,$\pm$\,0.08 & 21.03\,$\pm$\,0.03 \\
  ALESS 5.1$^{a}$ &  24.83\,$\pm$\,0.08 & 24.75\,$\pm$\,0.17 & ... & 23.15\,$\pm$\,0.01 & 22.11\,$\pm$\,0.01 & 21.33\,$\pm$\,0.01 & 20.70\,$\pm$\,0.01 & 20.60\,$\pm$\,0.01 & 20.26\,$\pm$\,0.00 & ... & 19.79\,$\pm$\,0.00 & 19.35\,$\pm$\,0.00 & 19.50\,$\pm$\,0.00 & 19.87\,$\pm$\,0.02 & 20.09\,$\pm$\,0.01 \\
  ALESS 6.1 &  24.58\,$\pm$\,0.07 & 24.48\,$\pm$\,0.13 & 24.52\,$\pm$\,0.01 & 22.51\,$\pm$\,0.01 & 21.70\,$\pm$\,0.00 & 20.88\,$\pm$\,0.00 & 19.90\,$\pm$\,0.00 & 19.83\,$\pm$\,0.00 & 19.73\,$\pm$\,0.00 & 19.32\,$\pm$\,0.01 & 19.81\,$\pm$\,0.00 & 20.07\,$\pm$\,0.00 & 20.36\,$\pm$\,0.01 & 20.45\,$\pm$\,0.04 & 20.81\,$\pm$\,0.02 \\
  ALESS 7.1 &  25.97\,$\pm$\,0.22 & $>$25.29 & 26.09\,$\pm$\,0.04 & 24.63\,$\pm$\,0.05 & 23.81\,$\pm$\,0.03 & 23.06\,$\pm$\,0.03 & 21.89\,$\pm$\,0.02 & 21.81\,$\pm$\,0.03 & 21.13\,$\pm$\,0.01 & 20.68\,$\pm$\,0.03 & 20.16\,$\pm$\,0.01 & 19.65\,$\pm$\,0.00 & 19.54\,$\pm$\,0.00 & 19.47\,$\pm$\,0.02 & 19.76\,$\pm$\,0.01 \\
  ALESS 9.1 &  $>$26.18 & $>$25.29 & $>$28.14 & $>$26.53 & $>$26.32 & $>$25.53 & $>$24.68 & $>$24.29 & $>$24.88 & ... & 23.57\,$\pm$\,0.14 & 21.76\,$\pm$\,0.02 & 21.35\,$\pm$\,0.02 & 21.03\,$\pm$\,0.06 & 20.85\,$\pm$\,0.02 \\
  ALESS 10.1 &  25.80\,$\pm$\,0.19 & ... & 25.58\,$\pm$\,0.03 & 25.28\,$\pm$\,0.09 & 25.35\,$\pm$\,0.17 & 24.77\,$\pm$\,0.14 & 24.36\,$\pm$\,0.20 & $>$24.29 & 23.70\,$\pm$\,0.10 & ... & 22.74\,$\pm$\,0.07 & 21.82\,$\pm$\,0.03 & 21.39\,$\pm$\,0.03 & 21.28\,$\pm$\,0.09 & 21.06\,$\pm$\,0.04 \\
  ALESS 11.1 &  $>$26.18 & $>$25.29 & 28.06\,$\pm$\,0.25 & $>$26.53 & $>$26.32 & $>$25.53 & $>$24.68 & $>$24.29 & $>$24.88 & $>$23.06 & 23.03\,$\pm$\,0.09 & 21.78\,$\pm$\,0.02 & 21.26\,$\pm$\,0.02 & 20.89\,$\pm$\,0.06 & 20.75\,$\pm$\,0.02 \\
  ALESS 13.1 &  $>$26.18 & $>$25.29 & $>$28.14 & $>$26.53 & 26.09\,$\pm$\,0.22 & 25.35\,$\pm$\,0.23 & $>$24.68 & $>$24.29 & 24.11\,$\pm$\,0.14 & $>$23.06 & 22.54\,$\pm$\,0.06 & 21.78\,$\pm$\,0.02 & 21.51\,$\pm$\,0.02 & 21.30\,$\pm$\,0.08 & 21.14\,$\pm$\,0.03 \\
  ALESS 14.1$^{a}$ &  $>$26.18 & $>$25.29 & ... & $>$26.53 & $>$26.32 & $>$25.53 & $>$24.68 & $>$24.29 & 24.00\,$\pm$\,0.13 & ... & 23.15\,$\pm$\,0.10 & 22.14\,$\pm$\,0.03 & 21.50\,$\pm$\,0.02 & 21.16\,$\pm$\,0.07 & 20.74\,$\pm$\,0.02 \\
  ALESS 15.1 &  $>$26.18 & $>$25.29 & ... & $>$26.53 & $>$26.32 & $>$25.53 & $>$24.68 & $>$24.29 & $>$23.22$^{**}$ & ... & $>$22.41$^{**}$ & 21.54\,$\pm$\,0.02 & 20.93\,$\pm$\,0.01 & 20.72\,$\pm$\,0.05 & 20.64\,$\pm$\,0.02 \\
  ALESS 15.3 &  $>$26.18 & $>$25.29 & ... & $>$26.53 & $>$26.32 & $>$25.53 & $>$24.68 & $>$24.29 & $>$23.22$^{**}$ & ... & $>$22.41$^{**}$ & 22.76\,$\pm$\,0.06 & 22.93\,$\pm$\,0.09 & $>$22.44 & $>$23.38 \\
  ALESS 17.1 &  24.80\,$\pm$\,0.08 & 25.02\,$\pm$\,0.21 & 24.84\,$\pm$\,0.01 & 24.20\,$\pm$\,0.03 & 24.14\,$\pm$\,0.04 & 23.72\,$\pm$\,0.06 & 23.00\,$\pm$\,0.06 & 22.83\,$\pm$\,0.08 & 21.70\,$\pm$\,0.02 & 21.14\,$\pm$\,0.05 & 20.77\,$\pm$\,0.01 & 19.98\,$\pm$\,0.00 & 19.77\,$\pm$\,0.01 & 19.96\,$\pm$\,0.02 & 20.35\,$\pm$\,0.01 \\
  ALESS 18.1 &  25.62\,$\pm$\,0.17 & $>$25.29 & 25.54\,$\pm$\,0.03 & 25.24\,$\pm$\,0.09 & 25.06\,$\pm$\,0.09 & 24.99\,$\pm$\,0.17 & 24.31\,$\pm$\,0.20 & 24.15\,$\pm$\,0.24 & 22.64\,$\pm$\,0.04 & 21.66\,$\pm$\,0.08 & 21.13\,$\pm$\,0.02 & 20.01\,$\pm$\,0.00 & 19.68\,$\pm$\,0.00 & 19.61\,$\pm$\,0.02 & 20.28\,$\pm$\,0.01 \\
  ALESS 19.1 &  $>$26.18 & $>$25.29 & 26.81\,$\pm$\,0.09 & $>$26.53 & 26.12\,$\pm$\,0.23 & $>$25.53 & $>$24.68 & $>$24.29 & $>$24.88 & ... & 23.60\,$\pm$\,0.14 & 22.35\,$\pm$\,0.04 & 21.85\,$\pm$\,0.03 & 21.43\,$\pm$\,0.09 & 21.58\,$\pm$\,0.04 \\
  ALESS 19.2 &  26.04\,$\pm$\,0.24 & $>$25.29 & 25.81\,$\pm$\,0.03 & 25.05\,$\pm$\,0.07 & 24.71\,$\pm$\,0.07 & 24.48\,$\pm$\,0.11 & 23.97\,$\pm$\,0.15 & 24.02\,$\pm$\,0.21 & 23.15\,$\pm$\,0.06 & ... & 22.29\,$\pm$\,0.04 & 21.62\,$\pm$\,0.02 & 21.68\,$\pm$\,0.03 & 21.46\,$\pm$\,0.09 & 21.60\,$\pm$\,0.05 \\
  ALESS 22.1 &  $>$26.18 & $>$25.29 & ... & $>$26.53 & 25.96\,$\pm$\,0.20 & 25.29\,$\pm$\,0.22 & $>$24.68 & $>$24.29 & ... & ... & ... & 20.11\,$\pm$\,0.00 & 19.79\,$\pm$\,0.01 & 19.64\,$\pm$\,0.02 & 20.15\,$\pm$\,0.01 \\
  ALESS 23.1 &  ... & ... & ... & ... & ... & ... & ... & ... & $>$24.20 & ... & $>$23.74 & 23.11\,$\pm$\,0.08 & 22.55\,$\pm$\,0.06 & 21.78\,$\pm$\,0.12 & 21.43\,$\pm$\,0.04 \\ ALESS 23.7 &  ... & ... & ... & ... & ... & ... & ... & ... & ... & ... & ... & 22.90\,$\pm$\,0.06 & 22.70\,$\pm$\,0.07 & $>$22.44 & $>$23.38 \\
  ALESS 25.1 &  $>$26.18 & $>$25.29 & 26.66\,$\pm$\,0.07 & 25.74\,$\pm$\,0.14 & 25.19\,$\pm$\,0.10 & 24.73\,$\pm$\,0.14 & 24.12\,$\pm$\,0.17 & 23.95\,$\pm$\,0.20 & 23.01\,$\pm$\,0.05 & ... & 21.52\,$\pm$\,0.02 & 20.66\,$\pm$\,0.01 & 20.35\,$\pm$\,0.01 & 20.19\,$\pm$\,0.03 & 20.44\,$\pm$\,0.02 \\
  ALESS 29.1 &  $>$26.18 & $>$25.29 & ... & $>$26.53 & $>$26.32 & $>$25.53 & $>$24.68 & $>$24.29 & $>$23.22$^{**}$ & ... & $>$22.41$^{**}$ & 21.85\,$\pm$\,0.02 & 21.33\,$\pm$\,0.02 & 20.86\,$\pm$\,0.05 & 20.79\,$\pm$\,0.02 \\
  ALESS 31.1 &  $>$26.18 & $>$25.29 & ... & $>$26.53 & $>$26.32 & $>$25.53 & $>$24.68 & $>$24.29 & $>$24.88 & ... & 23.58\,$\pm$\,0.14 & 22.30\,$\pm$\,0.04 & 21.76\,$\pm$\,0.03 & 21.27\,$\pm$\,0.08 & 21.25\,$\pm$\,0.03 \\
  ALESS 37.1 &  $>$26.18 & $>$25.29 & ... & $>$26.53 & 25.55\,$\pm$\,0.14 & 24.29\,$\pm$\,0.09 & 23.50\,$\pm$\,0.10 & ... & 23.07\,$\pm$\,0.24$^{**}$ & ... & 21.43\,$\pm$\,0.12$^{**}$ & 20.56\,$\pm$\,0.01 & 20.30\,$\pm$\,0.01 & 20.46\,$\pm$\,0.04 & 20.60\,$\pm$\,0.02 \\
  ALESS 37.2 &  $>$26.18 & $>$25.29 & ... & $>$26.53 & 26.10\,$\pm$\,0.22 & 24.62\,$\pm$\,0.12 & ... & ... & $>$23.22$^{**}$ & $>$23.06 & $>$22.41$^{**}$ & 22.33\,$\pm$\,0.04 & 22.13\,$\pm$\,0.04 & 21.73\,$\pm$\,0.12 & 21.72\,$\pm$\,0.05 \\
  ALESS 39.1 &  25.08\,$\pm$\,0.10 & $>$25.29 & ... & 25.06\,$\pm$\,0.07 & 24.65\,$\pm$\,0.06 & 23.98\,$\pm$\,0.07 & 23.75\,$\pm$\,0.12 & 23.20\,$\pm$\,0.10 & $>$23.22$^{**}$ & ... & 22.04\,$\pm$\,0.04 & 21.24\,$\pm$\,0.01 & 20.90\,$\pm$\,0.01 & 20.66\,$\pm$\,0.05 & 20.67\,$\pm$\,0.02 \\
  ALESS 41.1 &  ... & ... & ... & ... & ... & ... & ... & ... & ... & ... & ... & 20.12\,$\pm$\,0.01 & 19.83\,$\pm$\,0.01 & 19.55\,$\pm$\,0.02 & 19.51\,$\pm$\,0.01 \\ ALESS 41.3 &  ... & ... & ... & ... & ... & ... & ... & ... & ... & ... & ... & 22.39\,$\pm$\,0.04 & 22.95\,$\pm$\,0.09 & $>$22.44 & 22.45\,$\pm$\,0.10 \\
  ALESS 43.1 &  $>$26.18 & $>$25.29 & 28.10\,$\pm$\,0.26 & $>$26.53 & $>$26.32 & $>$25.53 & $>$24.68 & $>$24.29 & 24.10\,$\pm$\,0.14 & ... & 22.48\,$\pm$\,0.05 & 21.10\,$\pm$\,0.01 & 20.67\,$\pm$\,0.01 & 20.69\,$\pm$\,0.05 & 21.33\,$\pm$\,0.04 \\
  ALESS 45.1 &  $>$26.18 & $>$25.29 & $>$28.14 & $>$26.53 & $>$26.32 & $>$25.53 & $>$24.68 & $>$24.29 & $>$24.88 & ... & 22.67\,$\pm$\,0.06 & 21.24\,$\pm$\,0.01 & 20.80\,$\pm$\,0.01 & 20.55\,$\pm$\,0.04 & 20.75\,$\pm$\,0.02 \\
  ALESS 49.1 &  $>$26.18 & $>$25.29 & ... & 24.61\,$\pm$\,0.05 & 24.35\,$\pm$\,0.05 & 24.10\,$\pm$\,0.08 & 23.80\,$\pm$\,0.13 & 24.07\,$\pm$\,0.22 & 23.32\,$\pm$\,0.07 & ... & 22.63\,$\pm$\,0.06 & 21.77\,$\pm$\,0.02 & 21.49\,$\pm$\,0.02 & 21.26\,$\pm$\,0.08 & 21.27\,$\pm$\,0.03 \\
  ALESS 49.2 &  26.11\,$\pm$\,0.25 & $>$25.29 & ... & 25.25\,$\pm$\,0.09 & 25.23\,$\pm$\,0.10 & 24.53\,$\pm$\,0.11 & 23.98\,$\pm$\,0.15 & 24.25\,$\pm$\,0.26 & 23.60\,$\pm$\,0.09 & ... & 21.94\,$\pm$\,0.03 & 20.88\,$\pm$\,0.01 & 20.59\,$\pm$\,0.01 & 21.01\,$\pm$\,0.06 & 21.32\,$\pm$\,0.03 \\
  ALESS 51.1 &  $>$26.18 & $>$25.29 & 25.63\,$\pm$\,0.03 & 24.97\,$\pm$\,0.07 & 24.63\,$\pm$\,0.06 & 23.65\,$\pm$\,0.05 & 22.29\,$\pm$\,0.03 & 22.09\,$\pm$\,0.04 & 21.15\,$\pm$\,0.01 & 20.73\,$\pm$\,0.03 & 20.38\,$\pm$\,0.01 & 19.47\,$\pm$\,0.00 & 19.61\,$\pm$\,0.00 & 19.80\,$\pm$\,0.02 & 20.17\,$\pm$\,0.01 \\
  ALESS 55.1 &  24.66\,$\pm$\,0.07 & 24.60\,$\pm$\,0.15 & 24.49\,$\pm$\,0.01 & 24.25\,$\pm$\,0.04 & 24.20\,$\pm$\,0.04 & 24.05\,$\pm$\,0.07 & 23.74\,$\pm$\,0.12 & ... & 23.26\,$\pm$\,0.07 & $>$23.06 & 22.72\,$\pm$\,0.11 & 22.25\,$\pm$\,0.04 & 22.07\,$\pm$\,0.04 & 21.82\,$\pm$\,0.13 & 21.56\,$\pm$\,0.07 \\
  ALESS 55.2 &  $>$26.18 & $>$25.29 & $>$28.14 & $>$26.53 & $>$26.32 & $>$25.53 & $>$24.68 & $>$24.29 & $>$24.88 & $>$23.06 & $>$24.35 & $>$24.45 & $>$24.09 & $>$22.44 & $>$23.38 \\
  ALESS 55.5 &  25.31\,$\pm$\,0.13 & 25.15\,$\pm$\,0.24 & 25.21\,$\pm$\,0.02 & 24.74\,$\pm$\,0.06 & 24.77\,$\pm$\,0.07 & 24.75\,$\pm$\,0.14 & $>$24.68 & ... & 24.42\,$\pm$\,0.18 & $>$23.06 & 23.64\,$\pm$\,0.15 & 22.70\,$\pm$\,0.05 & 22.50\,$\pm$\,0.06 & 22.35\,$\pm$\,0.20 & 22.04\,$\pm$\,0.07 \\
  ALESS 57.1 &  $>$26.18 & $>$25.29 & 26.38\,$\pm$\,0.06 & 24.86\,$\pm$\,0.06 & 25.10\,$\pm$\,0.09 & 24.65\,$\pm$\,0.13 & $>$24.68 & $>$24.29 & 23.99\,$\pm$\,0.12 & 22.66\,$\pm$\,0.19 & 22.55\,$\pm$\,0.06 & 21.64\,$\pm$\,0.02 & 21.26\,$\pm$\,0.02 & 20.95\,$\pm$\,0.06 & 20.15\,$\pm$\,0.01 \\
  ALESS 59.2 &  $>$26.18 & $>$25.29 & 26.90\,$\pm$\,0.09 & 26.21\,$\pm$\,0.20 & 26.29\,$\pm$\,0.26 & $>$25.53 & $>$24.68 & $>$24.29 & 24.24\,$\pm$\,0.15 & ... & 23.51\,$\pm$\,0.13 & 22.92\,$\pm$\,0.06 & 22.54\,$\pm$\,0.06 & $>$22.44 & 22.31\,$\pm$\,0.09 \\
  ALESS 61.1$^{a}$ &  $>$26.18 & $>$25.29 & ... & $>$26.53 & $>$26.32 & $>$25.53 & 24.39\,$\pm$\,0.21 & 23.94\,$\pm$\,0.20 & 23.00\,$\pm$\,0.05 & - & 22.56\,$\pm$\,0.06 & 22.83\,$\pm$\,0.06 & 22.45\,$\pm$\,0.06 & $>$22.44 & 22.02\,$\pm$\,0.07 \\
  ALESS 63.1 &  $>$26.18 & $>$25.29 & ... & 25.96\,$\pm$\,0.17 & 25.52\,$\pm$\,0.14 & 24.79\,$\pm$\,0.14 & 23.64\,$\pm$\,0.11 & 23.23\,$\pm$\,0.11 & 22.14\,$\pm$\,0.02 & ... & 21.26\,$\pm$\,0.02 & 20.51\,$\pm$\,0.01 & 20.29\,$\pm$\,0.01 & 20.38\,$\pm$\,0.04 & 20.63\,$\pm$\,0.02 \\
  ALESS 65.1 &  $>$26.18 & $>$25.29 & ... & $>$26.53 & $>$26.32 & $>$25.53 & $>$24.68 & $>$24.29 & $>$24.88 & ... & $>$24.35 & 23.33\,$\pm$\,0.09 & 23.77\,$\pm$\,0.19 & $>$22.44 & 22.88\,$\pm$\,0.14 \\
  ALESS 66.1 &  20.91\,$\pm$\,0.00 & 20.85\,$\pm$\,0.01 & ... & 21.27\,$\pm$\,0.00 & 21.22\,$\pm$\,0.00 & 20.82\,$\pm$\,0.00 & 20.66\,$\pm$\,0.01 & 20.08\,$\pm$\,0.01 & 21.21\,$\pm$\,0.05$^{**}$ & ... & 20.28\,$\pm$\,0.04$^{**}$ & 19.41\,$\pm$\,0.00 & 19.26\,$\pm$\,0.00 & 19.08\,$\pm$\,0.01 & 19.10\,$\pm$\,0.00 \\
  ALESS 67.1 &  25.69\,$\pm$\,0.18 & $>$25.29 & 25.24\,$\pm$\,0.02 & 24.65\,$\pm$\,0.05 & 24.30\,$\pm$\,0.05 & 24.17\,$\pm$\,0.08 & 23.52\,$\pm$\,0.10 & 23.50\,$\pm$\,0.14 & 22.41\,$\pm$\,0.03 & ... & 21.09\,$\pm$\,0.01 & 20.26\,$\pm$\,0.01 & 19.93\,$\pm$\,0.01 & 19.80\,$\pm$\,0.02 & 20.36\,$\pm$\,0.01 \\
  ALESS 67.2 &  $>$26.18 & $>$25.29 & 26.09\,$\pm$\,0.04 & 25.57\,$\pm$\,0.12 & 25.43\,$\pm$\,0.13 & 25.10\,$\pm$\,0.19 & 24.59\,$\pm$\,0.25 & 24.12\,$\pm$\,0.23 & 24.09\,$\pm$\,0.14 & ... & 22.98\,$\pm$\,0.08 & 21.36\,$\pm$\,0.02 & 21.13\,$\pm$\,0.02 & 20.82\,$\pm$\,0.05 & 21.48\,$\pm$\,0.04 \\
  ALESS 68.1 &  $>$26.18 & $>$25.29 & $>$28.14 & $>$26.53 & $>$26.32 & $>$25.53 & $>$24.68 & $>$24.29 & $>$24.88 & ... & $>$24.35 & 23.34\,$\pm$\,0.09 & 22.77\,$\pm$\,0.08 & $>$22.44 & 22.09\,$\pm$\,0.07 \\
  ALESS 69.1 &  $>$26.18 & $>$25.29 & ... & $>$26.53 & $>$26.32 & $>$25.53 & $>$24.68 & $>$24.29 & $>$24.88 & ... & 22.74\,$\pm$\,0.07 & 21.49\,$\pm$\,0.02 & 21.08\,$\pm$\,0.02 & 20.72\,$\pm$\,0.05 & 21.02\,$\pm$\,0.03 \\
  ALESS 69.2 &  $>$26.18 & $>$25.29 & ... & $>$26.53 & $>$26.32 & $>$25.53 & $>$24.68 & $>$24.29 & $>$24.88 & ... & 24.29\,$\pm$\,0.25 & $>$24.45 & $>$24.09 & $>$22.44 & $>$23.38 \\
  ALESS 69.3 &  $>$26.18 & $>$25.29 & ... & $>$26.53 & $>$26.32 & $>$25.53 & $>$24.68 & $>$24.29 & $>$24.88 & ... & $>$24.35 & $>$24.45 & $>$24.09 & $>$22.44 & $>$23.38 \\
  ALESS 70.1$^{a}$ &  24.86\,$\pm$\,0.09 & 24.37\,$\pm$\,0.12 & 24.49\,$\pm$\,0.01 & 23.72\,$\pm$\,0.02 & 23.61\,$\pm$\,0.02 & 23.47\,$\pm$\,0.04 & 23.33\,$\pm$\,0.08 & 23.34\,$\pm$\,0.12 & 22.27\,$\pm$\,0.03 & ... & 21.16\,$\pm$\,0.02 & 20.25\,$\pm$\,0.01 & 20.03\,$\pm$\,0.01 & 19.87\,$\pm$\,0.02 & 20.21\,$\pm$\,0.01 \\
  ALESS 71.1 &  $>$26.18 & $>$25.29 & ... & 25.36\,$\pm$\,0.10 & 25.23\,$\pm$\,0.10 & 24.28\,$\pm$\,0.09 & 23.10\,$\pm$\,0.07 & 23.03\,$\pm$\,0.09 & 21.81\,$\pm$\,0.08$^{**}$ & ... & 20.73\,$\pm$\,0.06$^{**}$ & 18.76\,$\pm$\,0.00 & 18.08\,$\pm$\,0.00 & 17.66\,$\pm$\,0.00 & 17.79\,$\pm$\,0.00 \\
  ALESS 71.3 &  26.18\,$\pm$\,0.27 & $>$25.29 & ... & 25.03\,$\pm$\,0.07 & 25.13\,$\pm$\,0.10 & 25.10\,$\pm$\,0.19 & $>$24.68 & $>$24.29 & $>$23.22$^{**}$ & ... & $>$22.41$^{**}$ & 23.40\,$\pm$\,0.10 & 23.34\,$\pm$\,0.13 & $>$22.44 & $>$23.38 \\
  ALESS 72.1 &  $>$26.18 & $>$25.29 & $>$28.14 & $>$26.53 & $>$26.32 & $>$25.53 & $>$24.68 & $>$24.29 & $>$24.88 & $>$23.06 & $>$24.35 & 22.80\,$\pm$\,0.06 & 22.91\,$\pm$\,0.09 & $>$22.44 & 22.93\,$\pm$\,0.15 \\
  ALESS 73.1 &  $>$26.18 & $>$25.29 & $>$28.14 & $>$26.53 & $>$26.32 & 25.73\,$\pm$\,0.13 & 24.00\,$\pm$\,0.15 & $>$24.29 & 24.04\,$\pm$\,0.13 & ... & 23.57\,$\pm$\,0.14 & 22.53\,$\pm$\,0.05 & 22.41\,$\pm$\,0.06 & 21.92\,$\pm$\,0.14 & 21.59\,$\pm$\,0.04 \\
  ALESS 74.1$^{a}$ &  $>$26.18 & $>$25.29 & 27.55\,$\pm$\,0.16 & $>$26.53 & $>$26.32 & $>$25.53 & $>$24.68 & $>$24.29 & 23.51\,$\pm$\,0.08 & ... & 22.36\,$\pm$\,0.05 & 21.17\,$\pm$\,0.01 & 20.77\,$\pm$\,0.01 & 20.61\,$\pm$\,0.04 & 21.18\,$\pm$\,0.03 \\
  ALESS 75.1 &  24.65\,$\pm$\,0.07 & 24.44\,$\pm$\,0.13 & ... & 23.58\,$\pm$\,0.02 & 23.32\,$\pm$\,0.02 & 23.11\,$\pm$\,0.03 & 23.05\,$\pm$\,0.07 & 23.22\,$\pm$\,0.11 & 22.38\,$\pm$\,0.03 & ... & 22.01\,$\pm$\,0.03 & 20.71\,$\pm$\,0.01 & 20.06\,$\pm$\,0.01 & 19.39\,$\pm$\,0.01 & 18.68\,$\pm$\,0.00 \\
  ALESS 75.4$^{a}$ &  26.09\,$\pm$\,0.25 & $>$25.29 & ... & 25.41\,$\pm$\,0.10 & 25.21\,$\pm$\,0.10 & 24.94\,$\pm$\,0.16 & $>$24.68 & $>$24.29 & 24.43\,$\pm$\,0.18 & ... & $>$24.35 & 24.01\,$\pm$\,0.17 & $>$24.09 & ... & ... \\
  ALESS 76.1 &  $>$26.18 & $>$25.29 & ... & $>$26.53 & $>$26.32 & $>$25.53 & $>$24.68 & $>$24.29 & $>$23.22$^{**}$ & ... & $>$22.41$^{**}$ & 23.49\,$\pm$\,0.11 & 23.06\,$\pm$\,0.10 & $>$22.44 & 22.68\,$\pm$\,0.12 \\
  ALESS 79.1 &  $>$26.18 & $>$25.29 & $>$28.14 & $>$26.53 & $>$26.32 & $>$25.53 & $>$24.68 & $>$24.29 & $>$24.88 & ... & 23.03\,$\pm$\,0.09 & 21.70\,$\pm$\,0.02 & 21.21\,$\pm$\,0.02 & 20.90\,$\pm$\,0.06 & 21.11\,$\pm$\,0.03 \\
  ALESS 79.2 &  26.08\,$\pm$\,0.25 & $>$25.29 & 25.65\,$\pm$\,0.03 & 24.97\,$\pm$\,0.07 & 24.64\,$\pm$\,0.06 & 24.31\,$\pm$\,0.09 & 23.39\,$\pm$\,0.09 & 23.31\,$\pm$\,0.12 & 22.05\,$\pm$\,0.02 & ... & 20.89\,$\pm$\,0.01 & 19.95\,$\pm$\,0.00 & 19.75\,$\pm$\,0.00 & 19.91\,$\pm$\,0.02 & 20.45\,$\pm$\,0.02 \\
  ALESS 79.4 &  $>$26.18 & $>$25.29 & 27.73\,$\pm$\,0.19 & $>$26.53 & $>$26.32 & $>$25.53 & $>$24.68 & $>$24.29 & $>$24.88 & ... & $>$24.35 & $>$24.45 & $>$24.09 & $>$22.44 & $>$23.38 \\
  ALESS 80.1 &  $>$26.18 & $>$25.29 & 27.68\,$\pm$\,0.18 & $>$26.53 & 26.30\,$\pm$\,0.26 & $>$25.53 & 24.66\,$\pm$\,0.26 & $>$24.29 & 23.88\,$\pm$\,0.11 & ... & 22.28\,$\pm$\,0.04 & 21.45\,$\pm$\,0.02 & 21.12\,$\pm$\,0.02 & 20.81\,$\pm$\,0.05 & 21.34\,$\pm$\,0.04 \\
  ALESS 80.2 &  $>$26.18 & $>$25.29 & 27.00\,$\pm$\,0.10 & 26.31\,$\pm$\,0.22 & 25.83\,$\pm$\,0.18 & $>$25.53 & $>$24.68 & $>$24.29 & 23.90\,$\pm$\,0.12 & ... & 22.51\,$\pm$\,0.05 & 21.39\,$\pm$\,0.02 & 21.14\,$\pm$\,0.02 & 21.25\,$\pm$\,0.08 & 21.94\,$\pm$\,0.06 \\
  ALESS 82.1 &  $>$26.18 & $>$25.29 & $>$28.14 & $>$26.53 & $>$26.32 & $>$25.53 & $>$24.68 & $>$24.29 & 24.56\,$\pm$\,0.27$^{*}$ & ... & 23.48\,$\pm$\,0.13 & 22.19\,$\pm$\,0.03 & 21.79\,$\pm$\,0.03 & 21.61\,$\pm$\,0.11 & 21.71\,$\pm$\,0.05 \\
  ALESS 83.4$^{a}$ &  ... & ... & ... & ... & ... & ... & ... & ... & ... & ... & ... & 20.79\,$\pm$\,0.01 & 21.01\,$\pm$\,0.02 & 21.35\,$\pm$\,0.09 & 22.06\,$\pm$\,0.07  \\
  ALESS 84.1 &  25.81\,$\pm$\,0.20 & 25.27\,$\pm$\,0.26 & 25.30\,$\pm$\,0.02 & 24.71\,$\pm$\,0.05 & 24.60\,$\pm$\,0.06 & 24.40\,$\pm$\,0.10 & 24.03\,$\pm$\,0.15 & ... & 23.24\,$\pm$\,0.06 & 22.45\,$\pm$\,0.16 & 21.95\,$\pm$\,0.03 & 21.06\,$\pm$\,0.01 & 20.71\,$\pm$\,0.01 & 20.50\,$\pm$\,0.04 & 20.70\,$\pm$\,0.02 \\
  ALESS 84.2 &  $>$26.18 & $>$25.29 & 26.56\,$\pm$\,0.07 & 25.83\,$\pm$\,0.15 & 25.33\,$\pm$\,0.11 & 25.08\,$\pm$\,0.18 & 24.71\,$\pm$\,0.27 & 24.28\,$\pm$\,0.26 & 22.83\,$\pm$\,0.04 & 22.48\,$\pm$\,0.16 & 21.65\,$\pm$\,0.02 & 21.00\,$\pm$\,0.01 & 20.81\,$\pm$\,0.01 & 20.77\,$\pm$\,0.05 & 21.33\,$\pm$\,0.04  \\  [0.5ex] 
  \hline\hline
  \end{tabular}}
\refstepcounter{table}\label{table:observed}
  \end{sidewaystable}

\clearpage

  %
  %
  \begin{sidewaystable}
  \vspace{10cm}
  \centerline{\sc Table 2: Continued}
  \resizebox{\linewidth}{!}{%
  \tiny
  \begin{tabular}{lccccccccccccccccc}
  \hline
  \noalign{\smallskip}
  ID & MUSYC $U$ & MUSYC $U_{38}$ & VIMOS $U$ & $B$ & $V$ & $R$ & $I$ & $z$ & $J^{b}$ & $H$ & $K^{b}$ & 3.6\,$\mu$m & 4.5\,$\mu$m & 5.8\,$\mu$m & 8.0\,$\mu$m \\  [0.5ex] 
  \hline 
  ALESS 87.1 &  ... & ... & 25.33\,$\pm$\,0.02 & ... & ... & ... & ... & ... & ... & ... & ... & 20.92\,$\pm$\,0.01 & 20.68\,$\pm$\,0.01 & 20.62\,$\pm$\,0.04 & 20.50\,$\pm$\,0.02 \\ ALESS 87.3 &  ... & ... & $>$28.14 & ... & ... & ... & ... & ... & ... & ... & ... & $>$24.45 & $>$24.09 & $>$22.44 & $>$23.38 \\
  ALESS 88.1 &  25.65\,$\pm$\,0.17 & $>$25.29 & 25.51\,$\pm$\,0.03 & 24.93\,$\pm$\,0.07 & 24.65\,$\pm$\,0.06 & 24.46\,$\pm$\,0.11 & 23.75\,$\pm$\,0.12 & 23.78\,$\pm$\,0.17 & 22.91\,$\pm$\,0.05 & 22.98\,$\pm$\,0.25 & 21.83\,$\pm$\,0.03 & 20.93\,$\pm$\,0.01 & 20.64\,$\pm$\,0.01 & 20.48\,$\pm$\,0.04 & 20.82\,$\pm$\,0.02 \\
  ALESS 88.2 &  $>$26.18 & $>$25.29 & 27.68\,$\pm$\,0.18 & $>$26.53 & $>$26.32 & $>$25.53 & $>$24.68 & $>$24.29 & $>$24.88 & $>$23.06 & $>$24.35 & $>$24.45 & $>$24.09 & $>$22.44 & $>$23.38 \\
  ALESS 88.5 &  $>$26.18 & $>$25.29 & 28.11\,$\pm$\,0.26 & $>$26.53 & 26.07\,$\pm$\,0.22 & $>$25.53 & $>$24.68 & $>$24.29 & 23.77\,$\pm$\,0.10 & 22.12\,$\pm$\,0.12 & 22.31\,$\pm$\,0.04 & 21.52\,$\pm$\,0.02 & 21.17\,$\pm$\,0.02 & 20.93\,$\pm$\,0.06 & 21.27\,$\pm$\,0.03 \\
  ALESS 88.11 &  25.03\,$\pm$\,0.10 & 24.92\,$\pm$\,0.20 & 25.17\,$\pm$\,0.02 & 24.06\,$\pm$\,0.03 & 23.65\,$\pm$\,0.03 & 23.50\,$\pm$\,0.05 & 23.19\,$\pm$\,0.07 & 23.08\,$\pm$\,0.09 & 22.99\,$\pm$\,0.05 & $>$23.06 & 22.06\,$\pm$\,0.04 & 21.37\,$\pm$\,0.02 & 21.21\,$\pm$\,0.02 & 21.18\,$\pm$\,0.07 & 21.50\,$\pm$\,0.04 \\
  ALESS 92.2 &  25.56\,$\pm$\,0.16 & $>$25.29 & 25.55\,$\pm$\,0.03 & 25.28\,$\pm$\,0.09 & 24.82\,$\pm$\,0.07 & 24.42\,$\pm$\,0.10 & 24.48\,$\pm$\,0.23 & 23.63\,$\pm$\,0.15 & 23.83\,$\pm$\,0.11 & ... & 23.75\,$\pm$\,0.16 & 23.48\,$\pm$\,0.11 & 23.52\,$\pm$\,0.15 & $>$22.44 & $>$23.38 \\
  ALESS 94.1 &  $>$26.18 & $>$25.29 & ... & 26.03\,$\pm$\,0.18 & 25.92\,$\pm$\,0.19 & $>$25.53 & $>$24.68 & $>$24.29 & $>$24.88 & $>$23.06 & 23.33\,$\pm$\,0.11 & 22.06\,$\pm$\,0.03 & 21.66\,$\pm$\,0.03 & 21.33\,$\pm$\,0.08 & 21.47\,$\pm$\,0.04 \\
  ALESS 98.1 &  $>$26.18 & $>$25.29 & ... & $>$26.53 & $>$26.32 & $>$25.53 & 24.28\,$\pm$\,0.19 & ... & 22.71\,$\pm$\,0.04 & ... & 21.22\,$\pm$\,0.02 & 19.86\,$\pm$\,0.00 & 19.47\,$\pm$\,0.00 & 19.59\,$\pm$\,0.02 & 19.91\,$\pm$\,0.01 \\
  ALESS 99.1 &  $>$26.18 & $>$25.29 & 28.10\,$\pm$\,0.26 & $>$26.53 & $>$26.32 & $>$25.53 & $>$24.68 & $>$24.29 & $>$24.88 & $>$23.06 & $>$24.35 & $>$24.45 & $>$24.09 & $>$22.44 & $>$23.38 \\
  ALESS 102.1 &  $>$26.18 & $>$25.29 & ... & $>$26.53 & 26.27\,$\pm$\,0.22 & $>$25.53 & $>$24.68 & $>$24.29 & 22.79\,$\pm$\,0.19$^{**}$ & ... & 21.07\,$\pm$\,0.08$^{**}$ & 20.07\,$\pm$\,0.00 & 19.78\,$\pm$\,0.01 & 19.77\,$\pm$\,0.02 & 20.56\,$\pm$\,0.02 \\
  ALESS 103.3 &  $>$26.18 & $>$25.29 & ... & $>$26.53 & $>$26.32 & $>$25.53 & $>$24.68 & $>$24.29 & $>$23.22$^{**}$ & ... & $>$22.41$^{**}$ & $>$24.45 & $>$24.09 & $>$22.44 & $>$23.38 \\
  ALESS 107.1 &  $>$26.18 & $>$25.29 & ... & 25.63\,$\pm$\,0.12 & 24.62\,$\pm$\,0.06 & 23.66\,$\pm$\,0.05 & 22.61\,$\pm$\,0.04 & 22.61\,$\pm$\,0.06 & 21.83\,$\pm$\,0.02 & ... & 21.08\,$\pm$\,0.01 & 20.49\,$\pm$\,0.01 & 20.71\,$\pm$\,0.01 & 20.77\,$\pm$\,0.05 & 20.89\,$\pm$\,0.02 \\
  ALESS 107.3 &  $>$26.18 & $>$25.29 & ... & 25.82\,$\pm$\,0.15 & 25.52\,$\pm$\,0.14 & 25.52\,$\pm$\,0.26 & $>$24.68 & $>$24.29 & 24.44\,$\pm$\,0.18 & ... & 24.17\,$\pm$\,0.23 & $>$24.45 & $>$24.09 & $>$22.44 & $>$23.38 \\
  ALESS 110.1 &  ... & ... & ... & ... & ... & ... & ... & ... & $>$24.88 & ... & $>$24.35 & 22.72\,$\pm$\,0.05 & 22.04\,$\pm$\,0.04 & 21.53\,$\pm$\,0.10 & 21.32\,$\pm$\,0.04 \\ ALESS 110.5 &  ... & ... & ... & ... & ... & ... & ... & ... & $>$24.88 & ... & $>$24.35 & 22.49\,$\pm$\,0.04 & 23.01\,$\pm$\,0.10 & $>$22.44 & $>$23.38 \\
  ALESS 112.1 &  ... & ... & 26.79\,$\pm$\,0.08 & ... & ... & ... & ... & ... & ... & ... & ... & 20.56\,$\pm$\,0.01 & 20.22\,$\pm$\,0.01 & 20.03\,$\pm$\,0.03 & 20.66\,$\pm$\,0.02 \\
  ALESS 114.1 &  $>$26.18 & $>$25.29 & $>$28.14 & $>$26.53 & $>$26.32 & $>$25.53 & $>$24.68 & $>$24.29 & $>$24.88 & $>$23.06 & $>$24.35 & 23.99\,$\pm$\,0.16 & 23.21\,$\pm$\,0.11 & $>$22.44 & $>$23.38 \\
  ALESS 114.2 &  24.83\,$\pm$\,0.08 & 24.87\,$\pm$\,0.19 & 24.78\,$\pm$\,0.01 & 24.24\,$\pm$\,0.04 & 23.93\,$\pm$\,0.03 & 23.56\,$\pm$\,0.05 & 22.73\,$\pm$\,0.05 & 22.61\,$\pm$\,0.06 & 21.21\,$\pm$\,0.01 & 20.58\,$\pm$\,0.03 & 20.37\,$\pm$\,0.01 & 19.56\,$\pm$\,0.00 & 19.28\,$\pm$\,0.00 & 19.46\,$\pm$\,0.02 & 19.70\,$\pm$\,0.01 \\
  ALESS 116.1$^{a}$ &  $>$26.18 & $>$25.29 & $>$28.14 & $>$26.53 & $>$26.32 & $>$25.53 & $>$24.68 & $>$24.29 & $>$24.88 & $>$23.06 & 24.01\,$\pm$\,0.20 & 23.52\,$\pm$\,0.11 & 22.83\,$\pm$\,0.08 & $>$22.44 & 22.49\,$\pm$\,0.10 \\
  ALESS 116.2 &  $>$26.18 & $>$25.29 & $>$28.14 & $>$26.53 & $>$26.32 & $>$25.53 & $>$24.68 & $>$24.29 & $>$24.88 & $>$23.06 & 23.85\,$\pm$\,0.17 & 22.86\,$\pm$\,0.06 & 22.22\,$\pm$\,0.05 & 21.92\,$\pm$\,0.14 & 21.87\,$\pm$\,0.06 \\
  ALESS 118.1 &  ... & ... & ... & ... & ... & ... & ... & ... & 23.42\,$\pm$\,0.08 & ... & 22.67\,$\pm$\,0.06 & 21.84\,$\pm$\,0.02 & 21.29\,$\pm$\,0.02 & 21.07\,$\pm$\,0.07 & 21.35\,$\pm$\,0.04 \\
  ALESS 119.1 &  $>$26.18 & $>$25.29 & ... & $>$26.53 & 25.82\,$\pm$\,0.18 & 25.17\,$\pm$\,0.20 & $>$24.68 & 24.04\,$\pm$\,0.22 & 24.20\,$\pm$\,0.20$^{*}$ & ... & 23.41\,$\pm$\,0.15$^{*}$ & 22.78\,$\pm$\,0.06 & 22.19\,$\pm$\,0.05 & 21.66\,$\pm$\,0.11 & 21.89\,$\pm$\,0.06 \\
  ALESS 122.1 &  24.34\,$\pm$\,0.05 & 24.31\,$\pm$\,0.12 & 24.22\,$\pm$\,0.01 & 23.49\,$\pm$\,0.02 & 23.16\,$\pm$\,0.02 & 22.96\,$\pm$\,0.03 & 22.62\,$\pm$\,0.04 & 22.55\,$\pm$\,0.06 & 21.49\,$\pm$\,0.01 & ... & 20.68\,$\pm$\,0.01 & 19.88\,$\pm$\,0.00 & 19.51\,$\pm$\,0.00 & 19.19\,$\pm$\,0.01 & 19.27\,$\pm$\,0.01 \\
  ALESS 124.1$^{a}$ &  $>$26.18 & $>$25.29 & $>$28.14 & $>$26.53 & $>$26.32 & $>$25.53 & $>$24.68 & $>$24.29 & 24.63\,$\pm$\,0.22 & $>$23.06 & 23.73\,$\pm$\,0.16 & 21.92\,$\pm$\,0.03 & 21.49\,$\pm$\,0.02 & 21.44\,$\pm$\,0.09 & 21.18\,$\pm$\,0.03 \\
  ALESS 124.4 &  $>$26.18 & $>$25.29 & $>$28.14 & $>$26.53 & 25.96\,$\pm$\,0.20 & $>$25.53 & $>$24.68 & ... & $>$24.88 & $>$23.06 & $>$24.35 & $>$24.45 & $>$24.09 & $>$22.44 & $>$23.38 \\
  ALESS 126.1 &  $>$26.18 & $>$25.29 & 26.68\,$\pm$\,0.08 & 26.11\,$\pm$\,0.19 & 26.31\,$\pm$\,0.26 & $>$25.53 & $>$24.68 & ... & 23.79\,$\pm$\,0.10 & 22.33\,$\pm$\,0.14 & 22.28\,$\pm$\,0.04 & 20.95\,$\pm$\,0.01 & 20.74\,$\pm$\,0.01 & 20.86\,$\pm$\,0.06 & 21.42\,$\pm$\,0.04 \\  [0.5ex] 
  \hline\hline
  \end{tabular}}
  \vspace{-0.3cm}
  \begin{flushleft}
   \footnotesize{ $3\,\sigma$ upper limits are presented for non-detections, and the entry is left blank where a source is not covered by available imaging. $^a$ Source is within 4$''$ of a 3.6--$\mu$m source of comparable, or greater, flux  $^b$ We measure $J$ and $K_S$ photometry from three imaging surveys, but quote a single value, in order of $3\,\sigma$ detection limit (see Table~\ref{table:depths}). $^\ast$ Photometry measured from HAWK--I imaging. $^{\ast \ast}$ Photometry measured from MUSYC imaging, otherwise photometry measured from TENIS imaging. 

}
  \end{flushleft}
  \end{sidewaystable}

\newpage
 
\begin{table*}
\centering
\small
\centerline{\sc Table 3: Derived Properties}
\vspace{0.1cm}
{
\begin{tabular}{lcccccccc}
\hline
\noalign{\smallskip}
ID & RA  & Dec & $z_{phot}$ & $z_{spec}$ & $\chi^{2}_{red}$ & Filters & $M_{H}$ & $M\,/\,L_{H}$  \\
 & (J2000) & (J2000) &  & &  & (Det [Obs]) & (AB) &  ($M_{\odot} L_{\odot}^{-1}$)  \\  [0.5ex] 
\hline\\ [-1.5ex] 
ALESS 001.1  & 03:33:14.46 & $-$27:56:14.5 & 4.34$^{+2.66}_{-1.43}$ & & 0.91 & 5 [15] & $-$24.90 & 0.14  \\
ALESS 001.2  & 03:33:14.41 & $-$27:56:11.6 & 4.65$^{+2.34}_{-1.02}$ & & 1.04 & 5 [15] & $-$24.79 & 0.29  \\
ALESS 001.3  & 03:33:14.18 & $-$27:56:12.3 & 2.85$^{+0.20}_{-0.30}$ & & 3.78 & 5 [15] & $-$23.83 & 0.29  \\
ALESS 002.1  & 03:33:02.69 & $-$27:56:42.8 & 1.96$^{+0.27}_{-0.20}$ & & 1.39 & 8 [14] & $-$23.24 & 0.17  \\
ALESS 003.1  & 03:33:21.50 & $-$27:55:20.3 & 3.90$^{+0.50}_{-0.59}$ & & 0.68 & 5 [14] & $-$25.51 & 0.20   \\
ALESS 005.1$^{a}$  & 03:31:28.91 & $-$27:59:09.0 & 2.86$^{+0.05}_{-0.04}$ & & 14.76 & 13 [13] & $-$25.71 & 0.04  \\
ALESS 006.1$^{a}$  & 03:32:56.96 &$-$28:01:00.7 & 0.45$^{+0.06}_{-0.04}$ & & 17.54 & 15 [15] & $-$21.95 & 0.25  \\
ALESS 007.1  & 03:33:15.42 & $-$27:45:24.3 & 2.50$^{+0.12}_{-0.16}$ & & 8.32 & 14 [15] & $-$25.96 & 0.04  \\
ALESS 009.1  & 03:32:11.34 & $-$27:52:11.9 & 4.50$^{+0.54}_{-2.33}$ & & 0.22 & 5 [14] & $-$25.98 & 0.25 \\
ALESS 010.1  & 03:32:19.06 & $-$27:52:14.8 & 2.02$^{+0.09}_{-0.09}$ & & 9.89 & 12 [13] & $-$23.57 & 0.15  \\
ALESS 011.1  & 03:32:13.85 & $-$27:56:00.3 & 2.83$^{+1.88}_{-0.50}$ & & 2.02 & 6 [15] & $-$24.95 & 0.36  \\
ALESS 013.1  & 03:32:48.99 & $-$27:42:51.8 & 3.25$^{+0.64}_{-0.46}$ & & 2.12 & 8 [15] & $-$25.05 & 0.11  \\
ALESS 014.1  & 03:31:52.49 & $-$28:03:19.1 & 4.47$^{+2.54}_{-0.88}$ & & 1.46 & 6 [13] & $-$26.13 & 0.06  \\
ALESS 015.1  & 03:33:33.37 & $-$27:59:29.6 & 1.93$^{+0.62}_{-0.33}$ & & 0.43 & 4 [13] & $-$23.94 & 0.21  \\
ALESS 017.1  & 03:32:07.30 & $-$27:51:20.8 & 1.51$^{+0.10}_{-0.07}$ & & 1.95 & 15 [15] & $-$24.46 & 0.18 \\
ALESS 018.1  & 03:32:04.88 & $-$27:46:47.7 & 2.04$^{+0.10}_{-0.06}$ & 2.25$^{b}$ & 5.44 & 14 [15] & $-$25.33 & 0.15 \\
ALESS 019.1  & 03:32:08.26 & $-$27:58:14.2 & 2.41$^{+0.17}_{-0.11}$ & & 8.40 & 7 [14] & $-$23.93 & 0.15  \\
ALESS 019.2  & 03:32:07.89 & $-$27:58:24.1 & 2.17$^{+0.09}_{-0.10}$ & & 2.04 & 13 [14] & $-$23.74 & 0.14 \\
ALESS 022.1  & 03:31:46.92 & $-$27:32:39.3 & 1.88$^{+0.18}_{-0.23}$ & & 3.50 & 6 [11] & $-$25.04 & 0.22  \\
ALESS 023.1  & 03:32:12.01 & $-$28:05:06.5 & 4.99$^{+2.01}_{-2.55}$ & & 0.35 & 4 [6] & $-$25.78 & 0.09  \\
ALESS 025.1  & 03:31:56.88 & $-$27:59:39.3 & 2.24$^{+0.07}_{-0.17}$ & & 1.65 & 12 [14] & $-$25.03 & 0.17  \\
ALESS 029.1  & 03:33:36.90 & $-$27:58:09.3 & 2.66$^{+2.94}_{-0.76}$ & & 0.10 & 4 [13] & $-$24.78 & 0.36 \\
ALESS 031.1  & 03:31:49.79 & $-$27:57:40.8 & 2.89$^{+1.80}_{-0.41}$ & & 1.09 & 5 [13] & $-$24.62 & 0.12 \\
ALESS 037.1  & 03:33:36.14 & $-$27:53:50.6 & 3.53$^{+0.56}_{-0.31}$ & & 2.07 & 9 [12] & $-$25.73 & 0.04 \\
ALESS 037.2  & 03:33:36.36 & $-$27:53:48.3 & 4.87$^{+0.22}_{-0.40}$ & & 0.95 & 6 [12] & $-$25.26 & 0.17 \\
ALESS 039.1  & 03:31:45.03 & $-$27:34:36.7 & 2.44$^{+0.17}_{-0.23}$ & & 9.14 & 11 [13] & $-$24.74 & 0.04  \\
ALESS 041.1  & 03:31:10.07 & $-$27:52:36.7 & 2.75$^{+4.25}_{-0.72}$ & & 0.00 & 4 [4] & $-$26.15 & 0.17  \\
ALESS 043.1  & 03:33:06.64 & $-$27:48:02.4 & 1.71$^{+0.20}_{-0.12}$ & & 4.33 & 7 [14] & $-$23.85 & 0.22  \\
ALESS 045.1  & 03:32:25.26 & $-$27:52:30.5 & 2.34$^{+0.26}_{-0.67}$ & & 0.32 & 5 [14] & $-$24.77 & 0.36  \\
ALESS 049.1  & 03:31:24.72 & $-$27:50:47.1 & 2.76$^{+0.11}_{-0.14}$ & & 1.56 & 11 [13] & $-$24.44 & 0.05  \\
ALESS 049.2  & 03:31:24.47 & $-$27:50:38.1 & 1.47$^{+0.07}_{-0.10}$ & & 3.99 & 12 [13] & $-$23.55 & 0.04  \\
ALESS 051.1  & 03:31:45.06 & $-$27:44:27.3 & 1.22$^{+0.03}_{-0.06}$ & & 7.76 & 13 [15] & $-$24.36 & 0.25  \\
ALESS 055.1  & 03:33:02.22 & $-$27:40:35.4 & 2.05$^{+0.15}_{-0.13}$ & & 7.04 & 13 [14] & $-$22.93 & 0.15  \\
ALESS 055.5  & 03:33:02.35 & $-$27:40:35.4 & 2.35$^{+0.11}_{-0.13}$ & & 6.89 & 12 [14] & $-$22.97 & 0.15  \\
ALESS 057.1  & 03:31:51.92 & $-$27:53:27.1 & 2.95$^{+0.05}_{-0.10}$ & 2.94$^{c}$ & 17.28 & 11 [15] & $-$24.91 & 0.15 \\
ALESS 059.2  & 03:33:03.82 & $-$27:44:18.2 & 2.09$^{+0.78}_{-0.29}$ & & 3.88 & 8 [14] & $-$22.55 & 0.15  \\
ALESS 061.1  & 03:32:45.87 & $-$28:00:23.4 & 6.52$^{+0.36}_{-0.34}$ & 4.44$^{d}$ & 3.97 & 7 [13] & $-$25.61 & 0.05  \\
ALESS 063.1  & 03:33:08.45 & $-$28:00:43.8 & 1.87$^{+0.10}_{-0.33}$ & & 3.07 & 11 [13] & $-$24.43 & 0.14  \\
ALESS 066.1  & 03:33:31.93 & $-$27:54:09.5 & 2.33$^{+0.05}_{-0.04}$ & 1.31$^{e}$ & 46.79 & 13 [13] & $-$26.24 & 0.15 \\
ALESS 067.1  & 03:32:43.20 & $-$27:55:14.3 & 2.14$^{+0.05}_{-0.09}$ & 2.12$^{f}$ & 3.31 & 13 [14] & $-$25.35 & 0.15  \\
ALESS 067.2  & 03:32:43.02 & $-$27:55:14.7 & 2.05$^{+0.06}_{-0.16}$ & & 7.42 & 12 [14] & $-$23.91 & 0.05  \\
ALESS 069.1  & 03:31:33.78 & $-$27:59:32.4 & 2.34$^{+0.27}_{-0.44}$ & & 0.51 & 5 [13] & $-$24.60 & 0.36  \\
ALESS 070.1  & 03:31:44.02 & $-$27:38:35.5 & 2.28$^{+0.05}_{-0.06}$ & & 2.47 & 14 [14] & $-$25.37 & 0.15  \\
ALESS 071.1  & 03:33:05.65 & $-$27:33:28.2 & 2.48$^{+0.21}_{-0.11}$ & & 7.65 & 11 [13] & $-$27.80 & 0.04  \\
ALESS 071.3  & 03:33:06.14 & $-$27:33:23.1 & 2.73$^{+0.22}_{-0.25}$ & & 2.87 & 6 [13] & $-$22.24 & 0.15  \\
ALESS 073.1  & 03:32:29.29 & $-$27:56:19.7 & 5.18$^{+0.43}_{-0.45}$ & 4.76$^{g}$ & 2.00 & 8 [14] & $-$25.61 & 0.07  \\
ALESS 074.1  & 03:33:09.15 & $-$27:48:17.2 & 1.80$^{+0.13}_{-0.13}$ & & 4.95 & 7 [14] & $-$23.90 & 0.19  \\
ALESS 075.1  & 03:31:27.19 & $-$27:55:51.3 & 2.39$^{+0.08}_{-0.06}$ & & 41.20 & 13 [13] & $-$25.97 & 0.05  \\
ALESS 075.4$^{a}$  & 03:31:26.57 & $-$27:55:55.7 & 2.10$^{+0.29}_{-0.34}$ & & 3.14 & 6 [11] & $-$20.94 & 0.09  \\
ALESS 079.1  & 03:32:21.14 & $-$27:56:27.0 & 2.04$^{+0.63}_{-0.31}$ & & 0.29 & 5 [14] & $-$23.88 & 0.36 \\
ALESS 079.2  & 03:32:21.60 & $-$27:56:24.0 & 1.55$^{+0.11}_{-0.18}$ & & 2.42 & 13 [14] & $-$24.56 & 0.18  \\
ALESS 080.1  & 03:31:42.80 & $-$27:48:36.9 & 1.96$^{+0.16}_{-0.14}$ & & 3.24 & 9 [14] & $-$23.77 & 0.15  \\
ALESS 080.2  & 03:31:42.62 & $-$27:48:41.0 & 1.37$^{+0.17}_{-0.08}$ & & 4.06 & 9 [14] & $-$22.81 & 0.15 \\
ALESS 082.1  & 03:32:54.00 & $-$27:38:14.9 & 2.10$^{+3.27}_{-0.44}$ & & 0.38 & 6 [14] & $-$23.34 & 0.14  \\
ALESS 083.4$^{a}$  & 03:33:08.71 & $-$28:05:18.5 & 0.57$^{+1.54}_{-0.50}$ & & 0.07 & 4 [4] & $-$21.51 & 0.11 \\ 
\hline\hline
\end{tabular}}
\refstepcounter{table}\label{table:derived}
\end{table*}

\begin{table*}
\centering
\small
\centerline{\sc Table 3: Continued}
\vspace{0.1cm}
{
\begin{tabular}{lcccccccc}
\hline
\noalign{\smallskip}
ID & RA & Dec & $z_{phot}$ & $z_{spec}$ & $\chi^{2}_{red}$ & Filters & $M_{H}$ & $M\,/\,L_{H}$ \\
 & (J2000) & (J2000) &  &  & & (Det [Obs]) & (AB) &  ($M_{\odot} L_{\odot}^{-1}$)  \\ [0.5ex] 
\hline\\ [-1.5ex] 
ALESS 084.1  & 03:31:54.50 & $-$27:51:05.6 & 1.92$^{+0.09}_{-0.07}$ & & 3.71 & 14 [14]  & $-$24.11 & 0.15 \\
ALESS 084.2  & 03:31:53.85 & $-$27:51:04.3 & 1.75$^{+0.08}_{-0.19}$ & & 1.70 & 13 [15]  & $-$23.77 & 0.20 \\
ALESS 087.1  & 03:32:50.88 & $-$27:31:41.5 & 3.20$^{+0.08}_{-0.47}$ & & 0.22 & 5 [5]      & $-$25.68 & 0.04 \\
ALESS 088.1  & 03:31:54.76 & $-$27:53:41.5 & 1.84$^{+0.12}_{-0.11}$ & 1.27$^{h}$ & 3.04 & 14 [15]  & $-$24.11 & 0.15  \\
ALESS 088.5  & 03:31:55.81 & $-$27:53:47.2 & 2.30$^{+0.11}_{-0.50}$ & & 3.69 & 9 [15]    &$-$24.34 & 0.25  \\
ALESS 088.11 & 03:31:54.95 & $-$27:53:37.6 & 2.57$^{+0.04}_{-0.12}$ & & 8.73 & 14 [15] & $-$24.32 & 0.07  \\
ALESS 092.2  & 03:31:38.14 & $-$27:43:43.4 & 1.90$^{+0.28}_{-0.75}$ & & 2.66 & 11 [14]  & $-$21.17 & 0.04  \\
ALESS 094.1  & 03:33:07.59 & $-$27:58:05.8 & 2.87$^{+0.37}_{-0.64}$ & & 3.98 & 7 [14]    &$-$24.46 & 0.15 \\
ALESS 098.1  & 03:31:29.92 & $-$27:57:22.7 & 1.63$^{+0.17}_{-0.09}$ & 1.48$^{b}$ & 2.65 & 7 [12]    &$-$24.97 & 0.20  \\
ALESS 102.1  & 03:33:35.60 & $-$27:40:23.0 & 1.76$^{+0.16}_{-0.18}$ & & 4.42 & 7 [13]    &$-$24.81 & 0.27   \\
ALESS 107.1  & 03:31:30.50 & $-$27:51:49.1 & 3.75$^{+0.09}_{-0.08}$ & & 3.55 & 11 [13]  & $-$25.49 & 0.04 \\
ALESS 107.3  & 03:31:30.72 & $-$27:51:55.7 & 2.12$^{+1.54}_{-0.81}$ & & 1.91 & 5 [13]    &$-$20.89 & 0.11  \\
ALESS 110.1  & 03:31:22.66 & $-$27:54:17.2 & 2.55$^{+0.70}_{-0.50}$ & & 0.78 & 4 [6]      &$-$24.01 & 0.36  \\
ALESS 112.1  & 03:32:48.86 & $-$27:31:13.3 & 1.95$^{+0.15}_{-0.26}$ & & 2.73 & 5 [5]      &$-$24.67 & 0.22  \\
ALESS 114.2  & 03:31:51.11 & $-$27:44:37.3 & 1.56$^{+0.07}_{-0.07}$ & 1.61$^{h}$ & 3.12 & 15 [15]  & $-$25.02 & 0.17  \\
ALESS 116.1  & 03:31:54.32 & $-$27:45:28.9 & 3.54$^{+1.47}_{-0.87}$ & & 0.82 & 4 [15]    &$-$23.84 & 0.25  \\
ALESS 116.2  & 03:31:54.44 & $-$27:45:31.4 & 4.02$^{+1.19}_{-2.19}$ & & 0.50 & 5 [15]    &$-$24.65 & 0.04  \\
ALESS 118.1  & 03:31:21.92 & $-$27:49:41.4 & 2.26$^{+0.50}_{-0.23}$ & & 3.85 & 6 [6]      &$-$24.15 & 0.04  \\
ALESS 119.1  & 03:32:56.64 & $-$28:03:25.2 & 3.50$^{+0.95}_{-0.35}$ & & 3.41 & 9 [13]    &$-$24.42 & 0.05   \\
ALESS 122.1  & 03:31:39.54 & $-$27:41:19.7 & 2.06$^{+0.05}_{-0.06}$ & 2.03$^{i}$ & 6.08 & 14 [14]  & $-$25.53 & 0.15  \\
ALESS 124.1  & 03:32:04.04 & $-$27:36:06.4 & 6.07$^{+0.94}_{-1.16}$ & & 0.80 & 6 [15]    &$-$26.22 & 0.16  \\
ALESS 126.1  & 03:32:09.61 & $-$27:41:07.7 & 1.82$^{+0.28}_{-0.08}$ & & 7.42 & 10 [14]  & $-$23.93 & 0.15  \\
\hline\hline
\end{tabular}}
\vspace{-0.3cm}
\begin{flushleft}
 \footnotesize{ $^a$ As discussed in \S\,\ref{subsubsec:caveats} these SMGs are potential gravitational lenses, or have significantly contaminated photometry. We advise that the photometric redshifts for these SMGs are treated with extreme caution. $^{b}$~\citet{Casey11} $^{c}$~\citet{Zheng04} $^{d}$~\citet{Swinbank12} $^{e}$~\citet{Silverman10} $^{f}$~\citet{Kriek08} $^{g}$~\citet{Coppin09} $^{h}$~\citet{Coppin12}; Danielson et al.\ in prep $^{i}$~\citet{Bonzini12}
}
  \end{flushleft}

\end{table*}

\clearpage

\appendix
\begin{figure*}[b]
\vspace{-2.cm}
\centerline{\sc APPENDIX:A}
\vspace{0.1cm}
\centerline{ \psfig{figure= 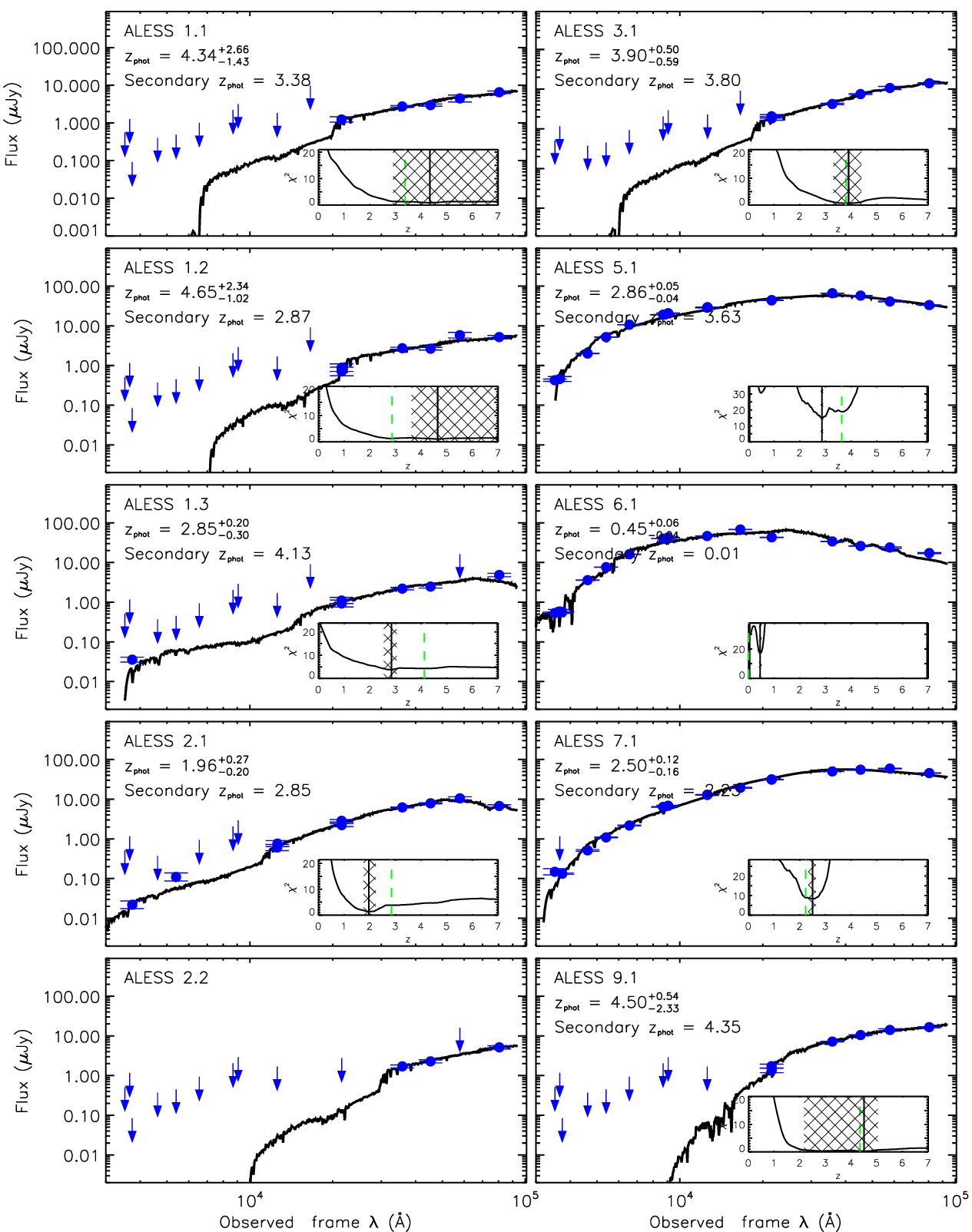,width=0.95\textwidth}}
\vspace{0.1cm}
\footnotesize{ {\bf Figure A1:} The photometry and best fit spectral energy distribution for all 96 ALESS SMGs we consider in this study. Data points and errors are observed photometry, and arrows indicate 3\,$\sigma$ detection limits. Although we present the photometry for all sources, only redshifts derived from $\ge\,4$ photometric detections are considered in our results. In the inset panel in each plot we show the $\chi^2$ distribution as a function of redshift, and indicate the best-fit photometric redshift with a solid line. The hatched region shows the uncertainty on the derived redshift. Secondary redshifts are returned by {\sc hyperz} when a secondary minima has a $>10$ per cent probability of being true, based on the reduced $\chi^2$, and where appropriate these are indicated with a green dashed line. We find the majority of our SMGs are well-fit by the best-fit SED template, and only six SMGs display evidence of an 8\,$\mu$m excess above the best-fit SED. 
}
\label{fig:sedsmain}
\vspace{-2.9cm}
\end{figure*}

%
%
%
%
%
\begin{figure*}
\centerline{ \psfig{figure= 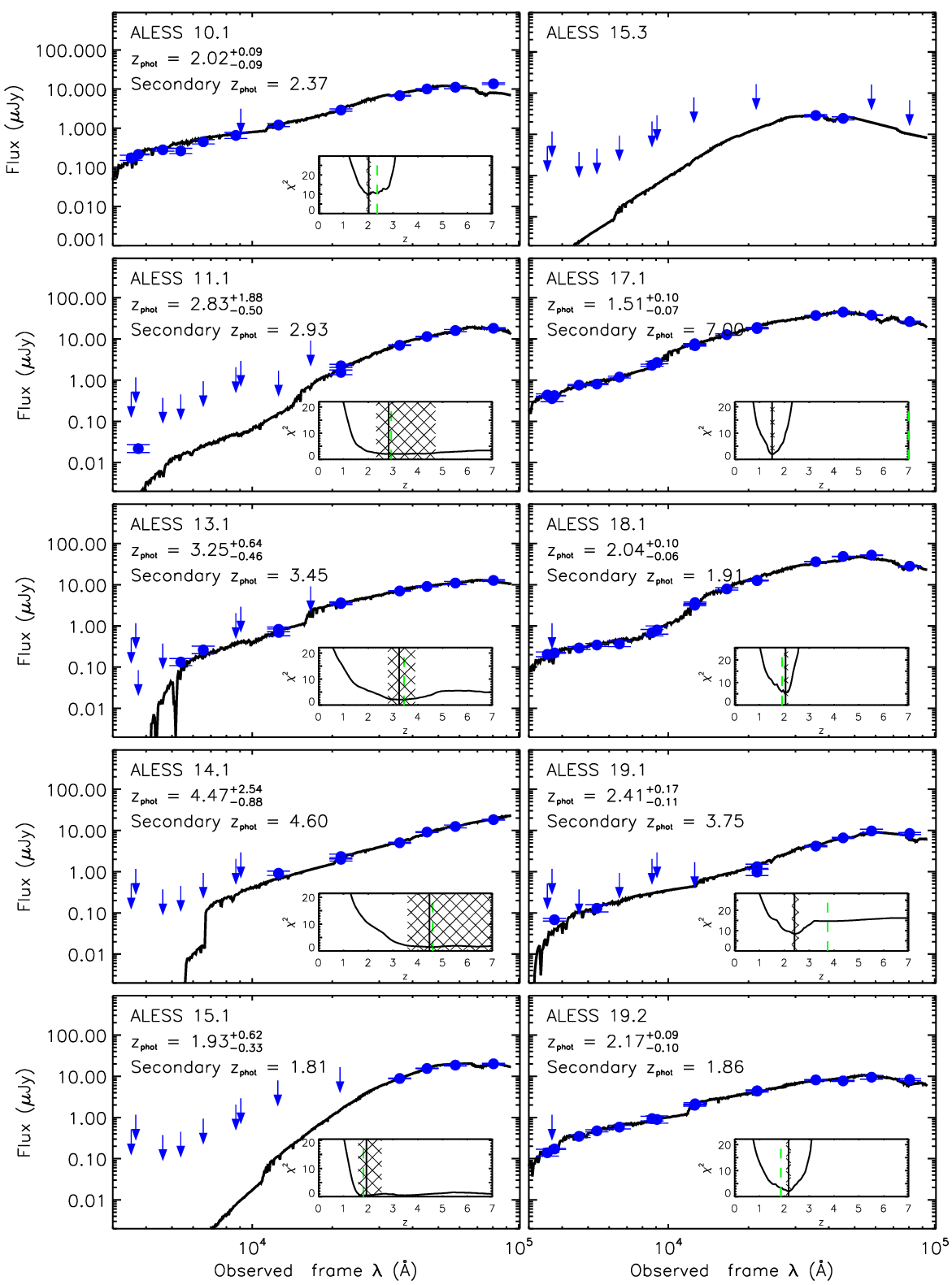,width=0.95\textwidth}}
\footnotesize{ {\bf Figure A1:} {\it cont.  } }
  \label{fig:seds10}
\end{figure*}

 %
 %
 %
 %
 %
 \begin{figure*}
 {\centerline{ \psfig{figure= 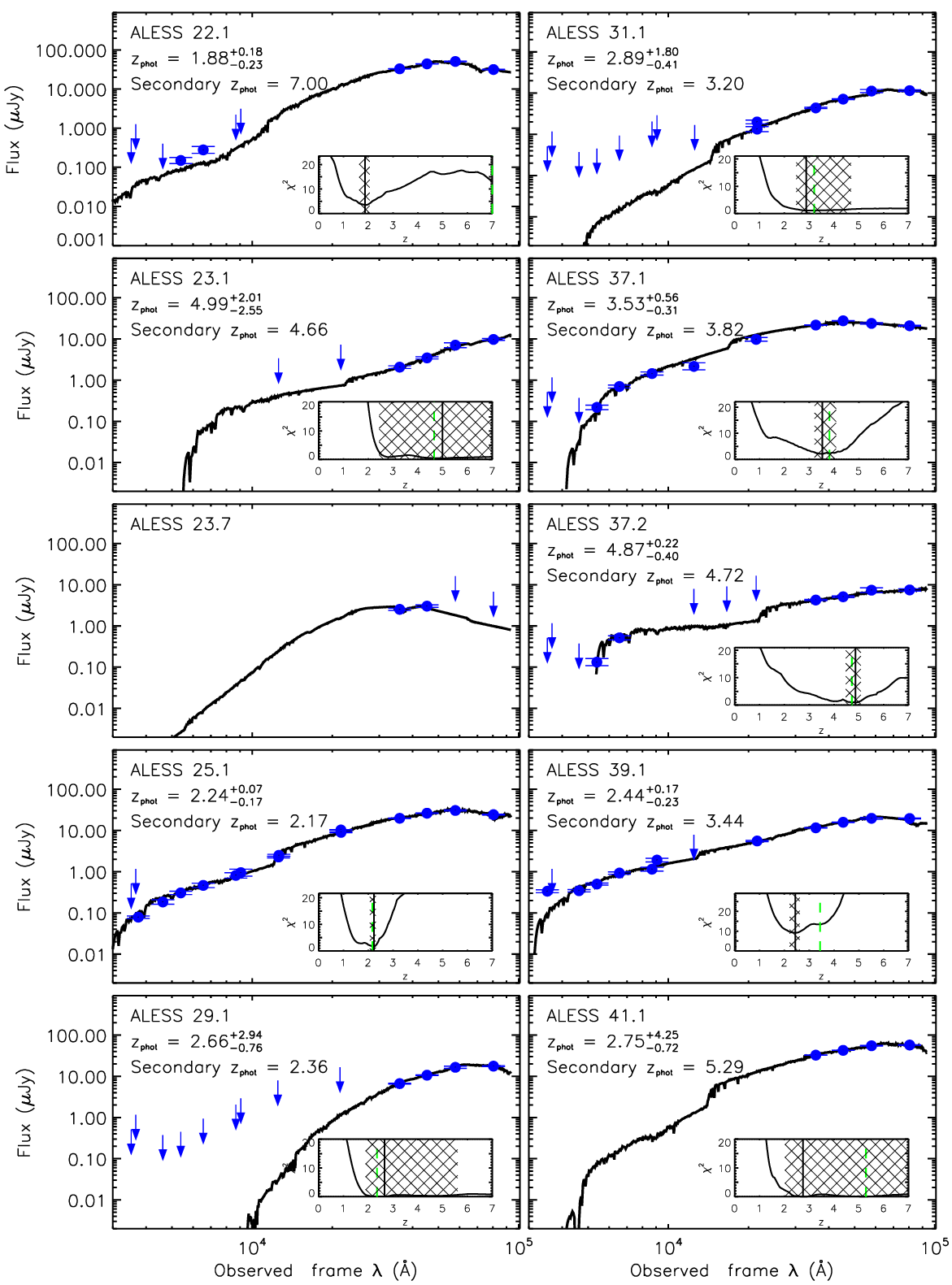,width=0.95\textwidth}} }
 \footnotesize{ {\bf Figure A1:} {\it cont.  } }
  \label{fig:seds10}
 \end{figure*}
 
 %
 %
 %
 %
 %
 \begin{figure*}
 \centerline{ \psfig{figure= 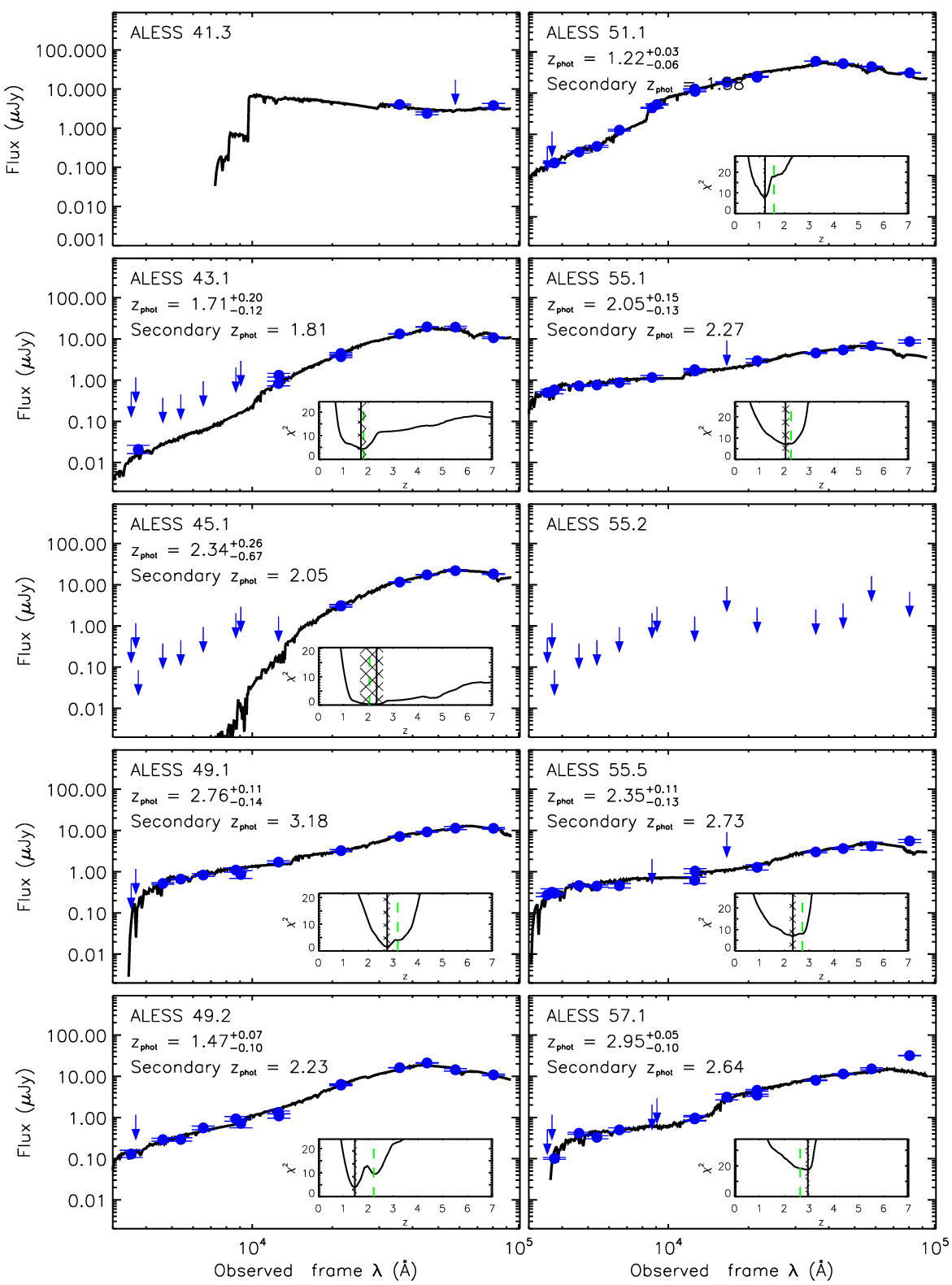,width=0.95\textwidth}}
 \footnotesize{ {\bf Figure A1:} {\it cont.  } }
  \label{fig:seds10}
 \end{figure*}
 
 %
 %
 %
 %
 %
 \begin{figure*}
 \centerline{ \psfig{figure= 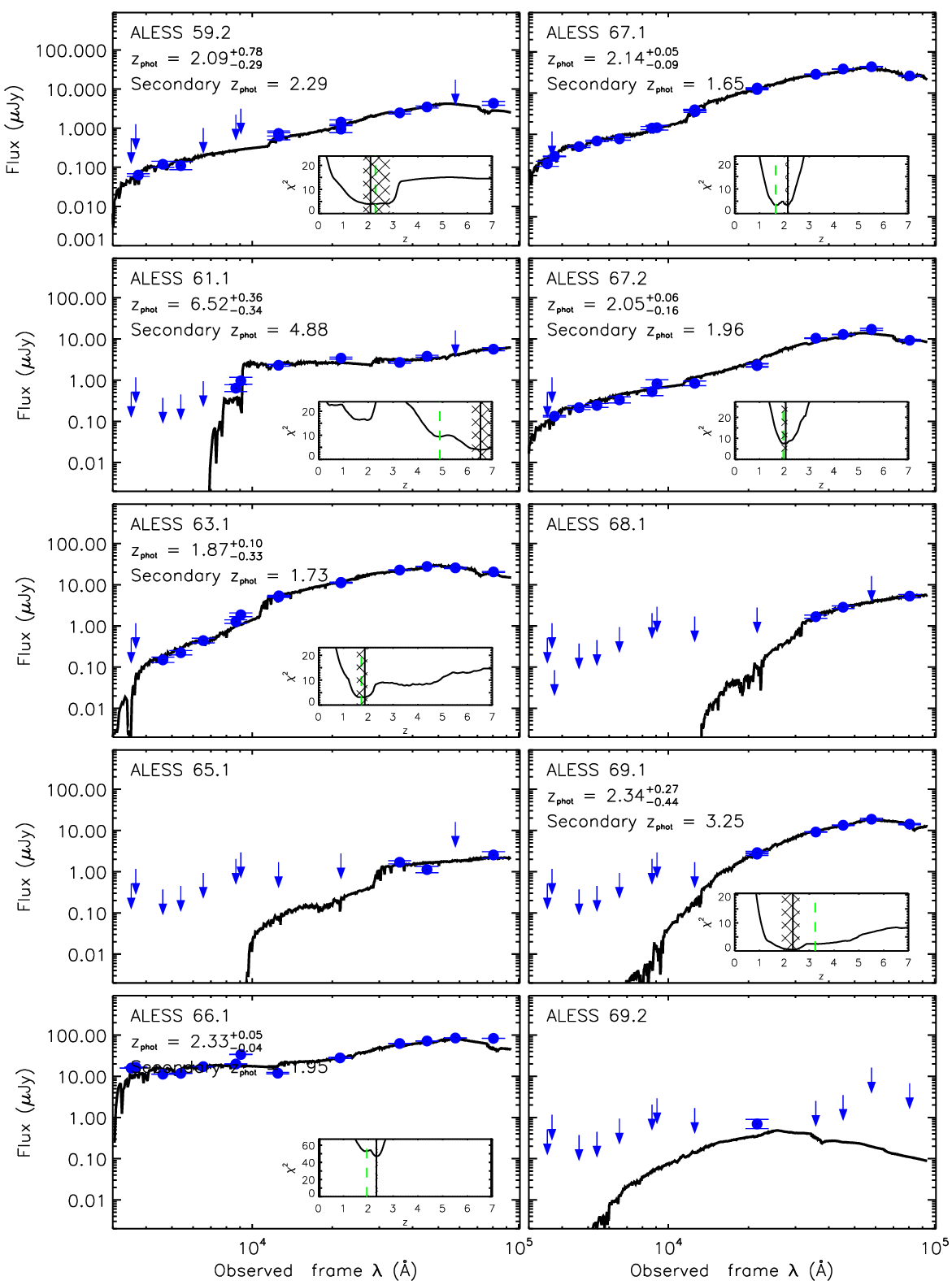,width=0.95\textwidth}}
 \footnotesize{ {\bf Figure A1:} {\it cont.} }
  \label{fig:seds10}
 \end{figure*}
 
 %
 %
 %
 %
 %
 \begin{figure*}
 \centerline{ \psfig{figure= 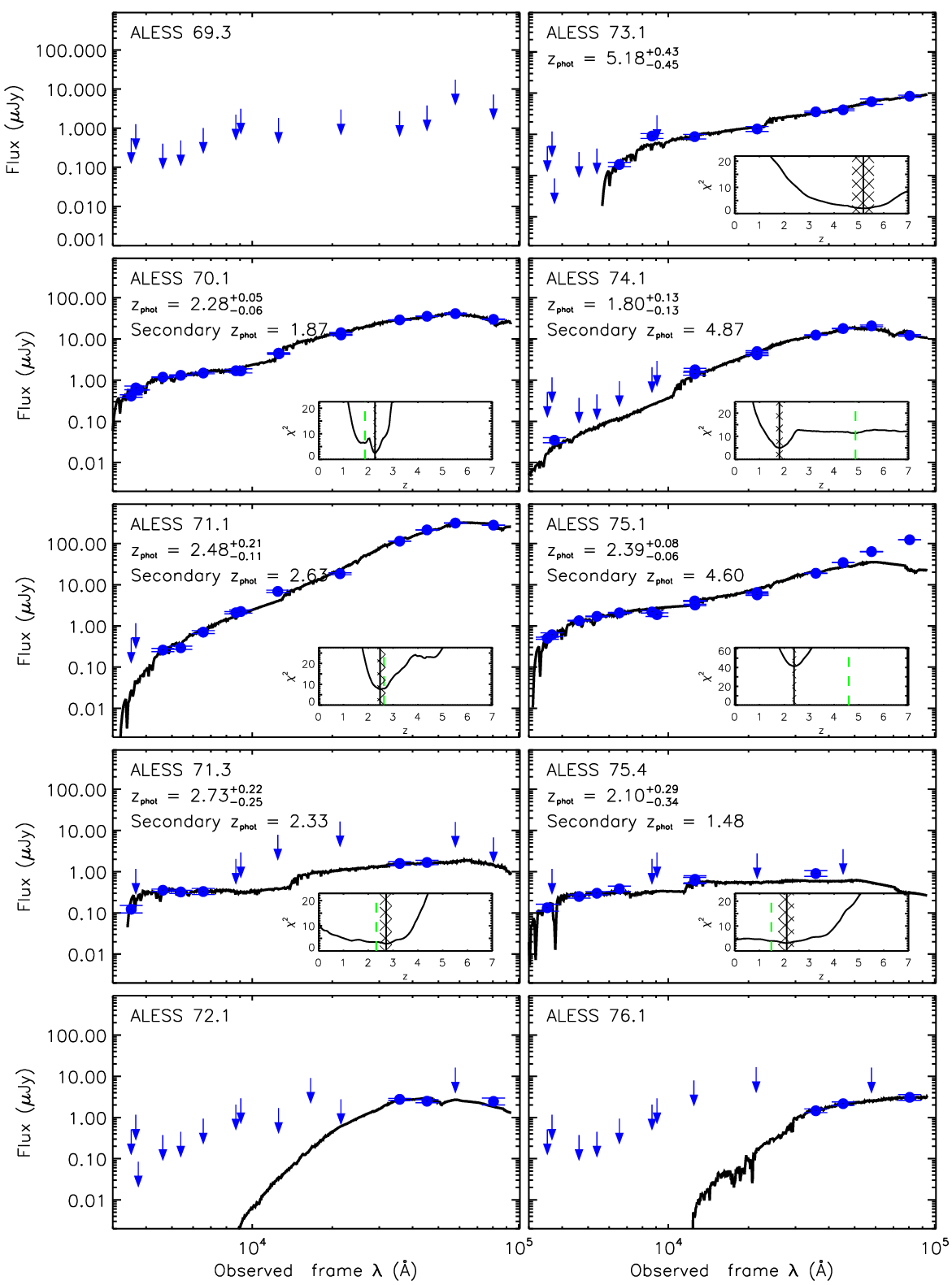,width=0.95\textwidth}}
 \footnotesize{ {\bf Figure A1:} {\it cont.} }
  \label{fig:seds10}
 \end{figure*}
 
 %
 %
 %
 %
 %
 %
 \begin{figure*}
 \centerline{ \psfig{figure= 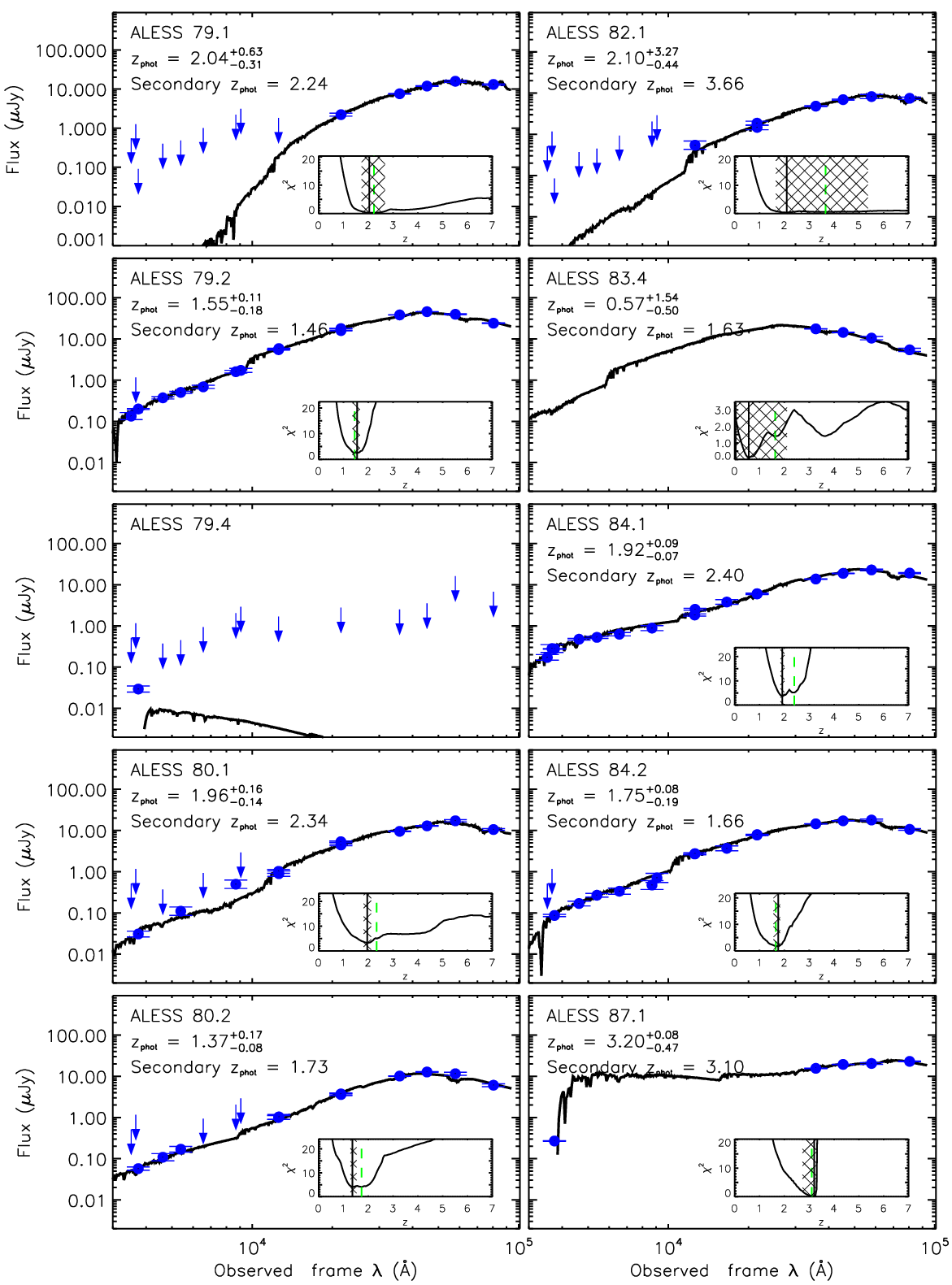,width=0.95\textwidth}}
 \footnotesize{ {\bf Figure A1:} {\it cont.} }
  \label{fig:seds10}
 \end{figure*}
 
 %
 %
 %
 %
 %
 %
 \begin{figure*}
 \centerline{ \psfig{figure= 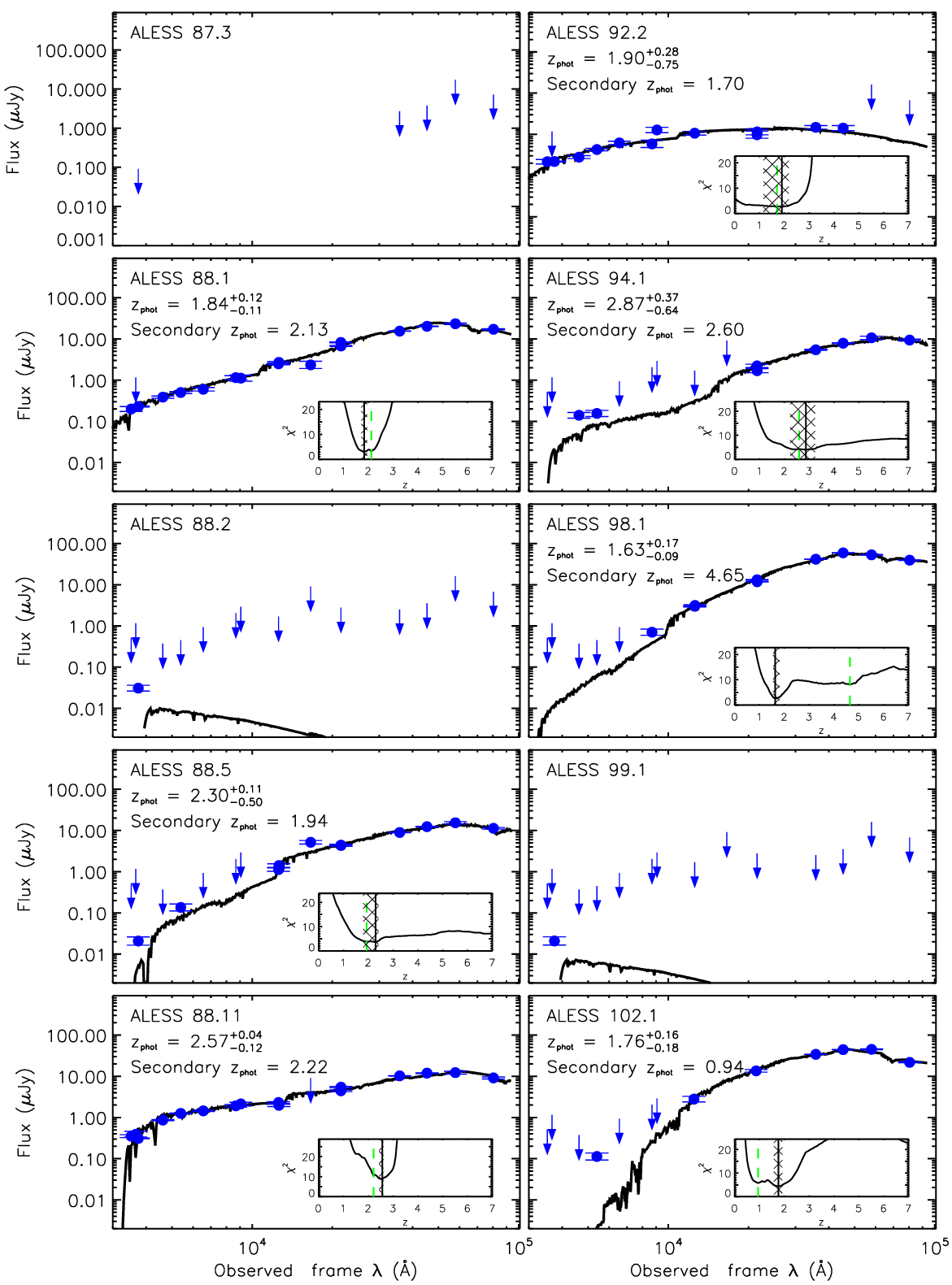,width=0.95\textwidth}}
 \footnotesize{ {\bf Figure A1:} {\it cont.} }
  \label{fig:seds10}
 \end{figure*}
 
 %
 %
 %
 %
 %
 %
 \begin{figure*}
 \centerline{ \psfig{figure= 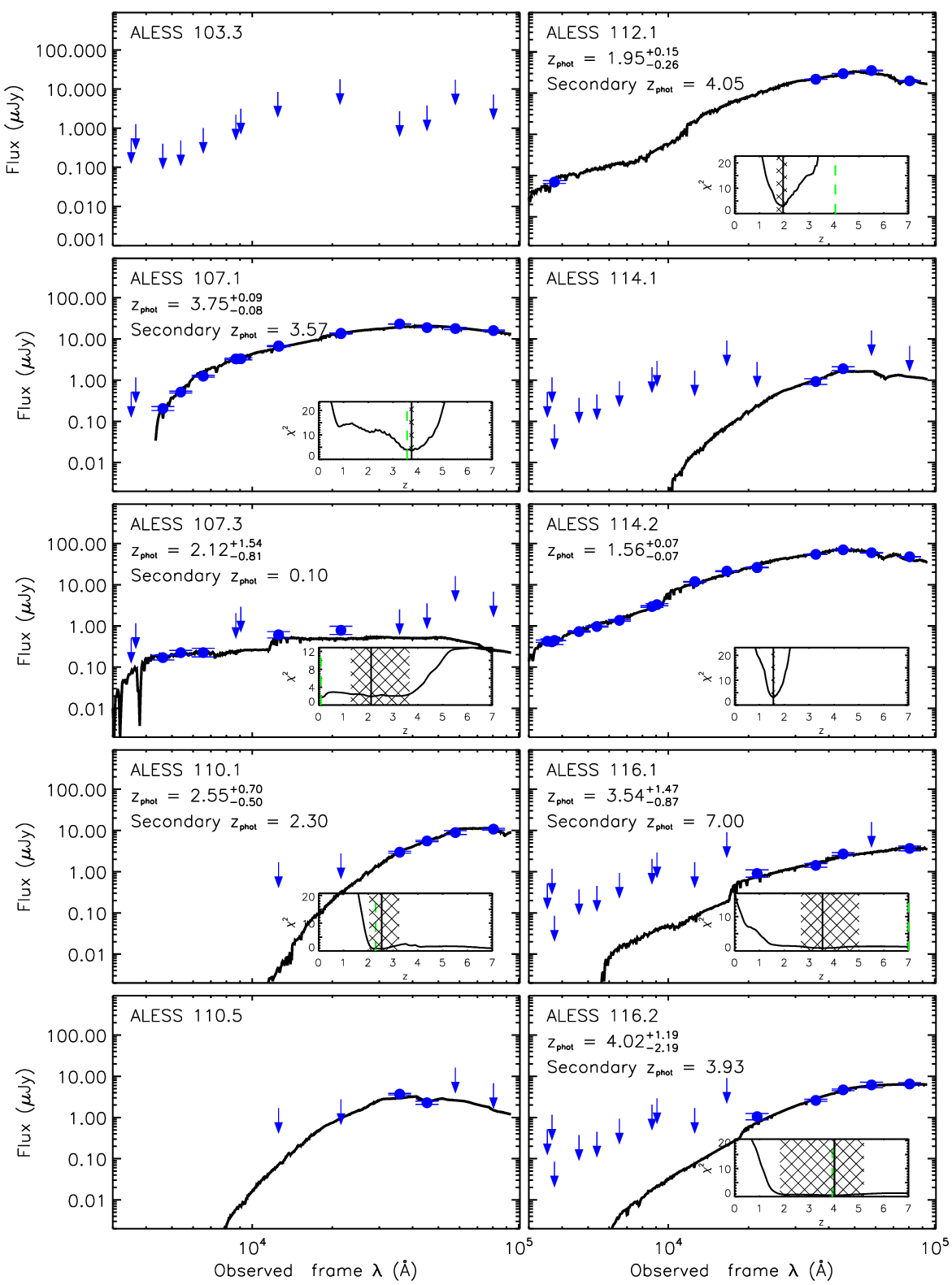,width=0.95\textwidth}}
 \footnotesize{ {\bf Figure A1:} {\it cont.  } }
  \label{fig:seds10}
 \end{figure*}
 
 %
 %
 %
 
 %
 %
 %
 \begin{figure*}
 \centerline{ \psfig{figure= 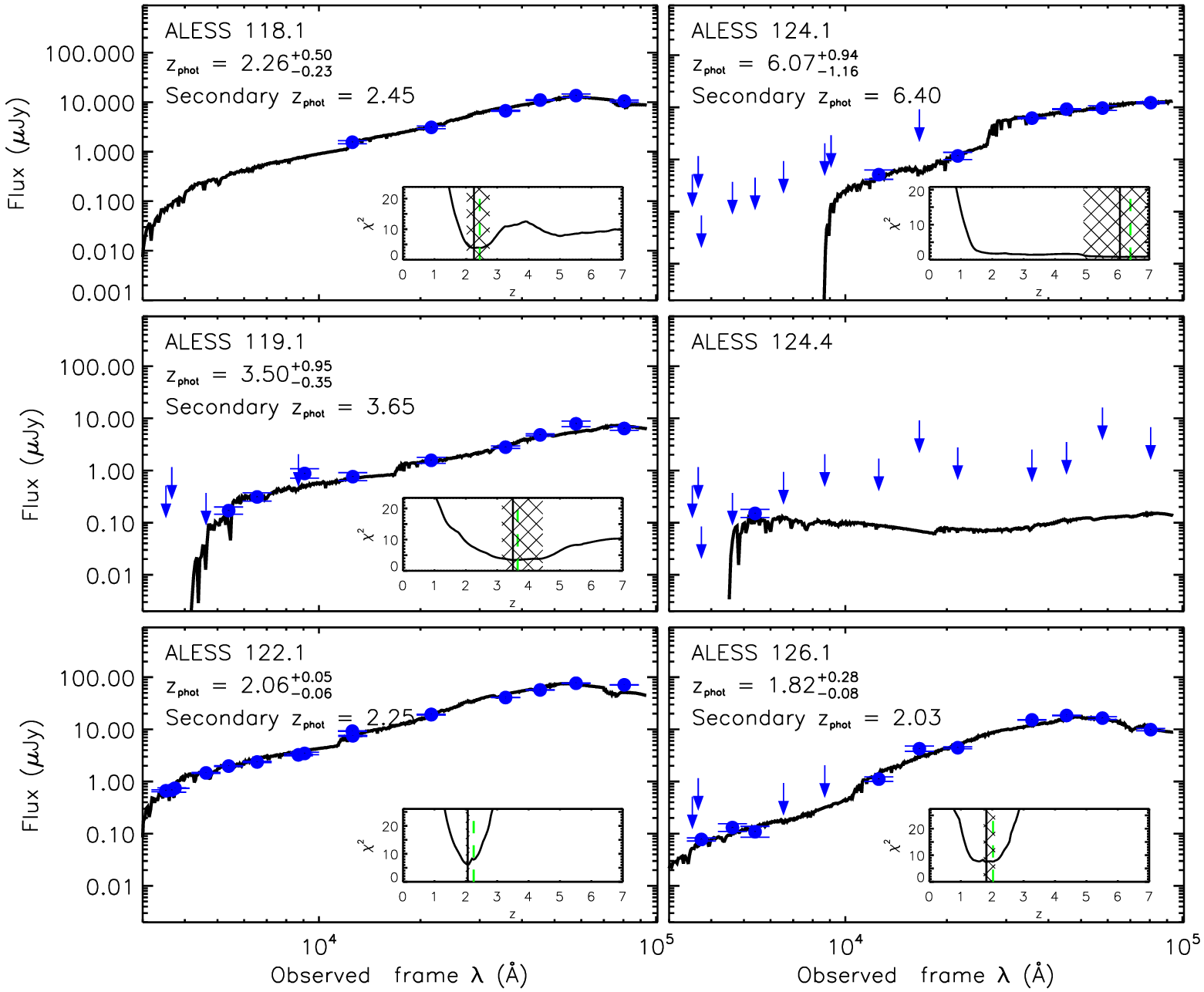,width=0.95\textwidth}}
 \footnotesize{ {\bf Figure A1:} {\it cont.  } }
  \label{fig:seds10}
 \end{figure*}

\section{}
\label{app:C}

\begin{figure*}
\centerline{\sc APPENDIX:B}
\vspace{0.2cm}
\centerline{ \psfig{figure= 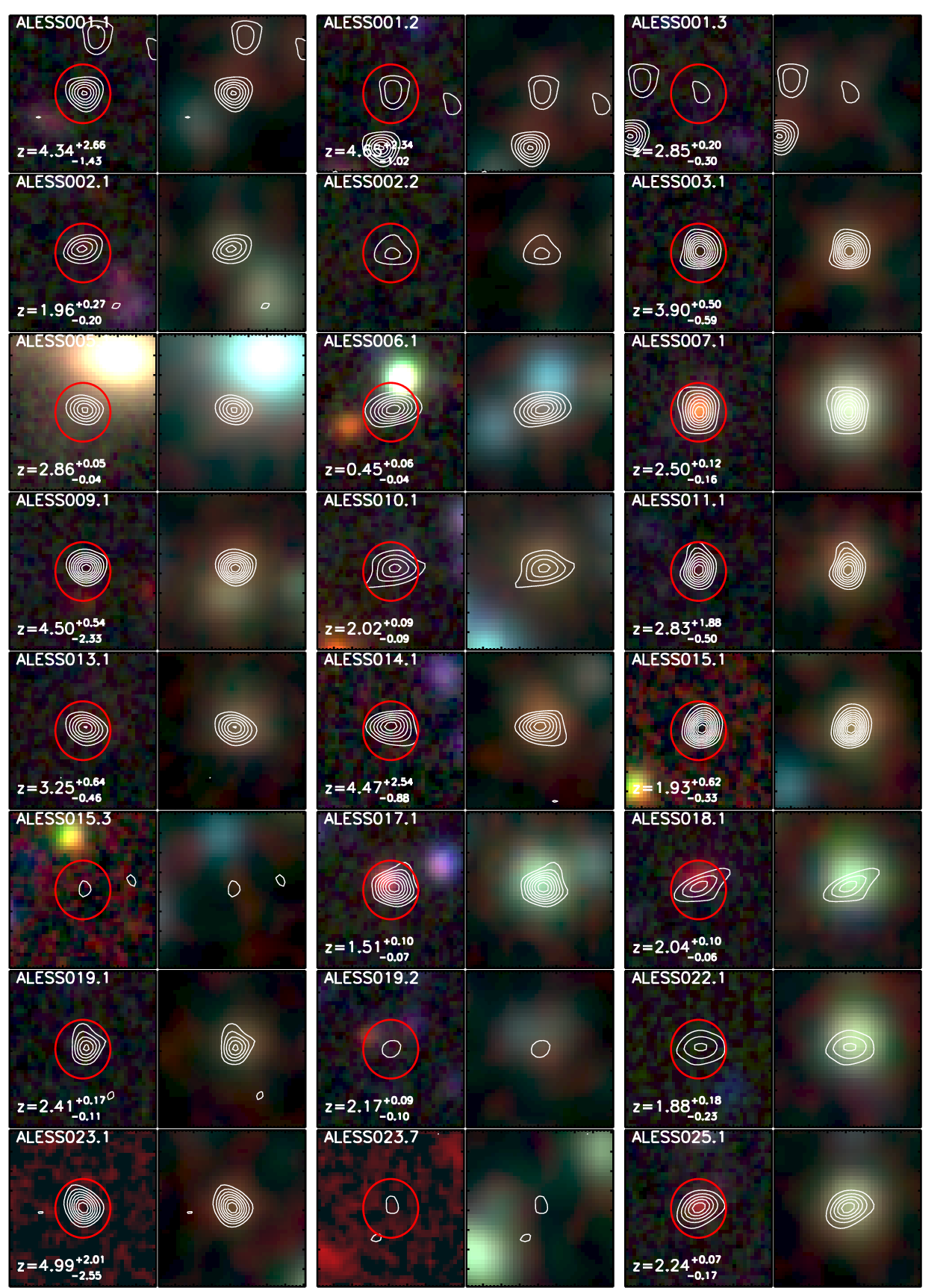,width=0.95\textwidth}}
\footnotesize{ {\bf Figure B1:} Here we show $8'' \times 8''$ $BIK_{s}$ (left) and 3.6$\,\mu m$, 4.5$\,\mu m$ and 8.0$\,\mu m$ (right) color images for each ALESS SMG. Contours indicate 870$\,\mu m$ detections at 3, 5, 7\,....\,$\times \sigma$. The ALESS SMGs are typically aligned with a near-infrared counterpart, and where detected appear red in the optical $BIK_{s}$ images. A red circle represents the aperture used to measure the photometry of each SMG.

}
 \label{fig:stmp1}
\end{figure*}

 \begin{figure*}
 \centerline{ \psfig{figure= 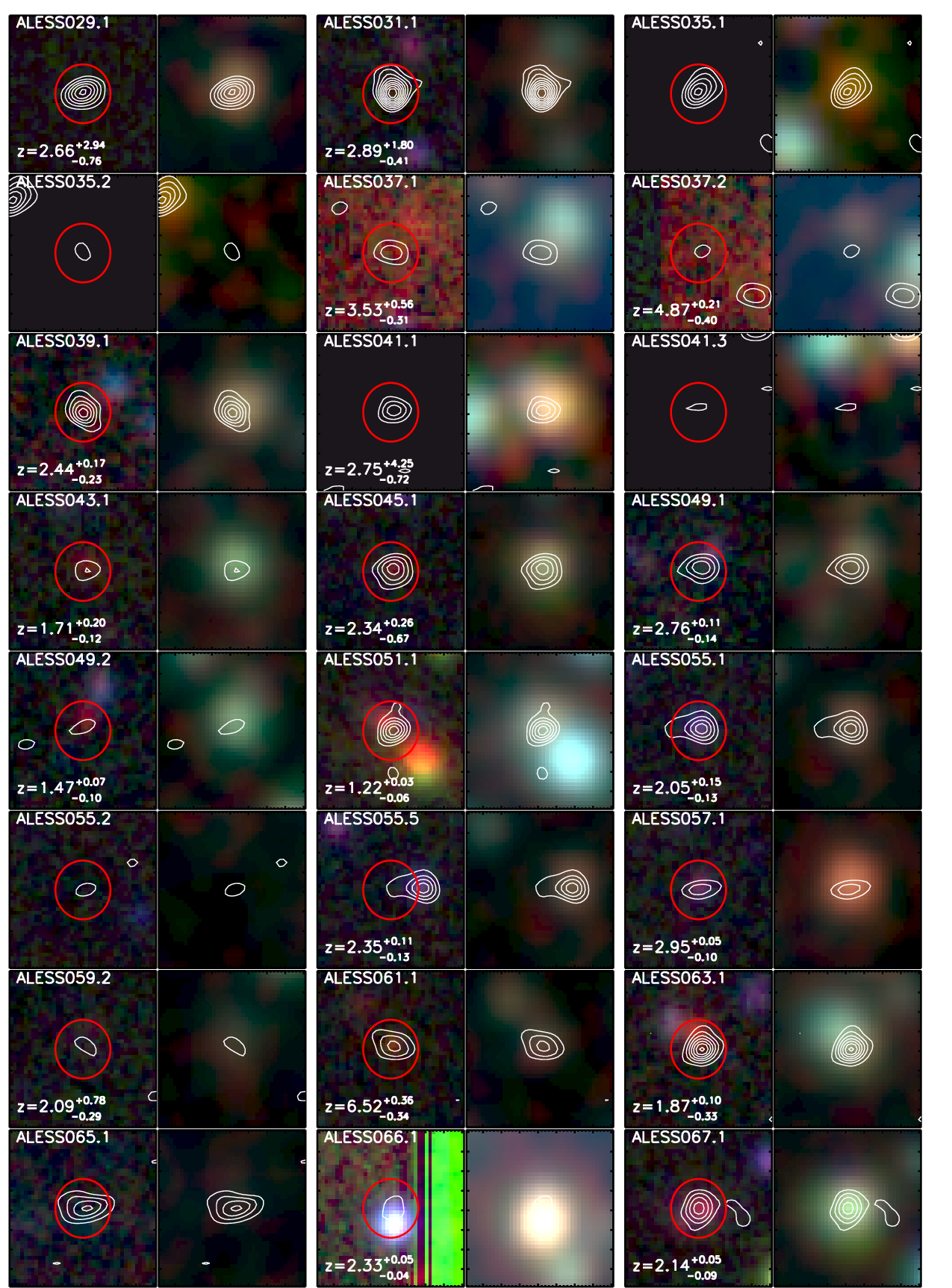,width=0.95\textwidth}}
 \footnotesize{ {\bf Figure B1:} {\it cont.} }
  \label{fig:stmp2}
 \end{figure*}

 \begin{figure*}
 \centerline{ \psfig{figure= 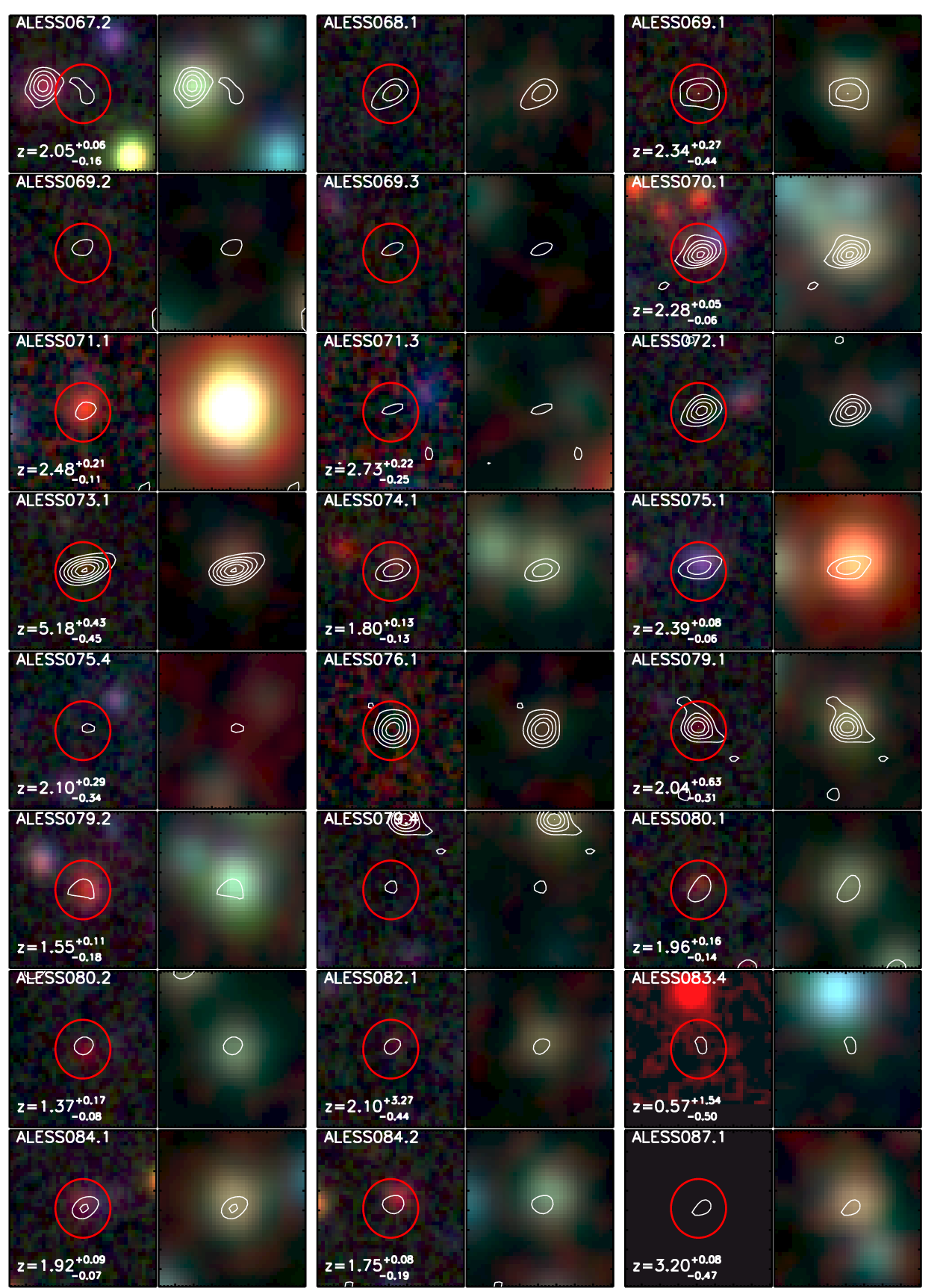,width=0.95\textwidth}}
 \footnotesize{ {\bf Figure B1:} {\it cont.} }
  \label{fig:stmp3}
 \end{figure*}

 \begin{figure*}
 \centerline{ \psfig{figure= 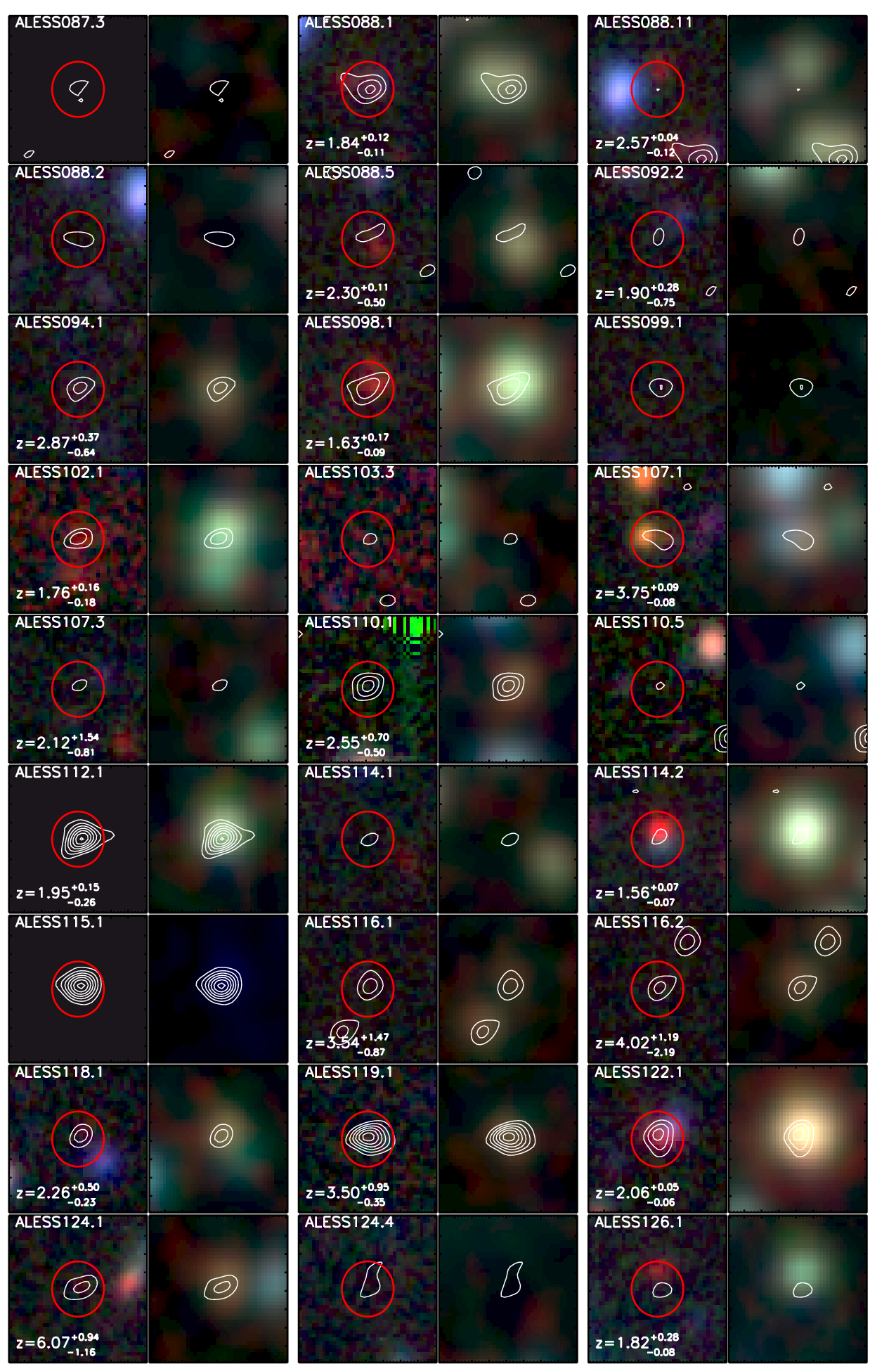,width=0.95\textwidth}}
 \footnotesize{ {\bf Figure B1:} {\it cont.} }
  \label{fig:stmp4}
 \end{figure*}

\clearpage

\section{Appendix:C}
\label{app:supp} Here we consider the {\sc supplementary} catalog of ALESS sources~\citep{Hodge13}. We present SED fits to the measured photometry of the 14 sources with $\ge 4$ detections, and provide the derived properties. As we expect a larger fraction of spurious sources in the {\sc supplementary} catalog, we do not perform any analysis on the non-detections.
%
\begin{figure*}[b]
\centerline{ \psfig{figure= 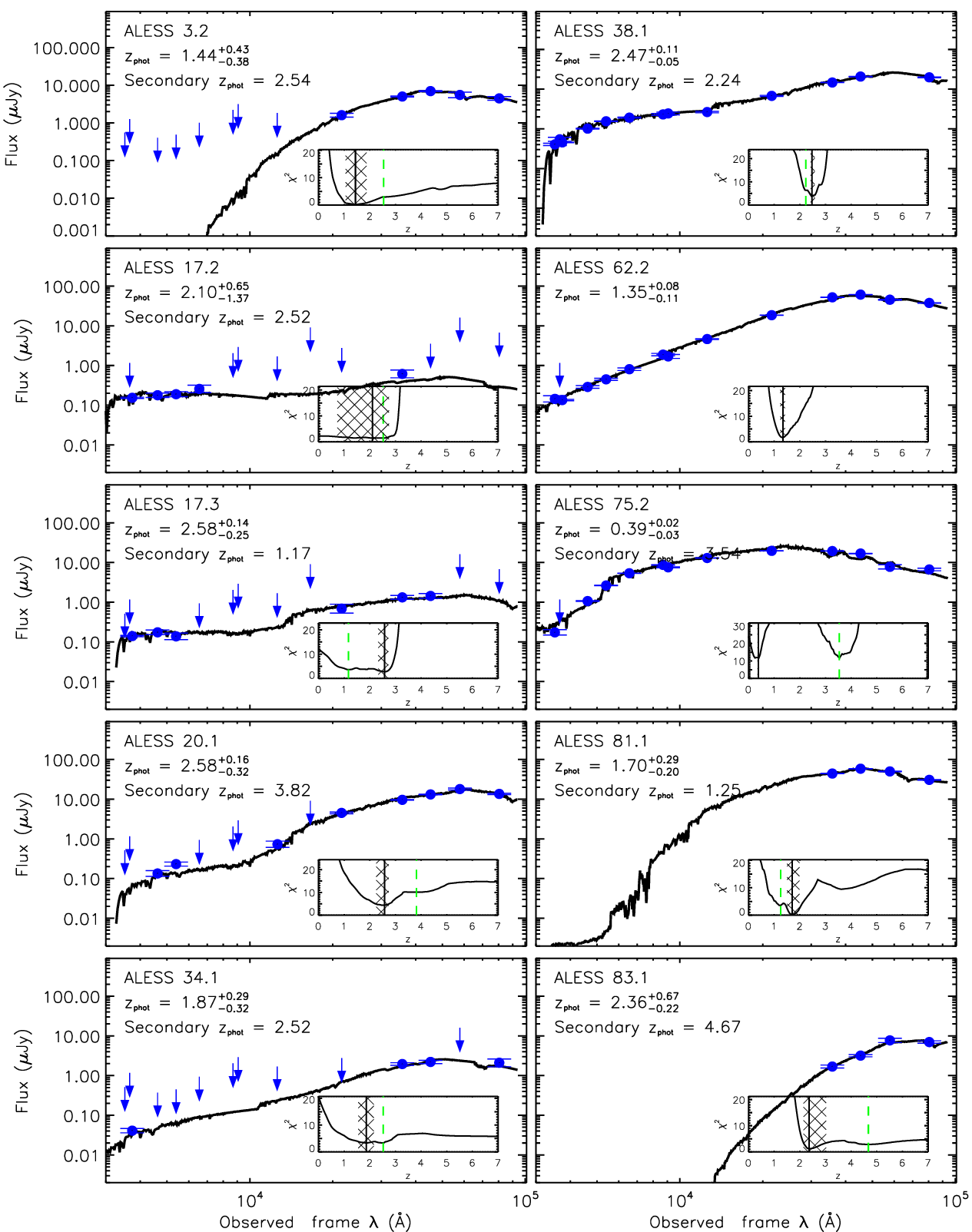,width=0.95\textwidth}}
\footnotesize{ {\bf Figure C1:} The photometry and best fit spectral energy distribution for 14 SMGs from the ALESS {\sc supplementary} catalog~\citep{Hodge13}. Data points and errors are observed photometry, and arrows indicate 3\,$\sigma$ detection limits. In the inset panel in each plot we show the $\chi^2$ distribution as a function of redshift, indicating the best-fit photometric redshift with a solid line, and where appropriate the secondary solution with a green dashed line. The hatched region shows the uncertainty on the derived redshift.
}
\label{fig:sedsmain}
\vspace{-2.cm}
\end{figure*}

\clearpage

%
%
\begin{figure*}
\centerline{ \psfig{figure= 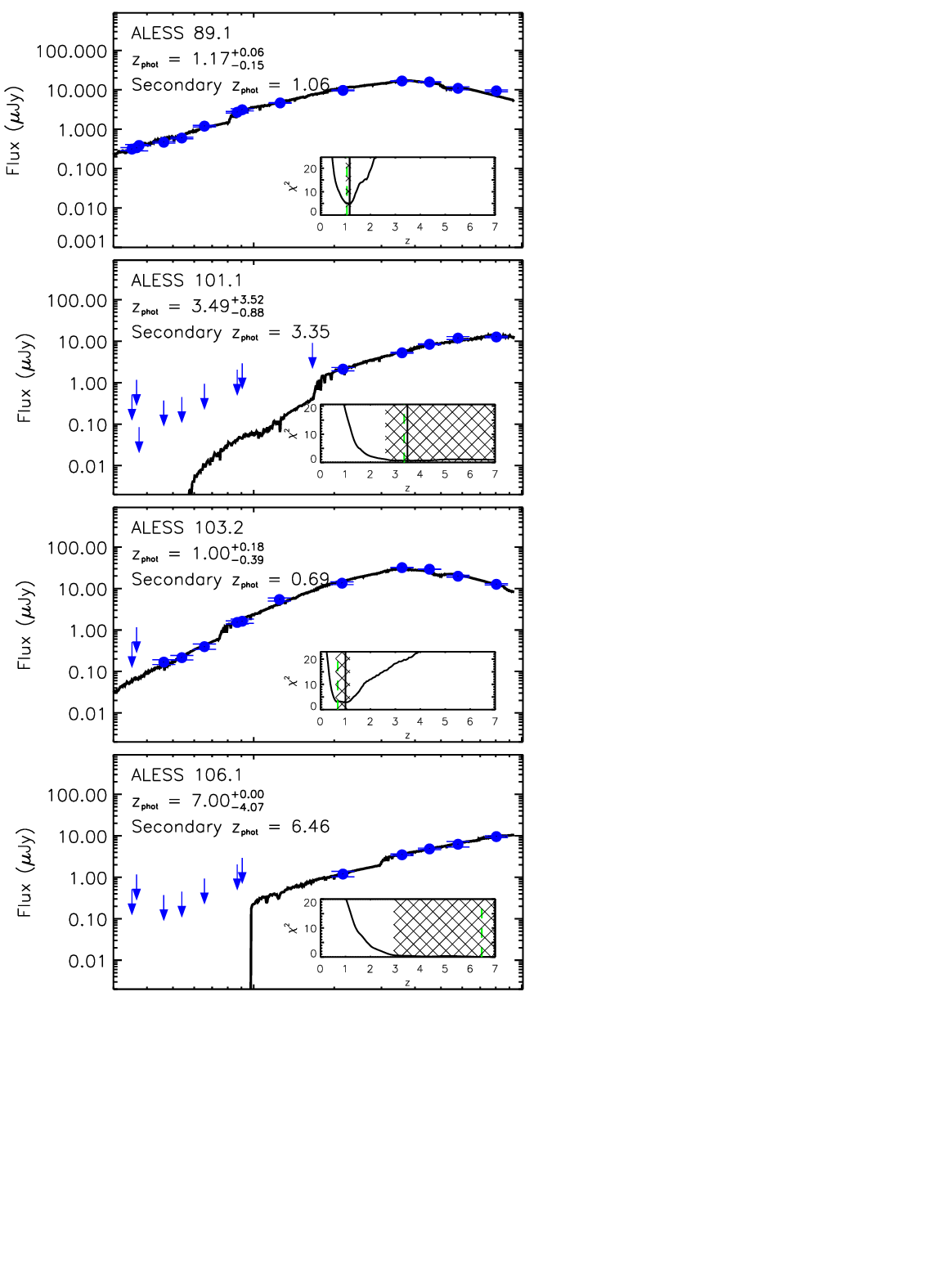,width=0.95\textwidth}}
\footnotesize{ {\bf Figure C1:} {\it cont.  } }
  \label{fig:seds10}
\end{figure*}

 \begin{sidewaystable}
 \centerline{\sc Table C1: Photometry: Supplementary Sources}
\vspace{0.1cm}
 \resizebox{\linewidth}{!}{%
 \tiny
 \begin{tabular}{lccccccccccccccccc}
 \hline
 \noalign{\smallskip}
 ID & MUSYC $U$ & MUSYC $U_{38}$ & VIMOS $U$ & $B$ & $V$ & $R$ & $I$ & $z$ & $J^{a}$ & $H$ & $K^{a}$ & 3.6\,$\mu$m & 4.5\,$\mu$m & 5.8\,$\mu$m & 8.0\,$\mu$m \\
\hline \\ [-1.5ex]  
 ALESS 3.2 &  $>$26.18 & $>$25.29 & ... & $>$26.53 & $>$26.32 & $>$25.53 & $>$24.68 & $>$24.29 & $>$24.88 & ... & 23.39\,$\pm$\,0.12 & 22.15\,$\pm$\,0.03 & 21.79\,$\pm$\,0.03 & 22.05\,$\pm$\,0.16 & 22.27\,$\pm$\,0.08 \\
 ALESS 17.2 &  ... & $>$25.29 & 25.94\,$\pm$\,0.04 & 25.77\,$\pm$\,0.14 & 25.71\,$\pm$\,0.16 & 25.38\,$\pm$\,0.24 & $>$24.68 & $>$24.29 & $>$24.88 & $>$23.06 & $>$24.35 & 24.43\,$\pm$\,0.24 & $>$24.09 & $>$22.44 & $>$23.38 \\
 ALESS 17.3 &  $>$26.18 & $>$25.29 & 26.04\,$\pm$\,0.04 & 25.80\,$\pm$\,0.14 & 26.05\,$\pm$\,0.21 & $>$25.53 & $>$24.68 & $>$24.29 & $>$24.88 & $>$23.06 & 24.30\,$\pm$\,0.26 & 23.60\,$\pm$\,0.12 & 23.51\,$\pm$\,0.15 & $>$22.44 & $>$23.38 \\
 ALESS 20.1 &  $>$26.18 & $>$25.29 & ... & 26.08\,$\pm$\,0.18 & 25.49\,$\pm$\,0.13 & $>$25.53 & $>$24.68 & $>$24.29 & 24.24\,$\pm$\,0.20$^{*}$ & $>$23.06 & 22.27\,$\pm$\,0.06$^{*}$ & 21.44\,$\pm$\,0.02 & 21.10\,$\pm$\,0.02 & 20.76\,$\pm$\,0.05 & 21.07\,$\pm$\,0.03 \\
 ALESS 34.1 &  $>$26.18 & $>$25.29 & 27.37\,$\pm$\,0.14 & $>$26.53 & $>$26.32 & $>$25.53 & $>$24.68 & $>$24.29 & $>$24.88 & ... & $>$24.35 & 23.17\,$\pm$\,0.08 & 23.04\,$\pm$\,0.10 & $>$22.44 & 23.09\,$\pm$\,0.17 \\
 ALESS 38.1 &  24.88\,$\pm$\,0.09 & 24.56\,$\pm$\,0.14 & 24.76\,$\pm$\,0.01 & 23.89\,$\pm$\,0.03 & 23.43\,$\pm$\,0.02 & 23.20\,$\pm$\,0.03 & 22.99\,$\pm$\,0.06 & 22.93\,$\pm$\,0.08 & 22.84\,$\pm$\,0.05 & ... & 21.83\,$\pm$\,0.03 & 20.98\,$\pm$\,0.01 & 20.61\,$\pm$\,0.01 & ... & 20.66\,$\pm$\,0.02 \\
 ALESS 62.2 &  26.00\,$\pm$\,0.23 & $>$25.29 & 26.08\,$\pm$\,0.04 & 25.25\,$\pm$\,0.09 & 24.77\,$\pm$\,0.07 & 24.13\,$\pm$\,0.08 & 23.21\,$\pm$\,0.08 & 23.33\,$\pm$\,0.12 & 22.24\,$\pm$\,0.03 & ... & 20.73\,$\pm$\,0.01 & 19.61\,$\pm$\,0.00 & 19.43\,$\pm$\,0.00 & 19.76\,$\pm$\,0.02 & 19.96\,$\pm$\,0.01 \\
 ALESS 75.2 &  25.80\,$\pm$\,0.19 & $>$25.29 & ... & 23.83\,$\pm$\,0.02 & 22.85\,$\pm$\,0.01 & 22.08\,$\pm$\,0.01 & 21.55\,$\pm$\,0.02 & 21.71\,$\pm$\,0.03 & 21.12\,$\pm$\,0.01 & ... & 20.67\,$\pm$\,0.01 & 20.68\,$\pm$\,0.01 & 20.83\,$\pm$\,0.01 & 21.67\,$\pm$\,0.11 & 21.84\,$\pm$\,0.06 \\
 ALESS 81.1 &  ... & ... & ... & ... & ... & ... & ... & ... & ... & ... & ... & 19.79\,$\pm$\,0.00 & 19.48\,$\pm$\,0.00 & 19.65\,$\pm$\,0.02 & 20.18\,$\pm$\,0.01 \\
 ALESS 83.1 &  ... & ... & ... & ... & ... & ... & ... & ... & ... & ... & ... & 23.33\,$\pm$\,0.09 & 22.64\,$\pm$\,0.07 & 21.68\,$\pm$\,0.11 & 21.79\,$\pm$\,0.05 \\
 ALESS 89.1 &  25.17\,$\pm$\,0.11 & 25.08\,$\pm$\,0.22 & 24.93\,$\pm$\,0.02 & 24.73\,$\pm$\,0.06 & 24.46\,$\pm$\,0.05 & 23.71\,$\pm$\,0.06 & 22.82\,$\pm$\,0.05 & 22.67\,$\pm$\,0.07 & 22.23\,$\pm$\,0.03 & ... & 21.44\,$\pm$\,0.02 & 20.84\,$\pm$\,0.01 & 20.90\,$\pm$\,0.01 & 21.30\,$\pm$\,0.08 & 21.47\,$\pm$\,0.04 \\
 ALESS 101.1 &  $>$26.18 & $>$25.29 & $>$28.14 & $>$26.53 & $>$26.32 & $>$25.53 & $>$24.68 & $>$24.29 & ... & $>$23.06 & 23.08\,$\pm$\,0.09 & 22.10\,$\pm$\,0.03 & 21.59\,$\pm$\,0.03 & 21.21\,$\pm$\,0.08 & 21.15\,$\pm$\,0.03 \\
 ALESS 103.2 &  $>$26.18 & $>$25.29 & ... & 25.84\,$\pm$\,0.15 & 25.57\,$\pm$\,0.14 & 24.91\,$\pm$\,0.16 & 23.45\,$\pm$\,0.09 & 23.37\,$\pm$\,0.12 & 22.07\,$\pm$\,0.10$^{**}$ & ... & 21.08\,$\pm$\,0.09$^{**}$ & 20.14\,$\pm$\,0.01 & 20.24\,$\pm$\,0.01 & 20.66\,$\pm$\,0.05 & 21.15\,$\pm$\,0.03 \\
 ALESS 106.1 &  $>$26.18 & $>$25.29 & ... & $>$26.53 & $>$26.32 & $>$25.53 & $>$24.68 & $>$24.29 & ... & ... & 23.71\,$\pm$\,0.16 & 22.55\,$\pm$\,0.05 & 22.19\,$\pm$\,0.05 & 21.91\,$\pm$\,0.14 & 21.45\,$\pm$\,0.04 \\
 \hline\hline
 \end{tabular}}
 \vspace{-0.3cm}
 \refstepcounter{table}\label{table:observed_supp}
 \begin{flushleft}
  \footnotesize{ $3\,\sigma$ upper limits are presented for non-detections, and photometry is left blank where a source is not covered by available imaging. $^a$ We measure J and K photometry from three imaging surveys, but quote a single value, in order of $3\,\sigma$ detection limit. $^*$ Photometry measured from HAWK-I imaging. $^{**}$ Photometry measured from MUSYC imaging. }
 \end{flushleft}
 
 \end{sidewaystable}

 \begin{table*}
 \centering
 \centerline{\sc Table C2: Derived Properties: Supplementary Sources}
\vspace{0.1cm}
 {%
 \small
 \begin{tabular}{lccccccccccccccccc}
 \hline
 \noalign{\smallskip}
ID & RA & Dec  & $z_{phot}$ & $\chi^{2}_{red}$ & Filters & $M_{H}$ & $M\,/\,L_{H}$  \\
 & (J2000) & (J2000) &  &  & (Det [Obs]) & (AB) &  ($M_{\odot} L_{\odot}^{-1}$) \\  [0.5ex]  
\hline  \\  [-1.5ex]  
ALESS 003.2  & 03:33:22.19 & $-$27:55:20.9 & 1.44$^{+0.43}_{-0.38}$ & 0.23 & 5 [13] & $-$22.34 & 0.36   \\
ALESS 017.2  & 03:32:08.26 & $-$27:51:19.7 & 2.10$^{+0.65}_{-1.37}$ & 1.43 & 5 [14] & $-$20.76 & 0.15  \\
ALESS 017.3  & 03:32:07.37 & $-$27:51:33.9 & 2.58$^{+0.14}_{-0.25}$ & 2.79 & 6 [15] & $-$21.97 & 0.15  \\
ALESS 020.1  & 03:33:16.76 & $-$28:00:16.0 & 2.58$^{+0.16}_{-0.32}$ & 4.38 & 8 [14] & $-$24.74 & 0.18 \\
ALESS 034.1  & 03:32:17.96 & $-$27:52:33.3 & 1.87$^{+0.29}_{-0.32}$ & 3.32 & 4 [14] & $-$21.70 & 0.15  \\
ALESS 038.1  & 03:33:10.84 & $-$27:56:40.2 & 2.47$^{+0.11}_{-0.05}$ & 3.79 & 13 [13] & $-$24.88 & 0.05  \\
ALESS 062.2  & 03:32:36.58 & $-$27:34:53.8 & 1.35$^{+0.08}_{-0.11}$ & 1.68 & 13 [14] & $-$24.59 & 0.05  \\
ALESS 075.2  & 03:31:27.67 & $-$27:55:59.2 & 0.39$^{+0.02}_{-0.03}$ & 11.44 & 12 [13] & $-$20.84 & 0.18  \\
ALESS 081.1  & 03:31:27.55 & $-$27:44:39.6 & 1.70$^{+0.29}_{-0.20}$ & 0.23 & 4 [4] & $-$25.04 & 0.36  \\
ALESS 083.1  & 03:33:09.42 & $-$28:05:30.6 & 2.36$^{+0.67}_{-0.22}$ & 1.22 & 4 [4] & $-$23.59 & 0.36  \\
ALESS 089.1  & 03:32:48.69 & $-$28:00:21.9 & 1.17$^{+0.06}_{-0.15}$ & 4.64 & 14 [14] & $-$22.88 & 0.17  \\
ALESS 101.1  & 03:31:51.60 & $-$27:45:53.0 & 3.49$^{+3.52}_{-0.88}$ & 0.59 & 5 [14] & $-$25.15 & 0.18 \\
ALESS 103.2  & 03:33:25.82 & $-$27:34:09.9 & 1.00$^{+0.18}_{-0.39}$ & 2.91 & 11 [13] & $-$23.24 & 0.08 \\
ALESS 106.1  & 03:31:39.64 & $-$27:56:39.2 & 7.00$^{+0.00}_{-4.07}$ & 0.02 & 5 [12] & $-$26.31 & 0.11  \\ [0.5ex]  
 \hline\hline \\ [-0.5ex]  
 \end{tabular}}
 \vspace{-0.3cm}
 \refstepcounter{table}\label{table:derived_supp}
 
 \end{table*}

\end{document}